\documentclass[a4paper,onecolumn,10pt,accepted=2025-07-26]{quantumarticle}
\pdfoutput=1

\usepackage[margin=0.9in]{geometry}

\usepackage{amsmath,amssymb,amsfonts,setspace,url,enumitem}

\usepackage{amsthm}
\usepackage{tikz}
\usepackage{color}
\usepackage{bigstrut}
\usepackage{multirow}
\usepackage{braket}
\usepackage{ascmac}
\usepackage{authblk}

\usepackage{algorithm}
\usepackage{algorithmicx}
\usepackage{algpseudocode}

\makeatletter
\newcommand{\newreptheorem}[2]{\newtheorem*{rep@#1}{\rep@title}\newenvironment{rep#1}[1]{\def\rep@title{#2 \ref*{##1}}\begin{rep@#1}}{\end{rep@#1}}}
\makeatother

\makeatletter
\newcommand{\newconjecture}[2]{\newconjecture*{rep@#1}{\rep@title}\newenvironment{rep#1}[1]{\def\rep@title{#2 \ref*{##1}}\begin{rep@#1}}{\end{rep@#1}}}
\makeatother
\usepackage{comment}
\usepackage{amsthm}
\usepackage{quantikz}
\usepackage{tikz}
\usepackage{mathabx}

\providecommand{\calA}{\ensuremath{\mathcal{A}}}

\providecommand{\calC}{\ensuremath{\mathcal{C}}}
\providecommand{\calD}{\ensuremath{\mathcal{D}}}

\providecommand{\calJ}{\ensuremath{\mathcal{J}}}

\providecommand{\calL}{\ensuremath{\mathcal{L}}}

\providecommand{\calN}{\ensuremath{\mathcal{N}}}
\providecommand{\calO}{\ensuremath{\mathcal{O}}}
\providecommand{\calP}{\ensuremath{\mathcal{P}}}

\providecommand{\calU}{\ensuremath{\mathcal{U}}}

\newcommand{\propref}[1]{\hyperref[#1]{Proposition~\ref{#1}}}
\newcommand{\corref}[1]{\hyperref[#1]{Corollary~\ref{#1}}}
\newcommand{\defref}[1]{\hyperref[#1]{Definition~\ref{#1}}}
\newcommand{\secref}[1]{\hyperref[#1]{Section~\ref{#1}}}
\newcommand{\appref}[1]{\hyperref[#1]{Appendix~\ref{#1}}}
\newcommand{\chapref}[1]{\hyperref[#1]{Chapter~\ref{#1}}}
\newcommand{\thmref}[1]{\hyperref[#1]{Theorem~\ref{#1}}}
\newcommand{\lemref}[1]{\hyperref[#1]{Lemma~\ref{#1}}}
\newcommand{\figref}[1]{\hyperref[#1]{Fig.~\ref{#1}}}
\renewcommand{\eqref}[1]{\hyperref[#1]{Eq. (\ref{#1})}}
\newcommand{\ineqref}[1]{\hyperref[#1]{Inequality (\ref{#1})}}
\newcommand{\tableref}[1]{\hyperref[#1]{Table (\ref{#1})}}
\renewcommand{\epsilon}{\varepsilon}

\newtheorem{innercustomgeneric}{\customgenericname}
\providecommand{\customgenericname}{}
\newcommand{\newcustomtheorem}[2]{%
  \newenvironment{#1}[1]
  {%
   \renewcommand\customgenericname{#2}%
   \renewcommand\theinnercustomgeneric{##1}%
   \innercustomgeneric
  }
  {\endinnercustomgeneric}
}

\newcustomtheorem{customthm}{Theorem}
\newcustomtheorem{customlemma}{Lemma}

\usepackage{booktabs}
\usepackage{algorithm}
\usepackage{amssymb}
\usepackage{amsmath}
\usepackage[colorlinks=true, urlcolor=blue, citecolor=blue]{hyperref}
\DeclareMathOperator{\Inf}{Inf}

\usepackage{xcolor}
\definecolor{armando}{rgb}{.2,.1,.5}

\usepackage[numbers]{natbib}

\theoremstyle{plain}
\newtheorem{thm}{Theorem}[section]
\theoremstyle{plain}
\newtheorem*{thm*}{Theorem}
\theoremstyle{plain}
\newtheorem{claim}{Claim}[section]
\theoremstyle{plain}

\theoremstyle{plain}
\newtheorem{prop}{Proposition}[section]
\theoremstyle{plain}
\theoremstyle{definition}
\newtheorem{definition}{Definition}[section]
\theoremstyle{plain}
\newtheorem{lem}{Lemma}[section]
\theoremstyle{plain}
\newtheorem{coro}{Corollary}[section]
\theoremstyle{plain}

\theoremstyle{definition}
\newtheorem{example}{Example}[section]
\theoremstyle{remark}
\newtheorem{remark}{Remark}[section]

\newcommand{\norm}[1]{\left\|#1\right\|} 

\newcommand{\Tr}{\mathrm{Tr}} 
\newcommand{\abs}[1]{\left|#1\right|}

\newcommand{\tr}{\mathrm{tr}}

\newcommand{\ketbra}[2]{|{#1}\rangle\langle{#2}|}

\newcommand{\poly}{\mathrm{poly}}

\begin{document}

\title{Learning unitaries with quantum statistical queries}
\author{Armando Angrisani}

\affil{LIP6, Centre National de la Recherche Scientifique (CNRS), Sorbonne Université, 75005 Paris, France
\\ Institute of Physics, Ecole Polytechnique Fédérale de Lausanne (EPFL),  Lausanne CH-1015, Switzerland \\ Center for Quantum Science and Engineering, Ecole Polytechnique Fédérale de Lausanne (EPFL), CH-1015 Lausanne, Switzerland }

\maketitle
\thispagestyle{empty}
\setcounter{page}{1}
\begin{abstract}
We propose several algorithms for learning unitary operators from quantum statistical queries 
with respect to their Choi-Jamiolkowski state.  
Quantum statistical queries capture the capabilities of a learner with limited quantum resources, which receives as input only noisy estimates of expected values of measurements. 
Our approach leverages quantum statistical queries to estimate the Fourier mass of a unitary on a subset of Pauli strings, generalizing previous techniques developed for uniform quantum examples.
Specifically, we show that the celebrated quantum Goldreich-Levin algorithm can be implemented with quantum statistical queries, whereas the prior version of the algorithm involves oracle access to the unitary and its inverse. As an application, we prove that quantum Boolean functions with constant total influence or with constant degree are efficiently learnable in our model.
Moreover, we prove that $\mathcal{O}(\log n)$-juntas are efficiently learnable and constant-depth circuits are learnable query-efficiently with quantum statistical queries. On the other hand, all previous algorithms for these tasks demand significantly greater resources, such as oracle access to the unitary or direct access to the Choi-Jamiolkowski state.
We also demonstrate that, despite these positive results, quantum statistical queries lead to an exponentially larger query complexity for certain tasks, compared to separable measurements to the Choi-Jamiolkowski state. In particular, we show an exponential lower bound for learning a class of phase-oracle unitaries and a double exponential lower bound for testing the unitarity of channels.
Taken together, our results indicate that quantum statistical queries offer a unified framework for various unitary learning tasks, with potential applications in quantum machine learning, many-body physics and benchmarking of near-term devices.
\end{abstract}

\newpage

{
  \hypersetup{linkcolor=blue}
  \tableofcontents
}

\section{Introduction}
Learning the dynamic properties of quantum systems is a fundamental problem at the intersection of machine learning (ML) and quantum physics. In the most general case, this task can be achieved under the broad framework of quantum process tomography (QPT)~\cite{chuang1997prescription, bisio2010optimal, caro2022learning}. However, QPT can be extremely resource-intensive, as learning the entire classical description of a unitary transformation requires exponentially many queries in the worst case~\cite{gutoski2014process}. This complexity can be significantly reduced if the unitary is not completely arbitrary, but instead it belongs to a specific class. For instance, this approach has been fruitfully adopted for quantum Boolean functions~\cite{montanaro2010quantum}, quantum juntas~\cite{chen2023testing, bao2025testing} and quantum circuits with bounded covering numbers~\cite{fanizza2024learning} or {\color{black} bounded gate complexity~\cite{zhao2024learning}.
Several studies have explored different classes of shallow circuits~\cite{nadimpalli2024pauli, huang2024learning, vasconcelos2024learning} and circuits with a limited number of non-Clifford gates~\cite{lai2022learning}. Alternatively, other works have investigated more heuristic approaches, including variational methods~\cite{xue2022variational, holmes2021barren} and tensor networks~\cite{torlai2023quantum}.}

Notably, the complexity of quantum process tomography can also be significantly reduced by focusing solely on local properties of the output state rather than requiring a complete description, as recently demonstrated in Ref.~\cite{huang2022learning}.
Another scenario of interest is property testing, where the goal is not to retrieve the classical description of the target process but rather to \emph{test} whether it satisfies a specific property~\cite{montanaro2013survey}.
A further figure of merit in quantum process learning is the type of resources that the learner is allowed to use.
For the special case of unitary transformations, the learner is usually given oracle access to the target unitary $U$ and its inverse $U^\dag$, or, alternatively, to the corresponding Choi-Jamiolkowski state.
In this paper we consider this latter approach and we ask the following question:
\begin{center}
   \emph{Which classes of unitaries are efficiently learnable with noisy \\ separable binary measurements of the Choi-Jamiolkowski state?} 
\end{center}

This question is motivated by near-term implementations of quantum algorithms, which are constrained by several sources of noise and severely limited entangling capacity~\cite{preskill2018quantum}. To address this, we adopt the model of \emph{quantum statistical queries} (QSQs), introduced in Ref.~\cite{arunachalam2020quantum} as a quantum analogue of the classical statistical query model~\cite{kearns1998efficient}. In the QSQ model, a learner without quantum memory can only access noisy estimates of the expected values of chosen observables on an unknown initial state. As noted in Ref.~\cite{arunachalam2023on}, this is essentially equivalent to performing noisy separable binary measurements.

The study of {classical} statistical query learning is primarily motivated by the need for noise tolerance in machine learning and data analysis. Likewise, in near-term quantum computing, various noise sources -- such as shot noise in measurements and hardware imperfections -- affect different stages of computation, highlighting the need for a quantum counterpart to classical statistical queries. Current implementations of quantum algorithms are further constrained by the absence of key features like complex entangling unitaries and the ability to perform measurements on multiple quantum states simultaneously to extract joint statistics. The quantum statistical query model, proposed by \citet{arunachalam2020quantum}, addresses both of these limitations.

Interestingly, several concept classes such as parities, juntas, and DNF formulae are efficiently learnable in the QSQ model, whereas the classical statistical query model necessitates an exponentially larger number of queries~\cite{arunachalam2020quantum}. Despite these positive results, resorting to quantum statistical queries can be considerably limiting for some tasks. In particular, the authors of~\cite{arunachalam2023on} have established an exponential gap between QSQ learning and learning with quantum examples in the presence of classification noise.
{\color{black}

Quantum statistical queries have also found practical applications beyond learning theory. They have been employed in classical verification of quantum learning~\cite{caro2024classical}, quantum neural networks~\cite{du2020learnability}, and quantum error mitigation models~\cite{quek2022exponentially, arunachalam2023on}. Specifically, they can capture interesting computational tasks such as \emph{weak} quantum error mitigation, where the goal is to estimate the expectation values of observables on a noisy quantum state up to an additive error with high probability. This task can be seen as simulating a quantum statistical query oracle for the state~\cite{quek2022exponentially}. Additionally, quantum statistical queries have been used to investigate tasks in quantum cryptanalysis, such as the forging attack proposed by \citet{wadhwa2023learning}, which breaks the security of a class of physical unclonable functions.

In previous works, the quantum statistical query framework has primarily been applied to learning classical functions encoded in quantum states and specific families of quantum states\ \cite{arunachalam2020quantum, arunachalam2023on}. In contrast, we explore the learnability of unitaries using quantum statistical queries, showcasing the broader applicability of this learning model. Going forward, even if fault-tolerant quantum devices become available, the storage and transmission of quantum data will likely remain constrained by hardware limitations. In this context, quantum statistical queries could serve as a powerful unifying framework for studying what can be learned by combining simple measurements with classical post-processing.}

\paragraph{Our contributions.}
In this paper we demonstrate that several classes of unitaries are efficiently learnable with quantum statistical queries with respect to their Choi state. In particular, we show our result for a natural distance over unitaries induced by the Choi-Jamiolkowski isomorphism and previously adopted in Refs.~\cite{montanaro2013survey, montanaro2010quantum, chen2023testing, bao2025testing}. We emphasize that this choice of distance allows to predict the action of the target unitary on a random input state sampled from a locally scrambled ensemble~\cite{caro2022outofdistribution}.
We now give an informal version of our upper bounds. When not explicitly stated, the tolerance of a quantum statistical query is at least polynomially small.
\begin{itemize}
    \item Quantum Boolean functions with constant degree (Theorem~\ref{thm:GL-low-d}) or constant total influence are efficiently learnable with polynomially many quantum statistical queries (Theorem~\ref{thm:gl-accuracy}). In order to prove these results, we show that the quantum Goldreich-Levin algorithm can be implemented with quantum statistical queries (Theorem~\ref{thm:GL}).
    \item Quantum $\mathcal{O}(\log n)$-juntas are efficiently learnable with polynomially many quantum statistical queries (Theorem~\ref{thm:juntas}).
    \item Constant-depth circuits are learnable with polynomially many quantum statistical queries  (Theorem~\ref{thm:constant}).
\end{itemize}
While these positive results show that a wide class of unitaries can be efficiently learned in our model, we also argue that resorting to quantum statistical queries leads to an exponentially larger query complexity for certain tasks. In particular, we give the following lower bound.
\begin{itemize}
    \item There is a class of phase oracle unitaries that requires exponentially many quantum statistical queries with polynomially small tolerance to be learnt below distance $0.05$ with high probability (Theorem~\ref{thm:hardness});
    \item Estimating the unitarity of a quantum channel with error smaller than $0.24$ and polynomially small tolerance requires double-exponentially many quantum statistical queries (Corollary~\ref{coro:unitarity}).
\end{itemize}
Moreover, prior results imply that both tasks can be efficiently performed with polynomially many copies of the associated Choi-Jamiolkowski state.
Finally, in Section~\ref{sec:surrogates} we suggest potential applications of our results to quantum machine learning, many-body physics and benchmarking of quantum devices. 

\paragraph{Related work.}
Our results generalize prior work in two ways. On one hand, we show that several classes of unitaries are learnable in the QSQ model, while all previous results involved the access to stronger oracles. The adoption of a weaker oracle is particularly advantageous for near-term implementation, since the definition of QSQs accounts for the measurement noise.
On the other hand, we demonstrate that prior QSQ algorithms for learning classical Boolean functions can be generalized to unitary learning.
In particular, \citet{chen2023testing} showed that $k$-junta unitaries are learnable with $\mathcal{O}(4^k)$ copies of the Choi state, and \citet{montanaro2010quantum} proposed the original version of the quantum Goldreich-Levin algorithm, requiring oracle access to the target unitary and its inverse. 

Furthermore, \citet{atici2007quantum} provided an algorithm for learning classical $k$-junta functions with $\mathcal{O}(2^k)$ uniform quantum examples, and \citet{arunachalam2020quantum} demonstrated that several classes of quantum Boolean functions are learnable with quantum statistical queries with respect to uniform quantum examples. In particular, they showed that classical $k$-junta functions are learnable with $\mathcal{O}(2^k+n)$ quantum statistical queries, and moreover that the (classical) Goldreich-Levin algorithm can be implemented in the QSQ model.
In a subsequent work, \citet{arunachalam2023on} showed that the output of constant-depth circuits is learnable with $\poly(n)$ quantum statistical queries and provided several hardness results for the QSQ model. Specifically, they showed an exponential lower bound for learning a class of classical Boolean functions, and a double exponential lower bound for testing the purity of a target state. 

\medskip

\noindent\emph{Added note.} A simultaneous work from \citet{wadhwa2023learning} devised a general QSQ oracle for learning quantum processes, where a learner can select both the input state and the measurement. Notably, the authors demonstrated that the algorithm for learning arbitrary quantum processes from Ref.~\cite{huang2022learning} can be implemented in their model. Such algorithm allows to estimate expectation values of typical quantum states evolved under an arbitrary unknown channel. In contrast, in our work we consider specific classes of unitaries, but we provide a stronger notion of unitary learning, which consists in learning  
the evolution of typical states with respect to the trace distance.
{\color{black} Shortly after the first version of the present work was released, \citet{nietner2023unifying} proposed a comprehensive framework encompassing a broad spectrum of (quantum) learning algorithms, including quantum statistical query and variational learning models.}

\paragraph{Open questions.}
We distil several open questions concerning quantum statistical queries and process learning.
\begin{enumerate}
    \item The main workhorse for QSQ learning classical Boolean functions is Fourier analysis. While Fourier analysis is usually cast under the uniform distribution, the $\mu$-\emph{biased} Fourier analysis can be applied to every product distribution. In particular, $\mu$-biased Fourier sampling can be used to learn linear functions~\cite{caro2020quantumlearning} and DNFs~\cite{kanade2019learning} under product distributions with quantum examples. Can we extend these results to the QSQ model?
    \item Which classes of channels can be learned with quantum statistical queries? 
    \item What is the power of quantum statistical queries for testing properties of unitaries (and more broadly channels)? While we provided a double exponential lower bound for testing unitarity, quantum statistical queries might suffice for testing other relevant properties.
    \item Following~\cite{hinsche2023one, nietner2023average}, we can restrict our model to \emph{diagonal} measurements. Which classes of channels are learnable under this restricted model?
\end{enumerate}

\section{Preliminaries}
\label{sec:prem}
We start by introducing the mathematical notation and the background. For $n \geq 1$, we will denote the set of integers $[n] \coloneqq \{1,2,\dots,n\}$. Given $T\subseteq [n]$, we will write $\overline{T}:=[n] \setminus T$. We will denote the $2^n \times 2^n$ identity matrix as $I_n$ and we may omit the index $n$ when is clear from the context. For a matrix $A$, we will denote as $A_{ij}$ the entry corresponding to the $i$-th row and the $j$-th column.
We will use the indicator string $(x_1,x_2,\dots,x_k,*,*\dots,*)$ to denote the set of $n$-element strings whose first $k$ elements are $x_1,x_2,\dots,x_k$, i.e. $\mathcal{S} = \{(t_1,t_2,\dots,t_n) | \;\forall i \in[k] : x_i = t_i\}$.
Given a random variable $X$ sampled according to a distribution $\nu$, we will denote by $\mathbb{E}_\nu[X]$ its expected value and its variance by $\mathbb{V}_\nu[X]$, and omit the index $\nu$ when it's clear from the context.
{\color{black}Given two positive values $a,b$ satisfying $b\leq a$, we denote the set $[a-b, a+b]$ as $a\pm b$.}

\subsection{Quantum information theory}
\label{sec:QI}
\noindent\textbf{Bra-ket notation.} Let $\{\ket{0},\ket{1}\}$ be the canonical basis of $\mathbb{C}^2$, and $\mathcal{H}_n =(\mathbb{C}^2)^{\otimes n}$ be the Hilbert space of $n$ qubits.
We use the bra-ket notation, where we denote a vector $v \in (\mathbb{C}^2)^{\otimes n}$ using the ket notation $\ket{v}$ and its adjoint using the bra notation $\bra{v}$. For $u,v\in \mathcal{H}_n$, we will denote by $\braket{u|v}$ the standard Hermitian inner product $u^\dag v$. A pure state is a normalized vector $\ket{v}$, i.e. $|\braket{v|v}|=1$.

\medskip

\noindent\textbf{Linear operators.} Let $\mathcal{L}_n$ be the subset of linear operators on $\mathcal{H}_n$ and let $\mathcal{O}_n \subset \mathcal{L}_n$ be the subset of self-adjoint linear operators on $\mathcal{H}_n$. We represent the $2^n\times 2^n$ identity operator as $I_n$ and we omit the index $n$ when it is clear from the context. We denote by $\mathcal{O}^T_n \subset \mathcal{O}_n $ be the subset of traceless self-adjoint linear operators on $\mathcal{H}_n$, by $\mathcal{O}^+_n \subset \mathcal{O}_n $ the subset of the positive semidefinite linear operators on $\mathcal{H}_n$ and  by $\mathcal{S}_n\subset\mathcal{O}^+_n$
the set of the quantum states of $\mathcal{H}_n$, i.e.  $\mathcal{S}_n := \{\rho \in \mathcal{L}_{n} : \rho \geq 0, \Tr[\rho]=1\}$. We denote by $\mathcal{U}_n$ the unitary group, that is the set linear operators $U\in\mathcal{L}_n$ satisfying $UU^\dag = U^\dag U = I$, and we denote by $\mathrm{Id}:\mathcal{L}_n \rightarrow \mathcal{L}_n$ the identity map.
For any two operators $A, B \in \mathcal{L}_n$, we their normalized Hilbert-Schmidt inner product is given by $\frac{1}{2^n} \Tr[AB]$.

\medskip

\noindent\textbf{Pauli basis and Pauli weight.} We denote by $\mathcal{P}_n:=\{I,X,Y,Z\}^{\otimes n}$ the \emph{Pauli basis}. Elements  of the Pauli basis are Hermitian, unitary, trace-less, they square to the identity and they are orthonormal to each other with respect the normalized Hilbert-Schmidt inner product. The Pauli basis forms an orthonormal basis for the set of linear operators $\mathcal{L}_n$ with respect to the normalized Hilbert-Schmidt inner product.
We will often denote the $2^n$-dimensional identity operator as $I_n \coloneqq I^{\otimes n}$.

\noindent Given a Pauli operator $P = P_1 \otimes P_2 \otimes \dots \otimes P_n \in \calP_n$, we define its \emph{Pauli weight} $\abs{P}$ as the number of non-identity component in the tensor decomposition of $P$, i.e. as the number of qubits on which $P$ acts non trivially on.

\medskip
{\color{black}
\noindent\textbf{Schatten norms.} Given $p\geq 1$ and a linear operator $T\in\mathcal{L}_n$, we define the associated Schatten $p$-norm of $T$ as $\norm{T}_p \coloneqq (\Tr[\abs{T}^p])^{1/p}$, where $\abs{T} = \sqrt{T^\dag T}$. When the subscript is omitted, we refer to the Schatten-\( \infty \) norm  (also known as the \emph{operator norm}), denoted by
\(
\|T\| \coloneqq \|T\|_\infty.
\)
The trace distance between two quantum states $\rho,\sigma \in \mathcal{S}_n$ is the normalized Schatten 1-norm of their difference:
\begin{align}
    \norm{\rho -\sigma}_{\tr} \coloneqq \frac{1}{2} \norm{\rho-\sigma}_1.
\end{align}
It is also customary to compare two quantum states with the distance induced by the squared Schatten 2-norm, i.e. $\norm{\rho -\sigma}_{2}^2$ -- also known as Hilbert-Schmidt distance.
Interestingly, if at least one of the states is pure, then the Hilbert-Schmidt distance and the squared trace distance are nearly equivalent, as shown by the following Lemma.
\begin{lem}[Adapted from Theorem 1 in Ref.~\cite{coles2019strong}]
\label{lem:strong}
Let $\rho, \sigma \in \mathcal{S}_n$ be quantum states. Moreover, assume that $\rho= \ketbra{\psi}{\psi}$ is a pure state.
It holds that
\begin{align}
    \frac{1}{2}\norm{\rho -\sigma}_{2}^2 \leq  \norm{\rho -\sigma}_{\tr}^2 \leq \norm{\rho -\sigma}_{2}^2.
\end{align}
\end{lem}
}

\medskip

\noindent\textbf{Other definitions.} 
We define the canonical maximally entangled state as $\ket{\Omega} = \frac{1}{\sqrt{2^n}} \sum_{i,j\in\{0,1\}^n}\ket{i,i}$.
Moreover, the \emph{identity} $\mathbb{I}$ and the \emph{flip operator} $\mathbb{F}$ associated to a tensor product of two Hilbert spaces $\mathcal{H}_n^{\otimes 2}$ are defined as
\begin{align}
    \mathbb{I}:=\sum_{i,j\in\{0,1\}^n} \ketbra{i,j}{i,j} = I^{\otimes 2}, \quad\quad\quad \mathbb{F}:=\sum_{i,j\in\{0,1\}^n} \ketbra{i,j}{j,i}.
\end{align}
Notably, they satisfy the following properties:
\begin{align}
\mathbb{I}\left(\ket{\psi}\otimes\ket{\phi}\right)=\ket{\psi}\otimes\ket{\phi}, \quad \quad
\mathbb{F}\left(\ket{\psi}\otimes\ket{\phi}\right)=\ket{\phi}\otimes\ket{\psi},
\end{align}
for all $\ket{\psi},\ket{\phi} \in \mathcal{H}_n$. 
We also define the single-qubit \emph{stabilizers states} as the eigenstates of single-qubit Pauli operators, i.e. $\mathsf{stab} := \{ |0\rangle, |1\rangle, |+\rangle, |-\rangle, |+y\rangle, |-y\rangle \}$.

\subsection{Ensembles of states and unitaries}

\noindent\textbf{Haar measure and $t$-designs.} We start by providing some essential notions about the Haar measure $\mu_n$, which can be thought as the uniform distribution over the unitary group $\mathcal{U}_n$. 
For a comprehensive introduction to the Haar measure and its properties, we refer to Ref.~\cite{mele2023introduction}.
The Haar measure on the unitary group $\mathcal{U}_n$ is the unique probability measure $\mu_n$
that is both left and right invariant over the set $\mathcal{U}_n$, i.e., for all measurable functions $F$ and for all $V \in \mathcal{U}_n$, we
have:
\begin{equation}
\mathbb{E}_{U\sim \mu_n} f(U)  = \mathbb{E}_{U\sim \mu_n} f(UV) = \mathbb{E}_{U\sim \mu_n} f(VU) .
\end{equation}
Given a state $\ket{\phi}$, we denote the $k$-th moment of a Haar random state as
\begin{equation}
\mathbb{E}_{\ket{\psi}\sim\mu_n}\left[\ket{\psi}^{\otimes k}\right]:= \mathbb{E}_{U\sim\mu_n}\left[U^{\otimes k}\ket{\phi}^{\otimes k}\right].
\end{equation}
Note that the right invariance of the Haar measure implies that the definition of $\mathbb{E}_{\ket{\psi}\sim\mu_n}\left[\ket{\psi}^{\otimes k}\right]$ does not depend on the choice of $\ket{\phi}$.
In many scenarios, random unitaries and states are sampled from distributions that match only the low-order moments of the Haar measure.
This leads to the definition of $t$-designs, for integers $t\geq 1$. 
Let $\nu$ be a probability distribution over the set of quantum states $\mathcal{S}_n$.
The distribution $\nu$ is said to be a state $t$-design if
\begin{equation}
    \mathbb{E}_{\ket{\psi}\sim\nu}\left[\ket{\psi}^{\otimes t}\right]=\mathbb{E}_{\ket{\psi}\sim\mu_n}\left[\ket{\psi}^{\otimes t}\right].
\end{equation}

\medskip

\noindent\textbf{Locally scrambled ensembles.} Along with $t$-designs and Haar random ensembles, another important family of states (and unitaries) is the one of \emph{locally scrambled ensembles}, introduced in Ref.~\cite{caro2022outofdistribution}.
An ensemble of $n$-qubit unitaries is called locally scrambled if it is invariant under pre-processing by tensor products of arbitrary local unitaries. That is a unitary ensemble $\mathcal{U}_{\mathrm{LS}}$ is locally scrambled if for, for all integrable functions $F:\mathcal{U}_n \rightarrow \mathbb{R}$ $U \sim \mathcal{U}_{\mathrm{LS}}$ and for any fixed
$U_1, \dots, U_n \in \mathcal{U}_1$, we have
\begin{align}
    \mathbb{E}_{U \sim \mathcal{U}_{\mathrm{LS}}} F(U)=\mathbb{E}_{U \sim \mathcal{U}_{\mathrm{LS}}} F(UV),
\end{align}
where we denoted $V\coloneqq \bigotimes^n_{i=1} U_i$.
Accordingly, an ensemble $\mathcal{S}_{\mathrm{LS}}$ of $n$-qubit quantum states is locally scrambled if it is of the form $\mathcal{S}_{\mathrm{LS}} = \mathcal{U}_{\mathrm{LS}}\ket{0^n}$ for some locally scrambled unitary ensemble $\mathcal{U}_{\mathrm{LS}}$.
Notable examples of locally scrambled ensembles are the products of random single-qubit stabilizer states and the products of Haar random $k$-qubit states, which, in particular, include Haar random $n$-qubit states the products of Haar random single-qubit states. 
We emphasize that the above families include both product states and highly entangled states.

\medskip

{\color{black} \noindent\textbf{Low-average ensembles.}
We also consider another wide class of ensembles of quantum states, recently introduced in Ref.~\cite{schuster2024polynomial}. An ensemble of quantum states $\nu$ is a \emph{low-average ensemble} with purity $c$ if
\begin{align}
   \norm{\mathbb{E}_{\rho \sim \nu}[\rho]} \leq \frac{c}{2^n}. \label{eq:low-a}
\end{align}
For ensembles of pure states and $c = 1$, this coincides with the definition of a 1-design of states.
The introduction of this family of ensembles is motivated by the following property.
\begin{lem}[Lemma 3 in Ref.~\cite{schuster2024polynomial}]
\label{lem:schuster}
Let $O$ be an observable and $\nu$ be a \emph{low-average ensemble} with purity $c$.  We have
    \begin{align}
    \mathbb{E}_{\rho\sim \nu} \Tr[O\rho]^2 \leq c \, \frac{\norm{O}_2^2}{2^n} \leq c \norm{O}^2.
\end{align}
\end{lem}
}

\subsection{The Choi-Jamiolkowski isomorphism}
Furthermore, we can represent a unitary $U\in\mathcal{U}_n$ with its dual pure state, known as Choi-Jamiolkowski state, or simply Choi state~\cite{choi1975completely, jamiolkowski1972linear}. 
The Choi state $\ket{v(U)}$ can be prepared by first creating the maximally entangled state on
$2n$ qubits, which we denoted by $\ket{\Omega}$, and then applying $U$ on half of the maximally entangled state. This is equivalent
to preparing $n$ Einstein–Podolsky–Rosen (EPR) pairs $\frac{1}{\sqrt{2}}(\ket{00}+\ket{11})$ (which altogether forms $2n$ qubits) and applying the unitary $U$ to the
$n$ qubits coming from the second half of each of the EPR pairs. We have
\begin{equation}
    \ket{v(U)} = (I_n\otimes U)\ket{\Omega} = \frac{1}{\sqrt{2^n}} \sum_{i \in \{0,1\}^n} \ket{i}\otimes U\ket{i} = \frac{1}{\sqrt{2^n}} \sum_{i \in \{0,1\}^n} U_{ji}\ket{i,j}
\end{equation}
This definition can be naturally extended to a general quantum channel $\mathcal{N}$:
\begin{equation}
    \mathcal{J}(\mathcal{N}) = \mathrm{Id}\otimes \mathcal{N}(\ketbra{\Omega}{\Omega}) = \frac{1}{2^n} \sum_{i,j \in \{0,1\}^n} \ketbra{i}{j}\otimes\mathcal{N} (\ketbra{i}{j}). 
\end{equation}
Clearly, $\mathcal{J}(U(\cdot)U^\dag) = \ketbra{v(U)}{v(U)}$.
Furthermore, we recall that each EPR pair can be prepared by a circuit of depth 2:
\begin{equation}
    \frac{1}{\sqrt{2}}(\ket{00}+\ket{11}) = \mathsf{CNOT} (H\otimes I) \ket{00}.
\end{equation}
Then that the $n$ EPR pairs may be prepared in parallel with a constant depth circuit.
The Choi states of Pauli strings is of particular interest:
\begin{align}
    \ket{v(I)}= \frac{1}{\sqrt{2}} (\ket{00} + \ket{11}), \; \ket{v(X)}= \frac{1}{\sqrt{2}} (\ket{01} + \ket{10})\\
    i\ket{v(Y)}= \frac{1}{\sqrt{2}} (\ket{01} - \ket{10}), \; \ket{v(Z)}= \frac{1}{\sqrt{2}} (\ket{00} - \ket{11}).
\end{align}
We note that the Choi states of the Pauli basis are proportional to the Bell basis.
This readily implies that the set $\left\{\ket{v(I)},\ket{v(X)},\ket{v(Y)},\ket{v(Z)}\right\}^{\otimes n}$ forms an orthonormal basis for $2n$-qubit pure states with respect to the standard Hermitian inner product, i.e. $\forall P,Q \in \mathcal{P}_n : |\braket{v(P)|v(Q)}| = \delta_{P,Q}$, where $\delta_{P,Q}$ is the Kronecker's delta function. 

{\color{black}
Thus, given a unitary $U = \sum_{P\in\mathcal{P}_n} \widehat{U}_P P$ expressed in the Pauli basis, its corresponding Choi state can be written as
\begin{align}
    \ket{v(U)} = \sum_{P\in\mathcal{P}_n} \widehat{U}_P \ket{v(P)}.
\end{align}

}

{
\color{black}
\subsection{Average-case distance between unitaries}
The Choi-Jamiolkowski isomorphism induces the following distance over unitaries:
\begin{equation}
    {D}(U,V) := \|\ketbra{v(U)}{v(U)} - \ketbra{v(V)}{v(V)} \|_\tr = \sqrt{1 - |\braket{v(U)|v(V)}|^2}.
\end{equation}
This distance was introduced in Ref.~\cite{low2009learning} and recently extended to general quantum channels in Ref.~\cite{bao2025testing}.
When the unitaries $U = \sum_{P\in\mathcal{P}_n} \widehat{U}_P P$  and  $V = \sum_{P\in\mathcal{P}_n} \widehat{V}_P P$ are expressed in the Pauli basis, their distance can be rewritten as
\begin{align}
    D(U,V) = \sqrt{1 - |\braket{v(U)|v(V)}|^2}  = \sqrt{1 - \abs{\sum_{P\in\mathcal{P}_n} \widehat{U}_P^*\widehat{V}_P }^2} \label{eq:dist-pauli}.
\end{align}
We remark that a closely related pseudo-distance also appeared in other works~\cite{wang2011property,chen2023testing}.
\begin{align}
    \mathrm{dist}(U,V) = \min_{\theta \in (0,2\pi]} \,\frac{1}{\sqrt{2^{n+1}}} \norm{U  - e^{i\theta} V }_2
\end{align}
As noted in Ref.~\cite{chen2023testing}, $D(U,V)$ and $\mathrm{dist}(U,V)$ are within a constant factor $\sqrt{2}$, and thus these two distances are nearly equivalent. 
\begin{lem}[\cite{chen2023testing}, Lemma 14]
\label{lem:chen-dist}
Given $U,V \in \calU_n$, we have
\begin{align}
    \mathrm{dist}(U,V) \leq D(U,V) \leq\sqrt{2} \,\mathrm{dist}(U,V).
\end{align}
\end{lem}
\begin{proof}
$U = \sum_{P\in\mathcal{P}_n} \widehat{U}_P P$  and  $V = \sum_{P\in\mathcal{P}_n} \widehat{V}_P P$, and let $\theta \in (0,2\pi]$. As a preliminary step, we compute the Hilbert-Schmidt distance between $U$ and $e^{i\theta} V$.
\begin{align}
    \frac{1}{2^n}\norm{U  - e^{i\theta} V }_2^2 = &\frac{1}{4^n} \sum_{P\in\calP_n} \Tr[(U  - e^{i\theta} V)P]^2
    = \sum_{P\in\calP_n} (\widehat{U}_P - e^{i\theta} \widehat{V}_P)^2
    \\= &\sum_{P\in\calP_n} \abs{\widehat{U}_P}^2 +\abs{\widehat{V}_P}^2 - 2 \mathfrak{Re}(e^{i\theta} \widehat{V}_P\widehat{U}_P^*)
    \\ = &2\left(1 - \sum_{P\in\calP_n}  \mathfrak{Re}(e^{i\theta} \widehat{V}_P\widehat{U}_P^*)\right).
\end{align}
In the last step, we used the fact that $\sum_{P\in\calP_n}\abs{\widehat{U}_P}^2 = 2^{-n}\norm{U}_2^2 = 1$ (and similarly for $V$).
Minimizing with respect to $\theta$, we have
\begin{align}
     \mathrm{dist}^2(U,V) =  \min_\theta\left(1 - \sum_{P\in\calP_n}  \mathfrak{Re}(e^{i\theta} \widehat{V}_P\widehat{U}_P^*)\right)
     = 1-\sum_{P\in\calP_n}\widehat{V}_P\widehat{U}_P^*.
\end{align}
Moreover, following Eq.\ \ref{eq:dist-pauli}, the square of the distance $D$ can be expressed as
\begin{align}
    D^2(U,V)  = {1 - \abs{\sum_{P\in\mathcal{P}_n} \widehat{U}_P^*\widehat{V}_P }^2}.
\end{align}
We conclude the proof by applying the inequalities $\sqrt{1-x} \leq \sqrt{1-x^2} \leq \sqrt{2}\sqrt{1-x}$, which hold for $x\in [0,1]$
\begin{align}
    \mathrm{dist}(U,V) \leq D(U,V) \leq\sqrt{2} \mathrm{dist}(U,V).
\end{align}
\end{proof}

In the following, we will argue that
${D}(U,V)$ is an ``average-case'' distance between unitaries, as it is closely related to task of learning the action of a unitary on a randomly sampled state.
\medskip

\noindent\textbf{Predicting the evolution of quantum states.}
We consider the following task: given a target unitary $U$ and an an ensemble of states $\nu$, we aim at estimating another unitary $V$ satisfying, with high probability
\begin{align}
    U\rho U^\dag \approx V\rho V^\dag, 
\end{align}
for a state $\rho$ randomly sampled from the ensemble $\nu$.

To formalize this problem, we need to pick a loss functions to compare $U$ and $V$. We consider the following two functions
\begin{align}
    &\calL_\nu^{\mathrm{(tr)}}(U,V):= \mathbb{E}_{\rho\sim\nu}\left[\left\|U\rho U^\dag - V\rho V^\dag\right\|_{\tr}^2\right],
    \\&\calL_\nu^{\mathrm{(obs)}}(U,V):=  \max_{\substack{O :  \norm{O}\leq 1}} \mathbb{E}_{\rho \sim \nu}\Tr[O (U\rho U^\dag - V\rho V^\dag)]^2.
\end{align}
Employing the variational characterization of trace distance, we can observe that $\calL_\nu^{\mathrm{(obs)}}(U,V)$ is upper bounded by $\calL_\nu^{\mathrm{(tr)}}(U,V)$.
\begin{align}
    \calL_\nu^{\mathrm{(obs)}}(U,V):=  &\max_{\substack{O :  \norm{O}\leq 1}} \mathbb{E}_{\rho \sim \nu}\Tr[O (U\rho U^\dag - V\rho V^\dag)]^2
    \\\leq &\mathbb{E}_{\rho \sim \nu}\max_{\substack{O :  \norm{O}\leq 1}} \Tr[O (U\rho U^\dag - V\rho V^\dag)]^2
    \\= &\mathbb{E}_{\rho\sim\nu}\left[\left\|U\rho U^\dag - V\rho V^\dag\right\|_{\tr}^2\right] := \calL_\nu^{\mathrm{(tr)}}(U,V),
\end{align}
where we applied Jensen's inequality in the second step.
We now rephrase a result of Ref.~\cite{montanaro2013survey} according to our notation.
\begin{lem}[\cite{montanaro2013survey}, Proposition 21]
\label{lem:cj-risk}
Let $\mu_n$ be the Haar measure over $n$-qubit states. For unitary operators $U,V\in \mathcal{U}_n$, it holds that
\begin{equation}
    \calL_{\mu_n}^{\mathrm{(tr)}}(U,V) = \frac{2^n}{2^n+1} D(U,V)^2
\end{equation}
\end{lem}


Moreover, the following result swiftly extends this guarantee to all locally scrambled ensembles of states.
\begin{lem}[\cite{caro2022outofdistribution}, Lemma 1] 
\label{lem:risk-ls}
Let $\nu$ a locally scrambled ensemble of states. We have,
\begin{equation}
    \frac{1}{2}\calL_{\mu_n}^{\mathrm{(tr)}}(U,V) \leq \frac{2^n}{2^n+1} \calL_{\nu}^{\mathrm{(tr)}}(U,V) \leq \calL_{\mu_n}^{\mathrm{(tr)}}(U,V).
\end{equation}
\end{lem}

We also provide the following result for low-average ensembles of states.

\begin{lem}
\label{lem:odg-1des}
We have
\begin{align}
     \max_{\rho:\norm{\rho}_2^2 = c/2^n}\norm{U\rho U^\dag - V\rho V^\dag }_{\mathrm{tr}}^2
     \leq 8 c D^2(U,V)
\end{align}
Given a positive value $c >0$, let $\nu$ be low-average ensemble of states with purity $c$, i.e.
\begin{align}
    \norm{\mathbb{E}_{\rho\sim\nu} [\rho]} \leq \frac{c}{2^n}
\end{align} 
Then we have
\begin{align}
    \calL_\nu^{\mathrm{(obs)}}(U,V) \leq 8 c D^2(U,V)
\end{align}
\end{lem}
\begin{proof}
Let $\theta^*$ be the angle satisfying the following
\begin{align}
     \theta^* = \arg \min_\theta \norm{U  - e^{i\theta} V }_2.
\end{align}
Then, for all $O$ satisfying $\norm{O}\leq 1$, we have
\begin{align}
    \norm{U^\dag O U  - V^\dag O V}_2 = 
    &\norm{U^\dag O U - e^{i\theta^* }U^\dag O V + e^{i\theta^* }U^\dag O V - V^\dag O V}_2
    \\ \leq &\norm{U^\dag (O U - e^{i\theta^* } O V)}_2 + \norm{(e^{i\theta^* }U^\dag O  - V^\dag O )V}_2
    \\ = &\norm{O U - e^{i\theta^* } O V}_2 + \norm{e^{i\theta^* }U^\dag O  - V^\dag O}_2
    \\ \leq & \norm{O}\left(\norm{U -e^{i\theta^* }V}_2 + \norm{e^{i\theta^* }U^\dag   - V^\dag }_2\right)
   \\ = & 2 \norm{U -e^{i\theta^* }V}_2  = \sqrt{2^{n+3}}\mathrm{dist}(U,V) \leq \sqrt{2^{n+3}}D(U,V)
\end{align}
The first part of the proof follows by the Cauchy-Schwarz inequality:
\begin{align}
     &\max_{\rho:\norm{\rho}_2^2 = c/2^n}\norm{U\rho U^\dag - V\rho V^\dag }_{\mathrm{tr}}^2
     = \max_{\substack{\rho:\norm{\rho}_2^2 = c/2^n\\O\in\calL_n, \norm{O}\leq 1}} \Tr[O(U\rho U^\dag - V\rho V^\dag)]^2
     \\= & \max_{\substack{\rho:\norm{\rho}_2^2 = c/2^n\\O\in\calL_n, \norm{O}\leq 1}} \Tr[(U^\dag O U  - V^\dag O V)\rho]^2
     \leq \max_{\substack{\rho:\norm{\rho}_2^2 = c/2^n\\O\in\calL_n, \norm{O}\leq 1}}\norm{U^\dag O U  - V^\dag O V}_2^2 \norm{\rho}_2^2
     \leq 8 c D^2(U,V)
\end{align}
We conclude the proof invoking Lemma~\ref{lem:schuster}:
\begin{align}
    \mathbb{E}_{\rho\sim\nu} \Tr[U^\dag O U  - V^\dag O V]^2 \leq c \, \frac{\norm{U^\dag O U  - V^\dag O V}_2^2}{2^n}
    \leq 8 c D^2(U,V).
\end{align}
\end{proof}

Lemma~\ref{lem:risk-ls} and Lemma~\ref{lem:odg-1des} instances of the phenomenon of \emph{out-of-distribution} generalization~\cite{caro2022outofdistribution}. Taken together, these results imply that learning the action of a unitary with respect to an arbitrary locally scrambled ensemble automatically brings accuracy guarantees on all low-average ensembles of states.
We emphasize that low-average ensembles of states encompass both random pure states, such as computational basis states sampled uniformly at random, as well as highly mixed states, such as those considered in the ``one clean qubit'' model of computation~\cite{knill1998power}. 
Furthermore, low-average states also include sparse states (as defined in Ref.~\cite{caro2022learning}) as a special case, as we show in the following Proposition.
\begin{prop}[Sparse states are low-average]
\label{prop:sparse-low}
Given a subset of Pauli operators $S\in\calP_n$, let $\rho = \frac{1}{2^n}\sum_{P\in S}a_P P$  be a quantum state.
Then $\{\rho\}$ is a low-average ensemble with purity $\abs{S}$. 
\end{prop}
\begin{proof}
The lemma follows by a simple application of Minkowski's inequality:
\begin{align}
    \norm{\rho} = \norm{\frac{1}{2^n}\sum_{P\in S}a_P P} \leq\frac{1}{2^n}\sum_{P\in S}\abs{a_P} \norm{P}
    \leq \frac{\abs{S}}{2^n}.
\end{align}
\end{proof}
}
\subsection{Fourier analysis on the unitary group}
Let $U \in \mathcal{U}_n$ a unitary and consider the Pauli expansion $U=\sum_{P\in\mathcal{P}_n} \widehat{U}_P P$. We observe that the corresponding Choi state $\ket{v(U)}$ admits an analogous expansion with the same coefficients:
\begin{equation}
        \ket{v(U)}=\left(I_n \otimes \sum_{P\in \mathcal{P}_n}\widehat{U}_P  P\right)\left({\frac{1}{\sqrt{2^n}}}\sum_{i\in \{0,1\}^n}\ket{i,i}\right) =
    \sum_{P\in \mathcal{P}_n}\widehat{U}_P\ket{v(P)}.
\end{equation}
We now recall the notion of influence of qubits on linear operators, introduced in Ref.~\cite{montanaro2010quantum} in the context of Hermitian operators and further developed in Ref.~\cite{chen2023testing,rouze2024quantum}. The related influence of variables is widely used in the analysis of Boolean functions~\cite{o2021analysis}.
We define the quantum analogue of the bit-flip map as superoperator on $\mathcal{L}_n$:
\begin{equation}
   d_j:= I^{\otimes (j-1)}\otimes \left(I - \frac{1}{2}\Tr\right) \otimes I^{\otimes (n-j)}. 
\end{equation}
Then for $P=\bigotimes_{i=1}^n P_i\in \mathcal{P}_n$, we have
\begin{equation}
    d_j P = \begin{cases}
    P &\;\;\text{ if $P_j\neq I, $}\\
    0 &\;\;\text{ if $P_j = I. $}
    \end{cases}
\end{equation}
For a linear operator $ A \in \mathcal{L}_n, A = \sum_{P\in\mathcal{P}_n} \widehat{A}_P P$, we have
\begin{equation}
  d_j A = \sum_{P : P_j \neq I}\widehat{A}_P P.  
\end{equation}

For $p\geq 1$ and $j=1$, we denote by $\Inf_j^p(A):= \|d_j A\|_p^p$ the $L^p$-influence of $j$-th qubit on the operator $A$. We also denote by $\Inf^p(A) := \sum_{j=1}^n \Inf^p_j(A)$ the associated total $L^p$-influence. We will often omit the index $p$ when $p=2$.
Following~\cite{chen2023testing}, we also define the influence of a subset of qubits $S\subseteq[n]$ as
\begin{equation}
\label{eq:influence}
    \Inf_S(A) = \sum_{\substack{P \in \mathcal{P}_n:\\\mathrm{supp}(P)\cap S \neq \emptyset}} |\widehat{A}_P|^2.
\end{equation}
We observe that $\Inf_j(A) = \Inf_{\{j\}}(A) = \sum_{\substack{P\in \mathcal{P}_n : P_j \neq I}} |\widehat{A}_P|^2$, as expected.
Intuitively, the influence of a unitary $U$ on a subset of qubits is a quantitative measure of the action of $U$ on such subset.

\section{Quantum statistical query learning}
We first give the definition of the QSQ oracle. For a state $\rho \in \mathcal{S}_n$, the $\mathsf{QStat}_\rho$ oracle receives as input an observable $O \in \mathcal{L}_n, \|O\|\leq 1$ and a tolerance parameter $\tau \geq 0$, and returns a $\tau$-estimate of $\Tr[O\rho]$, i.e.
\begin{equation}
    \mathsf{QStat}_\rho : (O,\tau) \mapsto v \in  \Tr[\rho O] \pm \tau.
\end{equation}
A typical choice of the target state is the uniform quantum example $\ket{\psi_{f}} := \sum_{x \in \{0,1\}^n} \frac{1}{\sqrt{2^n}}\ket{x,f(x)}$, for a suitable Boolean function $f:\{0,1\}^n \rightarrow \{0,1\}$, which was first introduced in Ref.~\cite{bshouty1995learning} and widely employed in previous works on quantum statistical query learning~\cite{arunachalam2020quantum, arunachalam2023on}. In this case, we will shorten the notation to $\mathsf{QStat}_f = \mathsf{QStat}_{\ketbra{\psi_{f}}{\psi_{f}}}$. 
To adapt their framework to our goal of learning unitaries, we need to devise an alternative input state. A natural choice is the Choi-Jamiolkowki state, which found many applications in prior work about unitary learning~\cite{chen2023testing}, and more broadly process learning~\cite{caro2022learning}, motivating its adoption in the context of quantum statistical query.
For brevity, we will write $\mathsf{QStat}_U$ instead of $\mathsf{QStat}_{\ketbra{v(U)}{v(U)}}$.
We now detail the mutual relationship between the oracle $\mathsf{QStat}_U$ and the previous oracles defined in terms of quantum examples. To this end, we consider two unitaries implementing $f$, notably the bit-flip oracle $U_f$ and the phase oracle $V_f$. We have,
\begin{align}
    &\forall x \in \{0,1\}^n, y \in \{0,1\} : U_f \ket{x,y} =\ket{x, y\oplus f(x)},
    \\&\forall x \in \{0,1\}^n : V_f \ket{x} =(-1)^{f(x)}\ket{x}
\end{align}
In particular we note that $\ket{\psi_{f}} = \frac{1}{\sqrt{2^n}}U_f\sum_{x\in\{0,1\}^n}\ket{x,0}$.
We show that $\mathsf{QStat}_f $ can be simulated by $\mathsf{QStat}_{U_f}$ and conversely $\mathsf{QStat}_{V_f}$ can be simulated by $\mathsf{QStat}_f $. The first result shows that our framework generalizes the previous one based on quantum examples, while the second one allows us to transfer lower bounds from classical Boolean functions to unitaries, as formalized in Theorem~\ref{thm:hardness}.
\begin{lem}[Relations between QSQ oracles]
\label{lem:equiv}
Let $f:\{0,1\}^n\rightarrow \{0,1\}$ a Boolean function and consider the bit-flip oracle $U_f$ and the phase oracle $V_f$. Then for every observable $A \in \mathcal{L}_{n+1}$, there exists an observable $A' \in \mathcal{L}_{2n+2}$ such that
\begin{equation}
    \braket{\psi_f |A|\psi_f} = \braket{v(U_f)|A'|v(U_f)} \quad \text{and} \quad \norm{A'} = \norm{A}.
\end{equation}
and, similarly, for every observable $B \in \mathcal{L}_{2n}$, there exists an observable $B' \in \mathcal{L}_{n+1}$ such that
\begin{equation}
     \braket{v(V_f)|B|v(V_f)} = \braket{\psi_f |B'|\psi_f}  \quad \text{and} \quad \norm{B'} = \norm{B}.
\end{equation}
\end{lem}

\begin{proof}
    The first result follows by selecting $A' = I_{n} \otimes \ketbra{0}{0} \otimes A$. As for the second result, we can write the following expansion $B=\sum_{P,Q\in\mathcal{P}_{n}}c_{P,Q}\ketbra{v(P)}{v(Q)}$. From (\cite{montanaro2010quantum}, Proposition 9), we know that $\ket{v(V_f)} = \sum_{x\in\{0,1\}^{n}} \widehat{f(x)}\ket{v(Z^x)}$, where we denoted $Z^x:= \bigotimes_{i\in[n]} Z^{x_i}$, with $Z^0 = I$ and $Z^1=Z$. Hence
    \begin{equation}
        \braket{v(V_f)|B|v(V_f)} = \sum_{x\in\{0,1\}^{n}} c_{Z^x,Z^x}^2 \widehat{f(x)}^2.
    \end{equation}
    Now, consider the observable $T = \sum_{x\in\{0,1\}^n}c_{Z^x,Z^x}\ketbra{x}{x} \in \mathcal{L}_{n}$  and define
    \begin{equation}
        B' = H^{\otimes (n+1)}(I_n \otimes \ketbra{1}{1}) \cdot T \cdot (I_n \otimes \ketbra{1}{1})H^{\otimes (n+1)},
    \end{equation}
    which is equivalent to perform the Fourier transform on $\ket{\psi_f}$, post-selecting on the last qubit being 1 and finally applying $T$ on $n$ qubits. The Fourier transform and the projection on $\ketbra{1}{1}$ give rise to 
    \begin{equation}
        \ket{\widehat{\psi}_f}=\sum_{x\in\{0,1\}^{n}} \widehat{f(x)} \ket{x}.
    \end{equation}
    Then the desired result follows by noting that
    \begin{equation}
        \braket{\psi_f |B'|\psi_f} = \braket{\widehat{\psi}_f |T|\widehat{\psi}_f} = \sum_{x\in\{0,1\}^{n}} c_{Z^x,Z^x}^2 \widehat{f(x)}^2.
    \end{equation}
\end{proof}

We also introduce the following notion of learnability of classes of unitaries with quantum statistical queries.
\begin{definition}[Unitary learning with QSQs]
Let $\varepsilon\in[0,1]$, $\mathcal{C}\subseteq \mathcal{U}_n$ a class of unitaries and $\nu$ an ensemble of $n$-qubit states. We say that $\mathcal{C}$ is efficiently $\varepsilon$-learnable with quantum statistical queries with respect to $\nu$ if there exists an algorithm $\mathcal{A}$ such that, for all $U\in\mathcal{C}$, $\calA$ runs in time $\poly(n)$, performs $\poly(n)$ queries to the oracle $\mathsf{QStat}_U$ with tolerance at least $1/\poly(n)$ and outputs an operator $V\in\mathcal{U}_n$ such that
{\color{black}
\begin{equation}
    D(U,V)\leq \varepsilon.
\end{equation}
}
\end{definition}

We emphasize  that all the algorithms proposed in this work are classical randomized algorithms equipped with a quantum oracle, e.g. they use no other quantum resource apart from the query access to $\mathsf{QStat}_U$.
The QSQ model is considerably more restrictive than the \emph{oracle access} model, where a learner has the freedom to implement the unitary $U$ and its inverse $U^\dag$ on an arbitrary input state.
Then, every algorithm implementable with QSQs can be also implemented with oracle access, but the converse it is not true in general.
In particular, we demonstrate in Theorem~\ref{thm:hardness} that there is a class of unitaries that is efficiently learnable with direct access to the Choi state, but requires exponentially many quantum statistical queries. 

{\color{black}
\subsection{Extension to near-unitary channels}
Although the present work is devoted to unitary learning, here we also show that all our results can be naturally extended to near-unitary quantum channels. This comes as a natural consequence of the inherent robustness of the quantum statistical query framework. 
A common way to quantify how close a quantum channel is to being unitary is through its \emph{unitarity}~\cite{wallman2015estimating, carignan2019bounding}, defined as:
\begin{equation}
u (\mathcal{N}) :=  \frac{2^n}{2^n - 1}  \mathbb{E}_{\ket{\psi}\sim \mu_n}\Tr\left[\mathcal{N}(\ketbra{\psi}{\psi})^2\right] - \frac{2^n}{2^n - 1}\Tr\left[\mathcal{N}\left(\frac{I_n}{2^n}\right)^2\right]
\end{equation}
We observe that the unitarity of a quantum channel $\mathcal{N}$ and the purity of its Choi state $\mathcal{J}(\mathcal{N})$ are within an inversely exponential additive term.
\begin{lem}
\label{lem:unitarity}
Let $\calN$ be a quantum channel. It holds that
\begin{equation}
    \frac{4^n}{4^n -1} \Tr[\calJ(\mathcal{N})^2] - \frac{1}{4^n - 1}\leq  u (\mathcal{N}) \leq \frac{4^n}{4^n -1} \Tr[\calJ(\mathcal{N})^2].
\end{equation}
\end{lem}
\begin{proof}
We start by rewriting the expected purity of the output state as follows:
    \begin{align}
     &\mathbb{E}_{\ket{\psi}\sim \mu_n}\Tr\left[\mathcal{N}(\ketbra{\psi}{\psi})^2\right] =  \mathbb{E}_{\ket{\psi}\sim \mu_n}\Tr\left[\mathbb{F}\mathcal{N}^{\otimes 2}(\ketbra{\psi}{\psi}^{\otimes 2})\right]
     \\= &\Tr\left[\mathbb{F}\mathcal{N}^{\otimes 2}(\mathbb{E}_{\ket{\psi}\sim \mu_n}\ketbra{\psi}{\psi}^{\otimes 2})\right] 
      = \Tr\left[\mathbb{F}\mathcal{N}^{\otimes 2}\left(\frac{\mathbb{I}+\mathbb{F}}{2^n(2^n+1)}\right)\right] 
      \\ = &\Tr\left[\mathbb{F}\mathcal{N}^{\otimes 2}\left(\frac{\mathbb{F}}{2^n(2^n+1)}\right)\right] + \Tr\left[\mathbb{F}\mathcal{N}^{\otimes 2}\left(\frac{\mathbb{I}} {2^n(2^n+1)}\right)\right] 
      \\ = &\Tr\left[\mathbb{F}\mathcal{N}^{\otimes 2}\left(\frac{\mathbb{F}}{2^n(2^n+1)}\right)\right] + \Tr\left[\mathcal{N}\left(\frac{I_n}{2^n}\right)^2\right]\frac{2^{n}}{(2^n+1)},
\end{align}
where we applied the identity $\Tr[AB] = \Tr[\mathbb{F}A\otimes B]$ and the third step follows from the Schur-Weyl duality (we refer to Corollary 13 in Ref.~\cite{mele2023introduction} for more details).
Then we can rearrange the unitarity as follows
\begin{equation}
    u (\mathcal{N}) =  \frac{1}{4^n - 1}  \Tr\left[\mathbb{F}\mathcal{N}^{\otimes 2}\left({\mathbb{F}}\right)\right] - \frac{1}{4^n - 1}\Tr\left[\mathcal{N}\left(\frac{I_n}{2^n}\right)^2\right] 
\end{equation}
We can also use the Kraus representation $\mathcal{N}(\cdot) = \sum_{\ell} K_\ell (\cdot) K_\ell^\dag$ and write
\begin{align}
     \Tr\left[\mathbb{F}\mathcal{N}^{\otimes 2}\left({\mathbb{F}}\right)\right] =\sum_{\ell,\ell'}  \Tr\left[\mathbb{F} (K_\ell \otimes K_{\ell'}) {\mathbb{F}}  (K_\ell^\dag \otimes K_{\ell'}^\dag)  \right] 
     \\ = \sum_{\ell,\ell'} |\Tr[K_\ell K_{\ell'}^\dag]|^2 = 4^n \Tr[J(\mathcal{N})^2],
\end{align}
where the last two identities are proven in (\cite{quek2022exponentially}, Eqs. 160-164). Putting all together, we obtain:
\begin{equation}
    \frac{4^n}{4^n -1} \Tr[J(\mathcal{N})^2] - \frac{1}{4^n - 1}\leq  u (\mathcal{N}) \leq \frac{4^n}{4^n -1} \Tr[J(\mathcal{N})^2]
\end{equation}
\end{proof}

Let $\mathsf{QStat}_{\calN} \coloneqq \mathsf{QStat}_{\calJ(\calN)}$ be the quantum statistical query oracle associated to the state $\calJ(\calN)$. We further observe that, if $\calN$ has large unitarity, then the oracle $\mathsf{QStat}_{\calN} $ approximates well $\mathsf{QStat}_{U}$, where $\ketbra{U}{U}$ is the principal component in the spectral decomposition of $\calJ(\calN)$.
\begin{prop}
Let $\calN$ be a quantum channel with unitarity $u(\calN)$, and let $\ketbra{v(U)}{v(U)}$ denote the eigenstate of $\calJ(\calN)$ with largest eigenvalue. Querying $\mathsf{QStat}_{\calN}$ with a tolerance of $\tau $ provides a valid answer to a query to $\mathsf{QStat}_U$ with a tolerance of $\tau + 1 - \frac{4^n - 1}{4^n}u(\calN)$.
\end{prop}
\begin{proof}
let $\calJ(\calN) = \sum_{i} \lambda_{i=1}^r \ketbra{v(U_i)}{v(U_i)}$ be the spectral decomposition of $\calJ(\calN)$, where $\lambda_1 \geq \lambda_2 \geq \dots \geq \lambda_r$. 
We observe that
\begin{align}
    \Tr\left[\calJ(\calN)^2\right] = \sum_{j=1}^r \lambda_j^2 \leq (\max_{i} \lambda_i) \left( \sum_{j=1}^r \lambda_j\right) = \max_{i} \lambda_i ,
\end{align}
where we used the fact that $\sum_{j=1}^r \lambda_j= \Tr[\calJ(\calN)]=1$.
Then we can upper bound the trace distance as follows
\begin{align}
    &\norm{\calJ(\calN) - \ketbra{v(U)}{v(U)}}_{\tr} = 1- \max_{i} \lambda_i \\\leq & 
    1 - \Tr\left[\calJ(\calN)^2\right]  \leq  1- \left(\frac{4^n - 1}{4^n}\right)u(\calN)  \leq 1 - \frac{4^n - 1}{4^n}u(\calN).
\end{align}
The desired result follows from Hölder's inequality:
\begin{align}
    \max_{O: \norm{O}\leq 1}\abs{\Tr[O(\calJ(\calN) - \ketbra{U_1}{U_1}]} = 
    & \norm{\calJ(\calN) - \ketbra{v(U)}{v(U)}}_{\tr}  \\\leq & 1 - \frac{4^n - 1}{4^n}u(\calN).
\end{align}
\end{proof}

}

\subsection{Useful primitives}
Our results are based on the following technical lemma, which extends (\cite{arunachalam2020quantum}, Lemma 4.1) to unitary operators.
In particular, this lemma allows us to estimate the influence of subset of qubits defined in Eq.~\ref{eq:influence}.
\begin{lem}[Learning the influence of a subset with a single QSQ]
\label{lem:qbf}
Let $A \in \mathcal{U}_n$ be a unitary operator and $\mathsf{QStat}_A$ be the quantum statistical query oracle associated to the Choi state $\ket{v(A)}$. There is a procedure that on input a subset of Pauli strings $T\subseteq\mathcal{P}_n$, outputs $\tau$-estimate of $\sum_{P \in T} |\widehat{A}_P|^2$ using one query to $\mathsf{QStat}_A$ with tolerance $\tau$.
\end{lem}
\begin{proof}
Let $M = \sum_{P\in T}\ket{v(P)}\bra{v(P)}$. We note that
\begin{align}
    \bra{v(A)}M\ket{v(A)}
     = &\left( \sum_{P\in \mathcal{P}_n}\widehat{A}^*_P\bra{v(P)}\right)\left(\sum_{Q\in T}\ket{v(Q)}\bra{v(Q)} \sum_{P\in \mathcal{P}_n}\widehat{A}_P\ket{v(P)}\right)\\=&\left( \sum_{P\in \mathcal{P}_n}\widehat{A}^*_P\bra{v(P)}\right)\left( \sum_{{Q}\in T}\widehat{A}_Q\ket{v(Q)}\right)= \sum_{P\in T}|\widehat{A}_{P}|^2.
\end{align}
Thus a single query to $\textsf{Qstat}_A$ with input $(M,\tau)$ yields the desired outcome.
\end{proof}
\begin{remark}[Computational efficiency]
We observe that the circuit implementing the measurement $M = \sum_{P\in T}\ket{v(P)}\bra{v(P)}$ can have exponential depth in the worst case. However, in some cases, even if the set $T$ has exponential size, we can implement $M$ with a $\poly(n)$ circuit. For instance, the influence of the $j$-th qubit $\Inf_j(A)$ can be expressed as
\begin{equation}
   \Inf_j(A) =  \sum_{\substack{P\in \mathcal{P}_n :\\ P_j \neq I}} |\widehat{A}_P|^2 = 1 - \sum_{\substack{P\in \mathcal{P}_n :\\ P_j = I}} |\widehat{A}_P|^2. 
\end{equation}
Thus it suffices to estimate the expected value of $\ket{v(I)}_j\bra{v(I)}_j\otimes I_{n-1}$. More generally, we can consider the indicator string $\mathbb{S} = (x_1,x_2,\dots,x_k,*,*\dots,*)$ to denote the set of length $n$ strings whose first $k$ elements are $x_1,x_2,\dots,x_k$, i.e. $\mathbb{S} = \{(t_1,t_2,\dots,t_n) \in \{0,1,2,3\}^n| \;\forall i \in[k] : x_i = t_i\}$. Then we have,
\begin{equation}
    \sum_{P\in \mathbb{S}}\ket{v(P)}\bra{v(P)} = \ket{v(\sigma_{x_1}\otimes \sigma_{x_2}\otimes \dots \otimes \sigma_{x_k})}\bra{v(\sigma_{x_1}\otimes \sigma_{x_2}\otimes \dots \otimes \sigma_{x_k})}\otimes I_{n-k},
\end{equation}
which again can be implemented by a $\poly(n)$ circuit.
\end{remark}

We will also need a further technical tool, which is an implementation of state tomography with quantum statistical queries, also previously exploited in Ref.~\cite{arunachalam2023on} for learning the output of shallow circuits. 
Here we propose a refined argument for the special case of pure states. Since the complexity is exponential in the number of qubits, this primitive can be used to efficiently estimate the reduced states of subsets of logarithmic size.
\begin{lem}[State tomography]
\label{lem:tomography}
Let $\rho\in \mathcal{S}_n$. There exists an algorithm that performs $4^n$ queries to the oracle $\mathsf{QStat}_\rho$ with tolerance at least $\Omega(\sqrt{\epsilon \cdot 2^{-n}})$ and returns a state $\widehat{\rho}$ such that
\begin{equation}
    \|\rho - \widehat{\rho}\|_{2}^2\leq \epsilon.
\end{equation}
Moreover, if $\rho=\ketbra{\psi}{\psi}$ is a pure state, there exists an algorithm that performs $4^n$ queries to the oracle $\mathsf{QStat}_\rho$ with tolerance at least $\Omega(\sqrt{\epsilon \cdot 2^{-n}})$ and returns a pure state $\ket{\widehat{\psi}}$ such that
\begin{equation}
    \|\rho - \ketbra{\widehat{\psi}}{\widehat{\psi}}\|_{\tr}^2\leq \epsilon.
\end{equation} 
\end{lem}
\begin{proof}    
We perform a state tomography by querying all $4^{n} - 1$ non-identity Pauli strings with tolerance $\tau=\widetilde\epsilon \cdot 4^{-n}$, for an appropriate $\widetilde\epsilon <1$. For all $P \in \mathcal{P}_{n}$, denote  the obtained outcome by 
\begin{align}
x_P = \Tr[P \rho] \pm \tau.
\end{align}
We define the following operator:
\begin{equation}
   \Lambda:=  \frac{1}{2^n}\left(I + \sum_{\substack{P\in\mathcal{P}_{n}\setminus I}} x_P P \right). 
\end{equation}
This allows to upper bound the distance between the partial state $\rho$ and its estimate $\widehat{\rho}$. 
\begin{align}
    \|\rho - \Lambda\|_2^2 = &\Tr\left[(\rho - \Lambda)^2\right] 
     = \frac{1}{4^n} \Tr\left[\left(\sum_{P \in \mathcal{P}_{n}\setminus \{I_n\}} ( \Tr[P \rho]- x_P)P\right)^2\right] 
    \\ = &\frac{1}{2^n} \sum_{P \in  \mathcal{P}_{n}\setminus \{I_n\}} ( \Tr[P \rho] - x_P)^2 \leq 2^n\tau^2 \coloneqq \widetilde\epsilon,
\end{align}
We observe that $\Lambda$ is a Hermitian operator, but it is not necessarily a quantum state, as it may not be positive semidefinite and it may have purity larger than one.
We distinguish between two cases.

\medskip

\noindent\underline{Case 1:} If $\norm{\Lambda}_2^2 \leq 1$, we let $\widehat{\Lambda} \coloneqq \Lambda$.

\medskip

\noindent\underline{Case 1:} If $\norm{\Lambda}_2^2 > 1$, then we define the rescaled operator $\widehat{\Lambda}$ as follows
\begin{equation}
   \widehat\Lambda:=  \frac{1}{2^n}\left(I + \sum_{\substack{P\in\mathcal{P}_{n}\setminus I}} \widehat{x}_P P \right), 
\end{equation}
where
\begin{align}
    \widehat{x}_P =  \frac{x_P}{\norm{\Lambda}_2^2}. 
\end{align}
By the triangle inequality, the purity of $\Lambda$ satisfies 
\begin{align}
    &\norm{\Lambda}_2^2 \leq \left(\norm{\rho}_2 + \norm{\rho - \Lambda}_2\right)^2 \leq (1+\widetilde\epsilon)^2
\end{align}

The distance between the new rescaled operator and the target state is
\begin{align}
    \|\rho - \widehat\Lambda\|_2^2  = 
    \frac{1}{2^n} \sum_{P \in  \mathcal{P}_{n}\setminus \{I_n\}} ( \Tr[P \rho] - \widehat{x}_P)^2
     \\\leq \frac{2}{2^n} \sum_{P \in  \mathcal{P}_{n}\setminus \{I_n\}} ( \Tr[P \rho] - x_P)^2
     + \frac{2}{2^n} \sum_{P \in  \mathcal{P}_{n}\setminus \{I_n\}} ( \widehat{x}_P - x_P)^2
     \\ \leq  2\widetilde\epsilon +2 \left({\norm{\Lambda}_2^2 -1}\right)^2 \leq 2\widetilde\epsilon + 2\widetilde\epsilon^2(2+\widetilde\epsilon)^2 \leq 16\widetilde\epsilon.
\end{align}

\medskip
Thus, in both cases we have that
\begin{align}
    \|\rho - \widehat\Lambda\|_2^2  \leq 16\widetilde\epsilon.
\end{align}

By Lemma\ \ref{lem:eigenstate}, the closest positive semidefinite operator to $\Lambda$ is its positive part, which we denote by $\Lambda_+$. 
As $\rho$ is also positive semidefinite, we have
\begin{align}
    \norm{\Lambda_+ - \Lambda}_2^2 \leq \norm{\rho - \Lambda}_2^2 \leq 16\widetilde\epsilon.
\end{align}
And therefore, by triangle inequality we have
\begin{align}
    \norm{\Lambda_+ - \rho}_2^2 \leq \left(\norm{\rho - \Lambda}_2 + \norm{\Lambda_+ - \Lambda}_2\right)^2\leq 
    2\norm{\rho - \Lambda}_2^2 + 2\norm{\Lambda_+ - \Lambda}_2^2 \leq 64 \widetilde\epsilon.
\end{align}

Thus, setting $\widehat{\rho} \coloneqq \Lambda_+$ and $\widetilde\epsilon = \epsilon/64$ proves the first part of the lemma.

\medskip

We now delve into the case where the input state is pure.
Thanks to Lemma\ \ref{lem:strong}, and since $\rho=\ketbra{\psi}{\psi}$ has rank 1, we obtain the following bound for the 1-distance:
\begin{equation}
    \|\rho- \Lambda_+\|_1^2 \leq  \|\rho - \Lambda_+\|_2^2 \leq 64 \widetilde\epsilon.
\end{equation}
We now consider the dominant eigenstate of $\Lambda_+$, denoted by $\ket{\widehat{\psi}}$, which can be computed in $\mathrm{poly}(2^n)$ time. By (\cite{melkani2020eigenstate}, Proposition 2) we know that $\ketbra{\widehat{\psi}}{\widehat{\psi}}$ is the unique closest pure state to $\Lambda_+$. Since $\rho$ is also a pure state, this immediately implies
\begin{align}
     \left\|\ketbra{\widehat{\psi}}{\widehat{\psi}} - {\rho}\right\|_{\tr}^2 \leq   &2\left\|\ketbra{\widehat{\psi}}{\widehat{\psi}} - \Lambda_+\right\|_{\tr}^2 + 2\|\rho - \Lambda_+\|_{\tr}^2 
     \\\leq &2  \|\rho- \widehat{\rho}\|_{\tr} \leq  256\widetilde\epsilon,
\end{align}
therefore setting $\widetilde\epsilon = \epsilon/256$ yields the desired result.
\end{proof}

\section{Quantum Boolean functions}
A quantum Boolean function \( A \) is defined as a Hermitian unitary operator~\cite{montanaro2010quantum}, meaning it satisfies the following condition:
\begin{equation}
    AA^\dag = A^\dag A = A^2 = I_n.
\end{equation}
Since quantum Boolean functions are Hermitian, their Fourier coefficients are all real. Formally, for any Pauli string \( P \in \mathcal{P}_n \), we have:
\begin{equation}
    \forall P \in \mathcal{P}_n : \widehat{A}_P \in \mathbb{R}.
\end{equation}
{\color{black}
We also prove some useful properties of quantum Boolean functions which we will employ in the following. 
\begin{lem}[Quantum Parseval's identity, adapted from Proposition 10 and Lemma 23 in Ref.\ \cite{montanaro2010quantum}]
\label{lem:parseval}
Let $A = \sum_{P \in \calP_n} \widehat{A}_P P$ be a quantum Boolean Function. It holds that
\begin{align}
    \sum_{P\in\calP_n} \widehat{A}_P^2  = 1.
\end{align}
As a consequence, given $\gamma > 0$, there are at most $\gamma^{-2}$ Pauli operators $P$ such that $\abs{\widehat{A}_P} \geq \gamma$.
\end{lem}
\begin{proof}
The first part of the Lemma can be proven as follows:
 \begin{align}
    \sum_{P\in\calP_n} \widehat{A}_P^2 = \frac{1}{2^n} \norm{A}_2^2 =\frac{1}{2^n} \norm{I_n}_2^2 = 1 
\end{align}
where the second identity follows from the unitarily invariance of the Schatten norms.    
As for the second part, let $S \subseteq \calP_n$ be the subset of Pauli operators $P$ such that $\abs{\widehat{A}_P} \geq \gamma$. Then we have
\begin{align}
    1 = \sum_{P \in \calP_n} \widehat{A}_P^2 \geq \abs{S} \gamma^2,
\end{align}
which implies that $\abs{S} \leq \gamma^{-2}$.
\end{proof}
}In the following, we propose an algorithm for learning quantum Boolean functions within the quantum statistical query model, inspired by the celebrated Goldreich-Levin algorithm~\cite{goldreich1989hard}.
\subsection{The quantum Goldreich-Levin algorithm}
We will now demonstrate that the quantum Goldreich-Levin (GL) algorithm\ \cite{montanaro2010quantum} 
 can be implemented via quantum statistical queries. Whereas the original algorithm requires oracles queries to the target unitary $U$ and its adjoint, we show that the weaker access to $\mathsf{QStat}_U$ suffices.
A similar result was also established for the \emph{classical} Goldreich-Levin algorithm, which can be implemented via quantum statistical queries with respect to uniform quantum examples, that are quantum encodings of \emph{classical} Boolean functions\ \cite{arunachalam2020quantum}. 
We remark that the quantum Goldreich-Levin algorithm does not require the target operator to be Hermitian and thus it could find broader applications for learning other classes of unitaries.

We start by presenting the QSQ implementation of the GL algorithm in Algorithm\ \ref{alg:GL}. 
Given some single-qubit Pauli operators $P_1, P_2, \dots, P_k \in \calP_1$, we use the indicator string  \begin{align}
    (P_1,P_2,\dots,P_k,*,*,\dots,*)
\end{align}
to denote the set of Pauli operators whose first $k$ components appearing in their tensor decomposition are $P_1,P_2,\dots,P_k$, i.e. 
\begin{align}
    \mathcal{S} = \{Q_1\otimes Q_2 \otimes \dots \otimes Q_n \in \calP_n| \;\forall i \in[k] : Q_i = P_i\}.
\end{align}
Hence, the empty indication string $(*,*,\dots,*)$ denotes the set of $n$-qubit Pauli operators $\calP_n$.

\begin{algorithm}[H]
\caption{Quantum Goldreich-Levin algorithm with statistical queries}
\label{alg:GL}
\begin{algorithmic}
    \State $L \leftarrow (*,*,\dots,*)$
    \For{$k=1$ to $n$}
    \For{each $\mathcal{S}\in L, \mathcal{S} = (P_{1},P_{2},\dots, P_{k-1},*,*,\dots, *)$}
    \For{$P_{k}$ in $\{I,X,Y,Z\}$}
        \State Let $\mathcal{S}_{P_{k}}  = (P_{1},P_{2},\dots,P_{k-1}, P_{k},*,*,\dots, *)$. 
        \State Estimate $\sum_{P\in \mathcal{S}_{P_{k}}} |\widehat{A}_{P}|^2$ to within $\gamma^2/4$ with a \textsf{QStat} query.
        \State Add $\mathcal{S}_{P_{k}}$ to $L$ if the estimate of $\sum_{P\in \mathcal{S}_{P_{k}}} |\widehat{A}_{P}|^2$ is at least $\gamma^2/2$.
        \EndFor
        \State Remove $\mathcal{S}$ from $L$.
    \EndFor
    \EndFor
    \State \textbf{return} $L$
\end{algorithmic}
\end{algorithm}

We now present the QSQ version of the Goldreich-Levin theorem, closely following the proof of Ref.\ \cite{montanaro2010quantum}. Unlike Ref.\ \cite{montanaro2010quantum}, our approach leverages \( \mathsf{QStat}_A \) queries instead  of oracle queries to \( A \) and \( A^\dag \).

\begin{thm}[Quantum Goldreich-Levin using QSQs]
\label{thm:GL}
Let $A \in \mathcal{U}_n$ be a unitary operator and $\mathsf{QStat}_A$ be the quantum statistical query oracle associated to the Choi state $\ket{v(A)}$. There is an algorithm running in time $\calO\left(n \gamma^{-2}\right)$ that accesses $A$ via queries to $\mathsf{QStat}_A$ with tolerance at least $\Omega\left(\gamma^2\right)$ and outputs a list $L=\{P^{(1)},P^{(2)},\dots,P^{(m)}\}\subseteq \mathcal{P}_n$ such that:
\begin{enumerate}
    \item if $|\widehat{A}_P|\geq\gamma$, then $P\in L$;
    \item and for all $P\in L$, $|\widehat{A}_P|\geq\gamma/2$.
\end{enumerate}
\end{thm}
\begin{proof}
{\color{black}
Algorithm\ \ref{alg:GL} operates similarly to a branch-and-bound approach. Initially, the list $L$ contains a single element, which is the entire set $\calP_n$ of all \( 4^n \) Pauli operators. We then divide this set into four equal parts and use Lemma\ \ref{lem:qbf} to efficiently estimate the total Fourier weight within each subset. The subsets with low Fourier weight are discarded, while those with high Fourier weight are added to $L$. Moreover, $\calP_n$ is removed from $L$. 

The process continues iteratively by further dividing the remaining subsets into four parts. Although the number of subsets currently in the list can increase up to a factor 4 during each iteration, by Parseval's identity (Lemma\ \ref{lem:parseval}) we know that there are at most $4\gamma^{-2}$ Pauli operators with Fourier weight larger than $\gamma/2$.
Thus, throughout the entire execution of the algorithm, we have that
\begin{align}
    \abs{L} \leq 4 \gamma^{-2}.
\end{align}

During each of the \( n \) iterations, Algorithm\ \ref{alg:GL} performs 4 queries to the oracle \( \mathsf{QStat}_A \) for each of the items in the list \( L \), which is at most \( 4 \gamma^{-2} \). Thus, the total number of queries made by the algorithm is at most \( \mathcal{O}(n \gamma^{-2}) \). Moreover, by Lemma\ \ref{lem:qbf}, the required tolerance is at least \( \Omega(\gamma^2) \).

}
\end{proof}

The GL algorithm returns a list of ``heavy-weight'' Fourier coefficients. If $A$ is a quantum Boolean function, we can easily recover the values of those coefficients via $\mathsf{QStat}_A$ queries, up to a global sign. We prove this result in the following lemma.

\begin{lem}[Estimating Fourier coefficients up to a global sign]
\label{lem:coeff}
Let $A = \sum_P \widehat{A}_P P$ a quantum Boolean function and let $L\subseteq \mathcal{P}_n$ a list of Pauli strings. Assume that $|\widehat{A}_P| > \tau/2$ for all $P$. There is a procedure running in time $\mathcal{O}(|L|)$ that accesses $A$ via $\mathsf{QStat}_A$ queries with tolerance at least $\tau^2/2$ and outputs some estimates $\left\{\widehat{B}_P \big| P \in L\right\}$ such that
\begin{enumerate}
    \item for all $P \in L$, $\left(\abs{\widehat{B}_P } - \abs{\widehat{A}_P}\right)^2 \leq \tau^2$,
    \item for all $P,Q \in L$, $\mathsf{sgn}\left(\widehat{B}_P\widehat{B}_Q\right) = \mathsf{sgn}\left(\widehat{A}_P\widehat{A}_Q\right)$,
\end{enumerate}
where $\mathsf{sgn}(\cdot)$ is that function that on input $x\in\mathbb{R}$ returns the sign of $x$. 
{\color{black} Moreover, the operator $B \coloneqq \sum_{P\in L} \widehat{B}_P P$ satisfies
\begin{align}
    \min_{\xi \in \{-1,1\}} \frac{1}{2^n}\norm{A^{(L)} - \xi B}_2^2 \leq \abs{L} \tau^2,
\end{align}
where we denote $A^{(L)} \coloneqq \sum_{P\in L} \widehat{A}_P P$.
}
\end{lem}
\begin{proof}
By Lemma~\ref{lem:qbf}, we can estimate the values of $\widehat{A}_P^2$ up to error $\tau^2$ via a  $\mathsf{QStat}$ query with tolerance $\tau^2$. Let $\widehat{B}_P^2$ be such estimates. Then we have that
\begin{equation}
   |\widehat{B}_P| \leq \sqrt{A_P^2 + \tau^2}\leq |\widehat{A}_P| +\tau, 
\end{equation}
which proves the first part of the lemma. It remains to estimate the signs of the coefficients, up to a global sign. Let $P^* = \arg\max \widehat{B}_{P^*}^2$, that is the largest estimated squared coefficient. We arbitrarily assign the positive sign to this coefficient, i.e. we let
$\widehat{B}_{P^*} = \sqrt{\widehat{B}_{P^*}^2}$.
For each other coefficient $P\neq P^*$, we assign the sign with the following procedure. We first define the following observables $M^+$ and $M^-$,
\begin{align}
&M^+:= \frac{1}{2}\left(\ket{v(P^*)} + \ket{v(P)}\right)\left(\bra{v(P^*)} + \bra{v(P)}\right),\\
&M^-:= \frac{1}{2}\left(\ket{v(P^*)} - \ket{v(P)}\right)\left(\bra{v(P^*)} - \bra{v(P)}\right). 
\end{align}
We now compute the expected value of $M^+$ with respect to $\ket{v(A)}$:
\begin{align}
  &\mu^+:= \bra{v(A)}M^+\ket{v(A)} 
   \\=&\frac{1}{2}\left( \sum_{Q\in\mathcal{P}_n}\widehat{A}_Q\bra{v(Q)}\left(\ket{v(P^*)} + \ket{v(P)}\right)\right)
   \left( \left(\bra{v(P^*)} + \bra{v(P)}\right)\sum_{Q\in\mathcal{P}_n}\widehat{A}_Q\ket{v(Q)}\right)
   \\= &\frac{1}{2}(\widehat{A}_{P^*} + \widehat{A}_P)^2,
\end{align}
and, similarly, for $M^-$,
\begin{align}
   &\mu^-:=\bra{v(A)}M^-\ket{v(A)} 
   \\=&\frac{1}{2}\left( \sum_{Q\in\mathcal{P}_n}\widehat{A}_Q\bra{v(Q)}\left(\ket{v(P^*)} - \ket{v(P)}\right)\right)
   \left( \left(\bra{v(P^*)} - \bra{v(P)}\right)\sum_{Q\in\mathcal{P}_n}\widehat{A}_Q\ket{v(Q)}\right)
   \\= &\frac{1}{2}(\widehat{A}_{P^*} - \widehat{A}_P)^2.
\end{align}
We notice that if $\widehat{A}_{P^*}$ and $\widehat{A}_{P}$ have the same sign, $\mu^+ > \mu^-$ and vice-versa. Moreover, $|\mu^+-\mu^-|= 2\left|\widehat{A}_P\widehat{A}_{P^*}\right|> \tau^2/2$. Then we can tell whether $\mu^+ > \mu^-$ by querying the oracle $\mathsf{QStat}_A$ with the observable $M^+ - M^-$ and tolerance $\tau^2/2$. If the output is positive, then we can conclude that $\mu^+ > \mu^-$ and assign $\widehat{B}_P$ positive sign, and vice-versa if the output is negative.
This proves the second part of the lemma.
{\color{black}
Moreover, we observe that
\begin{align}
    \min_{\xi  \in \{-1,1 \}} \frac{1}{2^n}\norm{A^{(L)} - \xi B}_2^2 = \min_{\xi  \in \{-1,1 \}} \sum_{P\in L} (\widehat{A}_P - \xi \widehat{B}_P)^2  =  \sum_{P\in L} \left(\abs{\widehat{A}_P} - \abs{\widehat{B}_P }\right)^2
    \leq  \abs{L} \tau^2,
\end{align}
where in the second step we used that $\mathsf{sgn}\left(\widehat{B}_P\widehat{B}_Q\right) = \mathsf{sgn}\left(\widehat{A}_P\widehat{A}_Q\right)$ for all $P,Q \in L$, and in the last step we used that $\left(\abs{\widehat{B}_P } - \abs{\widehat{A}_P}\right)^2 \leq \tau^2$ for all $P \in L$.
This concludes the proof the of the Lemma.
}
\end{proof}
{\color{black}
We note that learning a unitary operator \( U \) up to a global sign is sufficient for predicting the unitary evolution of the form \( U \rho U^\dagger \), as the global sign has no physical consequences -- i.e. \( (-U) \rho (-U)^\dagger = U \rho U^\dagger \).  

However, quantum Boolean functions are Hermitian and can also represent \emph{observables} rather than unitary evolution. 
Specifically, given a Pauli operator $P\in\calP_n$ and a unitary $U$, the Heisenberg-evolved observable $ O \coloneqq U^\dag P U$ is a quantum Boolean function, as it satisfies
\begin{align}
    & O^2 \coloneqq U^\dag  P UU^\dag  P U = U^\dag PPU = U^\dag U = I_n.
    \\ & O^\dag O \coloneqq (U P U^\dag )^\dag(U P U^\dag ) = U^\dag  P UU^\dag  P U = O^2 = I_n.
\end{align}
Let $O = \sum_{P\in\calP_n} \widehat{O}_P P$ be the Pauli expansion of $O$.
In order to estimate the expectation value $\Tr[O\rho]$, it is crucial to estimate also the global sign, as clearly \( \Tr[O \rho] \neq \Tr[(-O) \rho] \), unless \( \Tr[O \rho] = 0 \).  
Thus, order to recover the signed coefficients of $O$ via quantum statistical queries, we need to devise a refined version of Lemma\ \ref{lem:coeff}. To this end, we introduce the following unitary $\widetilde{O}$
\begin{align}
    \widetilde{O} \coloneqq & U^\dag \exp\left( i \frac{\pi}{6} P\right) U = \cos\left(\frac{\pi}{6}\right) I_n
    + i \sin\left(\frac{\pi}{6}\right)  U^\dag P U \\= &\frac{\sqrt{3}}{2} I_n
    + \frac{i}{2} U^\dag P U = \frac{\sqrt{3}}{2} I_n + \frac{i}{2}\sum_{P\in\calP_n}  \widehat{O}_P P \label{eq:expansion-signed}.
\end{align} 
While $\widetilde{O}$ is not a quantum Boolean function, its Pauli expansion exhibits some useful properties: the coefficient of the term $I_n$ is positive, and all the other coefficients are imaginary. Building on this simple observation, we implement a procedure analogous to that described in Lemma~\ref{lem:coeff} for retrieving the phases of the Pauli observables found by the Goldreich-Levin algorithm.
\begin{lem}[Estimating signed Fourier coefficients]
\label{lem:coeff2}
Let $ O \coloneqq U^\dag Q U = \sum_{P\in\calP_n} \widehat{O}_P P$ a Heisenberg-evolved Pauli observable and let $L\subseteq \mathcal{P}_n$ a list of Pauli strings. 
Denote $\widetilde{O} \coloneqq U^\dag \exp\left(i \frac{\pi}{6} Q \right) U$. 
There is a procedure running in time $\mathcal{O}(|L|)$ that accesses $\widetilde{O}$ via $\mathsf{QStat}_{\widetilde{O}}$ queries with tolerance at least $\Omega(\tau^2)$ and outputs some estimates $\left\{\widehat{B}_P \big| P \in L\right\}$ such that
for all $P \in L$, $\abs{\widehat{B}_P - \widehat{O}_P } \leq \tau$.
{\color{black} Moreover, the operator $B \coloneqq \sum_{P\in L} \widehat{B}_P P$ satisfies
\begin{align}
     \frac{1}{2^n}\norm{O^{(L)} - B}_2^2 \leq \abs{L} \tau^2,
\end{align}
where we denote $O^{(L)} \coloneqq \sum_{P\in L} \widehat{O}_P P$.
}
\end{lem}
\begin{proof}
As observed in Eq.~\ref{eq:expansion-signed}, the expansion of $\widetilde{O}$ in the Pauli basis is given by $\widetilde{O} = \frac{\sqrt{3}}{2} I_n + \frac{i}{2}\sum_{P\neq I_n}  \widehat{O}_P P$.
Therefore, the Choi state $\ket{\widetilde{O}}$ can be expressed as
\begin{align}
    \ket{v(\widetilde{O})} = \frac{\sqrt{3}}{2} \ket{v\left(I_n\right)} + \frac{i}{2}\sum_{P\neq I_n}  \widehat{O}_P \ket{v(P)}.
\end{align}
We first define the following observable $\widetilde{M}$,
\begin{align}
&\widetilde{M}:= \frac{1}{2}\left(\ket{v(I_n)} + i\ket{v(P)}\right)\left(\bra{v(I_n)} - i \bra{v(P)}\right). 
\end{align}
We now compute the expected values of $\widetilde{M}$ with respect to $\ket{v(\widetilde{O})}$:
\begin{align}
  \widetilde{\mu}:= &\bra{v(\widetilde{O})}\widetilde{M}\ket{v(\widetilde{O})} =
   \\= &\frac{1}{2}\left( \bra{v(\widetilde{O})}\left(\ket{v(I_n)} + i\ket{v(P)}\right)\right)
   \left( \left(\bra{v(I_n)} - i \bra{v(P)}\right)\ket{v(\widetilde{O})}\right)
   \\= & \frac{1}{4}\left(\sqrt{3} + \widehat{O}_P\right)^2,
\end{align}
Thus, we have
\begin{align}
    \widehat{O}_P = 2 \sqrt{\widetilde{\mu}} - \sqrt{3}
\end{align}
We also observe that, since $\abs{\widehat{O}_P} \leq \norm{O}_\infty \leq 1 $, we have
\begin{align}
    \sqrt{\widetilde{\mu}} \geq \frac{\sqrt{3}-1}{2} > \frac{1}{3}.
\end{align}
In order to estimate $\widehat{O}_P$, we perform a query to $\mathsf{QStat}_{\widetilde{O}}$ with input $\widetilde{M}$ and tolerance $\tau/6$. Let $x$ be the outcome and denote $\widehat{B}_P = 2\sqrt{x} - \sqrt{3}$. We have that
\begin{align}
    \abs{\widehat{O}_P - \widehat{B}_P} = 2\abs{\sqrt{x} - \sqrt{\widetilde\mu}} = 2 \frac{\abs{x^2 - \widetilde\mu^2}}{\sqrt{x} + \sqrt{\widetilde\mu}}\leq \frac{\tau}{3\sqrt{\widetilde\mu}} \leq {\tau} ,
\end{align}
which proves the first part of the Lemma. The second part can be proven via a simple triangle inequality:
\begin{align}
    \frac{1}{2^n}\norm{O^{(L)} - B}_2^2 =  \sum_{P \in L} (\widehat{O}_P - \widehat{B}_P)^2 \leq \abs{L}\tau^2.
\end{align}
\end{proof}

As a direct application, we show that ``sparse'' quantum Boolean functions -- that is quantum Boolean functions supported on few Pauli operators -- can be efficiently learned via quantum statistical queries.
\begin{coro}[Learning Sparse Quantum Boolean Functions]
{\label{cor:sparse}}
Given a subset of Pauli operators $S\subseteq \calP_n$, consider a quantum Boolean function $A = \sum_{P\in S} \widehat{A}_P P$.
\begin{enumerate}
    \item There is an algorithm running in time $\calO(n\abs{S}\epsilon^{-2})$ that accesses $A$ via $\mathsf{QStat}_A$ queries with tolerance at least $\Omega(\epsilon^2/\abs{S})$ and outputs a quantum Boolean function $B$ such that
    \begin{align}
        D(A,B) \leq \epsilon.
    \end{align}
    \item  If $A$ is a Heisenberg-evolved Pauli observable, i.e. if $A = U^\dag P U$ for some unitary $U$ and Pauli operator $P$, then there is an algorithm running in time $\calO(n\abs{S}\epsilon^{-2})$ that access the unitary $\widetilde{A} \coloneqq U^\dag \exp\left(i\frac{\pi}{6}P\right) U$ via $\mathsf{QStat}_{\widetilde{A}}$ queries with tolerance at least $\Omega(\epsilon^2/\abs{S})$ and outputs an operator $B$ such that
\begin{align}
    \frac{1}{\sqrt{2^n}}\norm{A-B}_2 \leq \epsilon.
\end{align}
\end{enumerate}
\end{coro}


\begin{proof}
The Corollary follows combining Theorem\ \ref{thm:GL} with the protocols for estimating Fourier coefficients described Lemmas\ \ref{lem:coeff} and\ \ref{lem:coeff2}. 
Given $\gamma >0$,  we can use the GL algorithm to output a list $L$ of Pauli operators such that:
\begin{enumerate}
    \item if $|\widehat{A}_P|\geq\gamma$, then $P\in L$;
    \item for all $P\in L$, $|\widehat{A}_P|\geq\gamma/2$.
\end{enumerate}
By Theorem\ \ref{thm:GL}, the list $L$ can be obtained in time $\calO\left(n \gamma^{-2}\right)$ via queries to $\mathsf{QStat}_A$ with tolerance at least $\Omega\left(\gamma^2\right)$.
Let $A^{(L)}= \sum_{P\in L}\widehat{A}_P P$. We have that
\begin{align}
    \frac{1}{2^n} \norm{A^{(L)} - A}_2^2 = \sum_{P \not \in L}\widehat{A}_P^2 \leq \abs{L} \gamma^2  \leq \abs{S} \gamma^2.
\end{align}
Thus, by Lemma\ \ref{lem:coeff} there is a procedure running in time $\mathcal{O}(|L|) \subseteq \calO(\abs{S})$ that accesses $A$ via $\mathsf{QStat}_A$ queries with tolerance at least $\tau^2/2$ and outputs an operator $B$ such that
\begin{align}
    \min_{\xi \in \{-1,1\}}\frac{1}{2^n}\norm{B - \xi A^{(L)}}_2^2 \leq \abs{L} \tau^2 \leq \abs{S}\tau^2.
\end{align}
The distance $D(A,B)$ can be upper bounded by triangle inequality:
\begin{align}
    & D(A,B) \leq D(A,A^{(L)}) + D(A^{(L)},B) \\\leq &\min_{\xi \in \{-1,1\}}\frac{1}{\sqrt{2^n}}\norm{B - \xi A^{(L)}}_2 + 
    \frac{1}{\sqrt{2^n}} \norm{A^{(L)} - A}_2 \\\leq &\sqrt{\abs{S}} \sqrt{\tau^2 + \gamma^2} \leq \sqrt{\abs{S}} (\tau + \gamma),
\end{align}
Thus, setting $\tau = \gamma = \epsilon/\left(2\sqrt{\abs{S}}\right)$ ensures that $D(A,B) \leq \epsilon$

The proof of the second part of the Corollary is analogous.
By Lemma\ \ref{lem:coeff2}, there is a procedure running in time $\mathcal{O}(|L|)$ that accesses $\widetilde{O}$ via $\mathsf{QStat}_{\widetilde{O}}$ queries with tolerance at least $\Omega(\tau^2)$ and outputs an operator $B$ such that
\begin{align}
    \frac{1}{2^n}\norm{B -  A^{(L)}}_2^2 \leq \abs{L} \tau^2 \leq \abs{S}\tau^2.
\end{align}
Hence applying the triangle inequality we obtain that
\begin{align}
    \frac{1}{\sqrt{2^n}}\norm{A-B}_2 \leq \frac{1}{\sqrt{2^n}}\norm{A^{(L)}-B}_2 + \frac{1}{\sqrt{2^n}}\norm{A-A^{(L)}}_2 \leq \sqrt{\abs{S}} (\tau + \gamma),
\end{align}
and thus also in this case the desired result follows by setting  $\tau = \gamma = \epsilon/\left(2\sqrt{\abs{S}}\right)$.
\end{proof}

We remark that the above result enables learning the action of sparse quantum Boolean functions on low-average states, which include sparse states as a special case. In particular, given 
two unitaries $A,B$ such that $D(A,B)\leq \epsilon$ and a state $\rho = \frac{1}{2^n}\sum_{P\in S_\rho} \widehat{\rho}_P P$, by Lemma\ \ref{lem:odg-1des} and Proposition\ \ref{prop:sparse-low}, it holds that
\begin{align}
    \norm{U\rho U^\dag - V\rho V^\dag}^2_{\tr} \in \calO(\abs{S}_\rho \epsilon^2 ).
\end{align}
It is insightful to compare this result with those of Ref.\ \cite{caro2022learning}, which investigated the learnability of quantum channels through Choi state access, assuming both the channels and the initial state are sparse in the Pauli basis. Notably, the query complexity of Corollary\ \ref{cor:sparse} matches that of Corollary 4.2 in Ref.\ \cite{caro2022learning}. However, the two results pertain to distinct scenarios: Ref.\ \cite{caro2022learning} applies to arbitrary sparse channels, while our result is restricted to quantum Boolean functions. Additionally, their approach relies on quantum memory, i.e., the ability to measure several Choi states in superposition, whereas our framework is based on quantum statistical queries -- an arguably weaker access model.

\subsection{Learning low-degree quantum Boolean functions}

We present an application of Theorem\ \ref{thm:GL} to learning quantum Boolean functions of low degree, that is quantum Boolean functions supported on low-weight Pauli operators. Recall that the weight of a Pauli operator is the number of qubits on which it acts non-trivially, i.e., the number of non-identity components in its tensor decomposition. To this end, we leverage the non-commutative Bohnenblust--Hille inequality presented in Ref.\ \cite{volberg2024noncommutative}.
\begin{lem}[Noncommutative Bohnenblust-Hille inequality, adapted from Theorem 1.2 in Ref.\ \cite{volberg2024noncommutative}]
\label{lem:BH}
Let $A \in \calL_{n}$ be a linear operator of the form $M = \sum_{\substack{P\in\calP_n : \abs{P}\leq k}} a_P P$. It holds that
\begin{align}
    \left(\sum_{\substack{P\in\calP_n : \abs{P}\leq k}} \abs{a_P}^{\frac{2k}{k+1}}\right)^{\frac{k+1}{2k}} \in \exp\left(\calO(k)\right) \norm{A}.
\end{align}
\end{lem}
We can now state our result on low-degree Quantum Boolean Functions.
\begin{thm}[Learning low-degree Quantum Boolean Functions]
\label{thm:GL-low-d}
Let $k \in \mathbb{N}$ and let $A = \sum_{\substack{P \in \calP_n: \abs{P} < k }} \widehat{A}_P P$ be a quantum Boolean function supported only on Pauli operators with weight smaller than $k$.
\begin{enumerate}
    \item There is an algorithm running in time $n\cdot \left(\epsilon^{-k}, 2^{k^2} \right)$ that access the unitary $A$ via $\mathsf{QStat}_A$ queries with tolerance at least $\poly\left( \epsilon^k, 2^{-k^2}\right)$ and outputs an operator $B$ such that
\begin{align}
    D(A,B) \leq \epsilon.
\end{align}
    \item If $A$ is a Heisenberg-evolved Pauli observable, i.e. if $A = U^\dag P U$ for some unitary $U$ and Pauli operator $P$, then there is an algorithm running in time $n\cdot \left(\epsilon^{-k}, 2^{k^2} \right)$ that access the unitary $\widetilde{A} \coloneqq U^\dag \exp\left(i\frac{\pi}{6}P\right) U$ via $\mathsf{QStat}_{\widetilde{A}}$ queries with tolerance at least $\Omega\left( \epsilon^k, 2^{-k^2}\right)$ and outputs an operator $B$ such that
\begin{align}
    \frac{1}{\sqrt{2^n}}\norm{A-B}_2 \leq \epsilon.
\end{align}
\end{enumerate}

\end{thm}

\begin{proof}
Given $\gamma >0$,  we can use the GL algorithm to output a list $L$ of Pauli operators such that:
\begin{enumerate}
    \item if $|\widehat{A}_P|\geq\gamma$, then $P\in L$;
    \item for all $P\in L$, $|\widehat{A}_P|\geq\gamma/2$.
\end{enumerate}
By Theorem\ \ref{thm:GL}, the list $L$ can be obtained in time $\calO\left(n \gamma^{-2}\right)$ via queries to $\mathsf{QStat}_A$ with tolerance at least $\Omega\left(\gamma^2\right)$. 
By Lemma\ \ref{lem:parseval}, the list $L$ contains at most $\calO(\gamma^{-2})$ elements.
Therefore, by Lemma\ \ref{lem:coeff}, we can estimate an operator $B$ in time $\calO(\gamma^{-2})$ via $\mathsf{QStat}_A$ queries with tolerance at least $\tau^2$, satisfying the following: 
\begin{align}
    \min_{\xi  \in \{-1,1 \}} \frac{1}{2^n}\norm{A^{(L)} - \xi B}_2^2 
    \leq  \abs{L} \tau^2 \leq 4 \gamma^{-2} \tau^2,
\end{align}
where $A^{(L)} \coloneqq \sum_{P\in L} \widehat{A}_P P$.
In particular, choosing an appropriate value of $\tau \in \calO(\gamma \epsilon)$, we obtain that
\begin{align}
    \min_{\xi  \in \{-1,1 \}} \frac{1}{\sqrt{2^n}}\norm{A^{(L)} - \xi B}_2 \leq \frac{\epsilon}{2} \label{eq:eps1}. 
\end{align}
Furthermore, we can upper bound the (normalized) Schatten 2-distance between $ A^{(L)}$ and $A$ as follows
    \begin{align}
        \frac{1}{2^n}\norm{A- A^{(L)}}_2^2
        = &\sum_{\substack{P \not\in L}} \widehat{A}_P^2
         = \sum_{\substack{P \not\in L}}\abs{\widehat{A}_P^{\frac{2}{k+1}}}\cdot\abs{\widehat{A}_P^{\frac{2k}{k+1}}} \\\leq &\gamma^{\frac{2}{k+1}} \sum_{\substack{P \in \calP_n: \\ \abs{P} < k}} \abs{\widehat{A}_P^{\frac{2k}{k+1}}}  \in \exp\left(\calO(k)\right) \gamma^{\frac{2}{k+1}}  \norm{A}^{\frac{k+1}{2k}} \\ = &\exp\left(\calO(k)\right) \gamma^{\frac{2}{k+1}} .
    \end{align}
where the inclusion follows from the Bohnenblust--Hille inequality (Lemma\ \ref{lem:BH}) and in the last step we used the fact that unitary operators have operator norm 1.
Therefore, choosing appropriate value of  $\gamma \in \exp\left(-\Omega(k^2) \right) \epsilon^{\calO(k)}$, we obtain that
\begin{align}
   \frac{1}{\sqrt{2^n}}\norm{A- A^{(L)}}_2 \leq \frac{\epsilon}{2} \label{eq:eps2}.
\end{align}
As a consequence, the procedure described runs in time $\calO\left(n \gamma^{-2}\right) = n\cdot \poly\left(\epsilon^{-k}, 2^{k^2} \right)$ and employs quantum statistical queries with tolerance at least $\poly\left( \epsilon^k, 2^{-k^2}\right)$.

In order to complete the proof, it remains to upper bound the distance $D(A,B)$ :
\begin{align}
    D(A,B) \leq \sqrt{2} \, \mathrm{dist}(A,B) \coloneqq & \min_{\theta \in (0,2\pi]} \frac{1}{\sqrt{2^n}}\norm{A - e^{i\theta} B}_2
    \leq \min_{\xi \in \{-1,1 \} } \frac{1}{\sqrt{2^n}}\norm{A - \xi B}_2
    \\ \leq & \frac{1}{\sqrt{2^n}}\left(\norm{A- A^{(L)}}_2 + \min_{\xi \in \{-1,1 \} } \norm{A^{(L)} -\xi B}_2\right) \leq \epsilon,
\end{align}
where the first inequality follows from Lemma\ \ref{lem:chen-dist}, the second-to-last step follows from the triangle inequality and in the last step we used Eqs.\ \ref{eq:eps1} and \ref{eq:eps2}.

\smallskip

The proof of the second part of the Theorem is analogous. Invoking Lemma\ \ref{lem:coeff2}, we can estimate an operator $B$ in time $\calO(\gamma^{-2})$ via $\mathsf{QStat}_{\widetilde{A}}$ queries with tolerance at least $\tau^2$, satisfying the following: 
\begin{align}
    \frac{1}{2^n}\norm{A^{(L)} - B}_2^2 \leq 4 \gamma^{-2} \tau^2.
\end{align}
In particular, choosing an appropriate value of $\tau \in \calO(\gamma \epsilon)$, we obtain that
\begin{align}
    \frac{1}{\sqrt{2^n}}\norm{A^{(L)} -  B}_2 \leq \frac{\epsilon}{2}. \label{eq:eps3}. 
\end{align}
Thus, the desired result can be obtained combining Eqs.\ \ref{eq:eps2} and\ \ref{eq:eps3} with the triangle inequality
\begin{align}
    \frac{1}{\sqrt{2^n}}\norm{A -  B}_2 \leq  \frac{1}{\sqrt{2^n}}\norm{A -  A^{(L)}}_2 \leq  + \frac{1}{\sqrt{2^n}}\norm{A^{(L)} -  B}_2 \leq {\epsilon}.
\end{align}
\end{proof}
}

\subsection{Learning quantum Boolean functions with low total influence}

In addition, we also demonstrate that quantum Boolean functions with constant total influence can be learnt efficiently with quantum statistical queries. 
{\color{black}
We establish this result leveraging the Quantum Friedgut’s Theorem provided in Ref.\ \cite{rouze2024quantum}.
\begin{lem}[Quantum Friedgut’s Theorem, Theorem 3.11 in Ref.\ \cite{rouze2024quantum}]
\label{lem:rouze}
 Let $A \in \calL_n$ be a linear operator and let $\epsilon \in (0,1]$. There exists a set $T \subseteq [n]$ of $n -k$ qubits, where
 \begin{align}
     k 
     \leq 
     2^{\frac{30 \mathrm{Inf}^2(A)}{\epsilon^2}}\frac{\mathrm{Inf}^1(A)^6}{\mathrm{Inf}^2(A)^5}\norm{A}^4,
 \end{align}
 such that
 \begin{align}
     \frac{1}{\sqrt{2^n}} \norm{A - \frac{I}{2^{\abs{T}}}\otimes \Tr_T(A) }_2 \leq \epsilon.
 \end{align}
\end{lem}
}
We can now state our result on Quantum Boolean Functions with constant total influence.

\begin{thm}[Learning Quantum Boolean Functions with constant total influence]
\label{thm:gl-accuracy}
Given a quantum Boolean function $A$ and a value $\epsilon \in (0,1]$, let $k$ be
\begin{align}
    k =  2^{\frac{270 \mathrm{Inf}^2(A)}{\epsilon^2}}\frac{\mathrm{Inf}^1(A)^6}{\mathrm{Inf}^2(A)^5}.
\end{align}
\begin{enumerate}
    \item There is an algorithm running in time $\poly(n,2^k)$ that accesses $A$ via $\mathsf{QStat}_A$ queries with tolerance at least $\Omega(4^{-k})$ and outputs a quantum Boolean function $B$ such that
    \begin{align}
        D(A,B) \leq \epsilon.
    \end{align}
    \item {\color{black} If $A$ is a Heisenberg-evolved Pauli observable, i.e. if $A = U^\dag P U$ for some unitary $U$ and Pauli operator $P$, then there is an algorithm running in time $\poly(n,2^k)$ that access the unitary $\widetilde{A} \coloneqq U^\dag \exp\left(i\frac{\pi}{6}P\right) U$ via $\mathsf{QStat}_{\widetilde{A}}$ queries with tolerance at least $\Omega(4^{-k})$ and outputs an operator $B$ such that
\begin{align}
    \frac{1}{\sqrt{2^n}}\norm{A-B}_2 \leq \epsilon.
\end{align}
}
\end{enumerate} 
\end{thm}
\begin{proof}
We can adapt the proof of  Proposition 6.7 in Ref.~\cite{rouze2024quantum} to the QSQ setting by replacing all the oracle access queries to $A$ with queries to $\textsf{QStat}_A$. 
{\color{black}
Given $\gamma >0$,  we can use the GL algorithm to output a list $L$ of Pauli operators such that:
\begin{enumerate}
    \item if $|\widehat{A}_P|\geq\gamma$, then $P\in L$;
    \item for all $P\in L$, $|\widehat{A}_P|\geq\gamma/2$.
\end{enumerate}
By Theorem\ \ref{thm:GL}, the list $L$ can be obtained in time $\calO\left(n \gamma^{-2}\right)$ via queries to $\mathsf{QStat}_A$ with tolerance at least $\Omega\left(\gamma^2\right)$. 
By Lemma\ \ref{lem:parseval}, the list $L$ contains at most $\calO(\gamma^{-2})$ elements.
Let $A^{(L)} \coloneqq \sum_{P\in L} \widehat{A}_P P$.  By Lemma\ \ref{lem:coeff}, we can estimate an operator $B$ in time $\calO(\gamma^{-2})$ via $\mathsf{QStat}_A$ queries with tolerance at least $\tau^2$, satisfying the following: 
\begin{align}
    \min_{\xi  \in \{-1,1 \}} \frac{1}{2^n}\norm{A^{(L)} - \xi B}_2^2 
    \leq 4 \gamma^{-2} \tau^2.
\end{align}
Moreover, by Lemma\ \ref{lem:coeff2}, we can estimate an operator $B$ in time $\calO(\gamma^{-2})$ via $\mathsf{QStat}_{\widetilde{A}}$ queries with tolerance at least $\tau^2$, satisfying the following: 
\begin{align}
    \frac{1}{2^n}\norm{A^{(L)} -  B}_2^2 
    \leq 4 \gamma^{-2} \tau^2.
\end{align}
It remains to show that $A^{(L)}$ and $A$ are close in normalized Schatten 2-distance.
By Lemma\ \ref{lem:rouze}, there exists a subset of qubits $T \subseteq [n]$ of size $\abs{T} \geq n- k$ such that
 \begin{align}
     \frac{1}{{2^n}} \norm{A - \frac{I^{\otimes \abs{T}}}{2^{\abs{T}}} \otimes \Tr_T(A) }_2^2 \leq \frac{\epsilon^2}{9}.
 \end{align}
Let $L'$ be the a subset of Pauli operators defined as follows
\begin{align}
    L' = \{ P \in \calP_n \,|\, \Tr_{{T}}(P) \neq 0\},
\end{align}
$L'$ is the set of Pauli operators that act non trivially only on the set $[n]-T$. Thus $L'$ contains $4^{n - \abs{T}} \leq 4^k$ elements and, moreover, 
\begin{align}
    A - \frac{I^{\otimes \abs{T}}}{2^{\abs{T}}} \otimes \Tr_T(A)  = \sum_{P\not\in L'} \widehat{A}_P P.
\end{align}
Following Ref.\ \cite{rouze2024quantum}, we can upper bound the normalized Schatten 2-distance between $A$ and $A^{(L)}$:
\begin{align}
    \frac{1}{4^n}\norm{A - A^{(L)}}_2^2 = &\sum_{P\not\in L} \widehat{A}_P^2
    =\sum_{\substack{P\in\calP_n:\\ P\not\in L, P\in L' }} \widehat{A}_P^2 
    + \sum_{\substack{P\in\calP_n:\\ P\not\in L, P\not\in L' }} \widehat{A}_P^2 
    \\ &\leq \sum_{\substack{P\in\calP_n:\\ P\not\in L, P\in L' }} \widehat{A}_P^2 
    + \sum_{\substack{P\in\calP_n:\\ P\not\in L' }} \widehat{A}_P^2  
    \leq 4^k \gamma^2 + \frac{\epsilon^2}{9}.
\end{align}
Putting all together, we obtain
\begin{align}
    &\frac{1}{\sqrt{2^n}}\norm{A - B}_2 \leq \frac{1}{\sqrt{2^n}}\norm{B- A^{(L)}}_2 + \frac{1}{\sqrt{2^n}}\norm{A - A^{(L)}}_2 \\\leq &4 \gamma^{-2} \tau^2+ \sqrt{4^k \gamma^2 + \frac{\epsilon^2}{3}} \leq 4 \gamma^{-2} \tau^2 + 2^k \gamma + \frac{\epsilon}{3},
\end{align}
therefore choosing $\gamma = \epsilon/(3\cdot2^{k}), \tau = {\epsilon^{3/2}}/({3\cdot2^{k+1}}) $ yields the desired result:
\begin{align}
    \frac{1}{\sqrt{2^n}}\norm{A - B}_2 \leq \epsilon.
\end{align}
}

\begin{remark}
The accuracy guarantees of Theorem~\ref{thm:gl-accuracy} are cast in terms of $\Inf^1(A), \Inf^2(A)$. In Ref.\ \cite{rouze2024quantum}, it was demonstrated that these parameters can be bounded for an observable evolved by a shallow circuit (in the Heisenberg picture),  by using a variant of the light-cone argument. We now introduce some further notation to state their claim. For any $j\in [n]$, let $N_j \subseteq [n] $ be the minimal set of qubits such that $\frac{\Tr_j}{2}\left(U \frac{\Tr_{N_j}}{2^{|N_j|}}(O)U^\dag\right) = U \frac{\Tr_{N_j}}{2^{|N_j|}}(O)U^\dag $ for any $O \in \mathcal{L}_{n}$ and denote $L := \max_i|\{j : i \in N_j\}|$. Then, if $O$ is a quantum Boolean function with $\Inf^1(O), \Inf^2(O),  \|O\|_2 = \mathcal{O}(1)$, and $U$ is a unitary with $L = \mathcal{O}(1)$, we can learn evolution in the Heisenberg picture $U^\dag O U$ by means of Theorem~\ref{thm:gl-accuracy} by picking $k = \mathcal{O}(1)$. This ensures that the algorithm runs in $\poly(n)$ time and that the statistical queries have constant tolerance.  
\end{remark}

\end{proof}

\section{Quantum juntas}
A unitary $U \in \mathcal{U}_{n}$ is a quantum $k$-junta if there exists $S \subseteq [n]$ with $|S| = k$ such that
\begin{align}
U = V_S \otimes I_{\overline{S}}
\end{align}
for some $V_S\in\mathcal{U}_{k}$. 
For a Pauli string $P = \bigotimes_{i \in [n] } P_i\in\mathcal{P}_n$, we denote the reduced string as $P_S=\bigotimes_{i \in S } P_i\in\mathcal{P}_k$.
We now consider the Pauli expansions $U = \sum_{P\in\mathcal{P}_n} \widehat{U}_P P$ and
$V_S = \sum_{P_S\in\mathcal{P}_k} \widehat{V}_{P_S} P_S$. Their coefficients satisfy the following relation.
\begin{align}
  \widehat{U}_P = \frac{1}{2^n} \Tr[UP] = \frac{1}{2^n} \Tr[V_S P_S] \Tr[P_{\overline{S}} I_{\overline{S}}] = \begin{cases}
   \widehat{V}_{P_S} & \text{ if $\mathrm{supp}(P) \in S$,}   \\
   0 & \text{ else. }
  \end{cases}
\end{align}
As for the Choi state, we have
\begin{align}
    \ket{v(U)} = \sum_{P\in\mathcal{P}_n} \widehat{U}_P \ket{v(P)} = \sum_{\mathrm{supp}(P) \in S}\widehat{V}_{P_S} \ket{v(P_S\otimes I_{\overline{S}})} = \ket{v(V_S)} \ket{v(I_{\overline{S}})}. 
\end{align}
We will now show that quantum $k$-juntas are efficiently learnable in our model. Our proof combines the techniques used in Ref.~\cite{chen2023testing} for learning quantum $k$-juntas from oracle access and the ones used in Ref.~\cite{arunachalam2020quantum} for learning (classical) $k$-juntas with quantum statistical queries.
Note that the algorithm given in (\cite{chen2023testing}, Theorem 28) has query complexity independent of $n$. Crucially, their algorithm involves a Pauli sampling as a subroutine to estimate the support of the Pauli strings with non-zero Fourier coefficients. We replaced this procedure by estimating the influences of each qubit by means of Lemma~\ref{lem:qbf}, introducing an additional factor $n$ in the query complexity.
\begin{algorithm}[H]
\caption{Learning quantum $k$-juntas with statistical queries}
\label{alg:juntas}
\begin{algorithmic}
    \For{$i=1$ to $n$}
    \State{Estimate $\Inf^2_i(U)$ with a quantum statistical query with accuracy $\varepsilon^2/(20k)$ and store the result in the variable $\alpha_i$.}
    \EndFor
    \State{Define the subset $T = \left\{ i \in [n] : \alpha_i \geq \varepsilon^2/(16k)\right\}$ and consider the set $T_2$, which includes the qubits in $T$ and the associated qubits in the dual space.}
    \For{ $P\in \mathcal{P}_{|T_2|}$}
    \State{Produce an estimate $o_p$ of 
    \begin{align}\Tr[P \cdot \ket{v(I^{\otimes (n-\ell)})}\bra{v(I^{\otimes (n-\ell)})} \cdot \left(\ket{v(U)}\bra{v(U)}\right)\cdot \ket{v(I^{\otimes (n-\ell)})}\bra{v(I^{\otimes (n-\ell)})} ]\end{align} with a quantum statistical query with tolerance $2^{-\ell} \epsilon/3$.}
    \State{Set $x_P = \min\{o_P,1\}$.}
    \EndFor
    \State{Reconstruct the density matrix $\widehat{\rho}_T = \frac{1}{2^{2\ell}}\left(I^{\otimes {2\ell}}     +\sum_{\substack{P\in\mathcal{P}_{2\ell}\setminus I^{\otimes 2\ell}}} x_P P\right)$ and compute its dominant eigenstate $\ket{\widehat{\psi}_T}$.}
    \State{Compute $W$ such that $\ket{v(W)}:=\ket{\widehat{\psi}_T}$}
    \State \textbf{return $W \otimes I^{\otimes (n-\ell)}$.} 
\end{algorithmic}
\end{algorithm}
\begin{thm}[Learning quantum $k$-juntas via QSQs]
\label{thm:juntas}
Let $U$ be a quantum $k$-junta. There is a $\poly(n,2^k, \varepsilon)$-time algorithm that accesses the state $\ket{v(U)}$ via $\mathsf{QStat}_U$ queries with tolerance $\poly(2^{-k},\varepsilon)$ and outputs a unitary $\widetilde{U}$ such that
\begin{equation}
    \mathrm{D}(U,\widetilde{U})\leq \varepsilon.
\end{equation}
\end{thm}
\begin{proof}
Throughout this proof, we will use the following notation to deal with the reduced Choi state with respect to a given subset of the qubits. Recall that the Choi state is a state over a set of $2n$ qubits, which we label as $\{i_1,i_2,\dots,i_n,i'_1,i'_2,\dots,i'_n\}$. For $S = \{i_{j+1}, i_{j+2}, \dots\}\subseteq \{i_1,i_2,\dots,i_n\}$ we will denote $S_2 := \{i_{j+1}, i_{j+2}, \dots\}\cup \{i'_{j+1}, i'_{j+2}, \dots\}$. Clearly, $|S_2| = 2|S|$.

Our algorithm consists in two separate steps: first we perform $n$ $\mathsf{QStat}_U$ queries with tolerance $\Theta(\varepsilon^2/k)$ to learn a subset $T\subseteq [n]$ containing all the variables $i$ for which $\Inf^2_i(U) \geq \varepsilon^2/(16k)$. Next we will define a reduced state on the subset $T_2$ and we will learn it by performing a state tomography with $4^{2|T|} - 1$ $\mathsf{QStat}_U$ queries with tolerance $\Omega(\varepsilon 4^{-2k})$. 

Let $U$ be a quantum $k$-junta over the subset $Q\subseteq [n]$. Then, it is not hard to see that $\Inf^2_i(U) = 0$ if $i \not\in Q$.
For each $j\in[n]$, we use Lemma~\ref{lem:qbf} to estimate $\Inf^2_j(U) \pm \varepsilon^2/(20k) $ via a single $\mathsf{Qstat}_U$ query. Suppose the outcomes of these queries are $\alpha_1,\dots \alpha_n$, and let 
\begin{align}
T = \left\{ i \in [n] : \alpha_i \geq \varepsilon^2/(16k)\right\}.
\end{align}
We observe that $T \subseteq Q$, as $\Inf^2_i(U) = 0$ implies that $\alpha_i \leq \varepsilon^2/(20k)$. On the other hand, for every $i \in Q \setminus T$, we have that $\Inf^2_i(U) < \varepsilon^2/(8k)$. Assume by contradiction that $i \not\in T$ and $\Inf^2_i(U) \geq \varepsilon^2/(4k)$. Then we have:
\begin{align}
\alpha_i \geq \Inf^2_i(U) - \frac{\varepsilon}{20k} > \frac{\varepsilon^2}{16k},
\end{align}
contradicting the fact that $i \not\in T$. As a consequence,
\begin{equation}
\label{eq:inf-junta}
   \sum_{i \in \overline{T}} \Inf^2_i(U) = \sum_{i \in Q \setminus T} \Inf^2_i(U) \leq k \cdot \frac{\varepsilon^2}{8k} = \frac{\varepsilon^2}{8}, 
\end{equation}

where the inequality follows from $|Q|\leq k$.
We now describe the second phase of the learning algorithm.
Let $|T|=\ell$ and consider the identity operator $I^{\otimes (n-\ell)}$  acting on the subset $\overline{T}$.
Let $\rho$ be the state obtained by measuring $\ket{v(U)}$ according to the projectors 
\begin{align}
    \left\{\ket{v(I^{\otimes (n-\ell)})}\bra{v(I^{\otimes(n-\ell)})},I^{\otimes (n-\ell)} - \ket{v(I^{\otimes (n-\ell)})}\bra{v(I^{\otimes(n-\ell)})}\right\},
\end{align} and then conditioning on the first outcome,
\begin{align}
  \ket{\psi}:= \frac{\left(I^{\otimes(\ell)}\otimes\ket{v(I^{\otimes (n-\ell)})}\bra{v(I^{\otimes (n-\ell)})}\right)\ket{v(U)}}{\left|\left(\Tr_{T_2} \bra{v(U)}\right)\ket{v(I^{\otimes (n-\ell)})}\right|}
  := \ket{v(V \otimes I^{\otimes (n - \ell)})},  
\end{align}
where in the last line we introduced the $\ell$-qubit unitary $V$ such that $\ket{\psi}$ is the state isomorphic to $V \otimes I^{\otimes (n - \ell)}$.
We make the following claim on the distance between $U$ and $V\otimes I^{\otimes (n - \ell)}$, which we will prove in the following.
\begin{claim}
\label{claim1}
$D(U,V\otimes I^{\otimes (n - \ell)})\leq {\varepsilon}/{2}.$
\end{claim}

Denote $\rho:= \ket{\psi}\bra{\psi}$. We will learn $\rho_{T_2} = \Tr_{\overline{T_2}}[\rho]$ by performing a state tomography via $\textsf{QStat}$ queries on a reduced state of $2\ell$ qubits. To this end, we query all $4^{2\ell} - 1$ non-identity Pauli strings with support on $T$ with tolerance $\tau=\epsilon 2^{-2\ell - 1}$. For all $P \in \mathcal{P}_{2\ell}=\{I,X,Y,Z\}^{\otimes 2\ell }$, denote  the obtained outcome by 
\begin{align}
o_P = \Tr[P \cdot \ket{v(I^{\otimes (n-\ell)})}\bra{v(I^{\otimes (n-\ell)})} \cdot \left(\ket{v(U)}\bra{v(U)}\right)\cdot \ket{v(I^{\otimes (n-\ell)})}\bra{v(I^{\otimes (n-\ell)})} ] \pm \tau
\end{align}
and set $x_P = \min\{o_P,1\}$.
Denote the estimated $2\ell$-qubit state by 
\begin{align}
\widehat{\rho}_T = \frac{1}{2^{2\ell}}\left(I^{\otimes {2\ell}}     +\sum_{\substack{P\in\mathcal{P}_{2\ell}\setminus I^{\otimes 2\ell}}} x_P P\right).\end{align}

Let $\ket{\widehat{\psi}_T}$ be the dominant eigenstate of $\widehat{\rho}_T $ and let $W$ be the unitary encoded by the state $\ket{\widehat{\psi}_T}$, i.e. let $\ket{v(W)}:=\ket{\widehat{\psi}_T}$.
We make a further claim and we delay its proof to the end.
\begin{claim}
\label{claim2}
$D(V,W)\leq {\varepsilon}/{2}$.
\end{claim}
Then the theorem follows by combining Claims~\ref{claim1} and~\ref{claim2} with the triangle inequality and letting $\widetilde{U} = W\otimes I^{\otimes (n-\ell)}$.
\end{proof}
We present the proofs of Claims~\ref{claim1} and~\ref{claim2} below.
\begin{proof}[Proof of Claim~\ref{claim1}]
Recall that $U = U_Q \otimes I_{\overline{Q}}$ is a $k$-junta which acts non trivially only on the set $Q$ and that $T\subseteq Q$ is the set of qubits with { large influence} learnt by the algorithm. It is sufficient to show that $\mathrm{dist}(U_Q,V)\leq \varepsilon/2$. First, we observe that 
$\ket{v(U)} = \ket{v(U_Q)}\otimes\ket{v(I^{\otimes(n-k)})}$.
We will need the following decomposition of $\ket{v(U_Q)}$:
\begin{align}
   \ket{v(U_Q)} &=\sum_{P_Q\in\mathcal{P}_k} \widehat{U}_{P_Q} \ket{v(P_Q)}    
   & = \sum_{\substack{P_Q\in\mathcal{P}_k\\\mathrm{supp}(P_Q)\cap \overline{T} = \emptyset}} \widehat{U}_{P_{Q}} \ket{v(P_Q)} + \sum_{\substack{P_Q\in\mathcal{P}_k :\\\mathrm{supp}(P_Q)\cap \overline{T} \neq \emptyset}} \widehat{U}_{P_Q } \ket{v(P_Q)},
\end{align}
where $\widehat{U}_{P_Q} = \widehat{U}_{P_Q\otimes I_{n-k}}$.
Similarly, we can expand  $\ket{v(V)}\otimes \ket{v(I^{\otimes (k-\ell)})}$ as follows
\begin{align}
    \ket{v(V)}\otimes \ket{v(I^{\otimes (k- \ell)})} &=\sum_{\substack{P_Q\in\mathcal{P}_k\\\mathrm{supp}(P_Q)\cap \overline{T} = \emptyset}} \widehat{U}_{P_{Q}} \ket{v(P_Q)} + \sum_{\substack{P_Q\in\mathcal{P}_k :\\\mathrm{supp}(P_Q)\cap \overline{T} \neq \emptyset}} \widehat{U}_{P_Q } \ket{v(I^{\otimes k})} 
\end{align}
Recall that the total influence of the qubits in $\overline{T}$ is at most $\varepsilon^2/8$. This immediately implies a lower bound on the inner product between $\ket{v(V)}\otimes \ket{v(I^{\otimes (k-\ell)})}$ and $\ket{v(U_Q)}$.
\begin{align}
  \left|\left(\bra{v(V)}\otimes \bra{v(I^{\otimes (k-\ell)})}\right)\ket{v(U_Q)}\right|^{ 2} =& \sum_{\substack{P_Q \in \mathcal{P}_k:\\\mathrm{supp}(P)\cap \overline{T} \neq \emptyset}} |\widehat{U}_{P_Q}|^2 
  \\=& 1 -  \sum_{\substack{P_Q \in \mathcal{P}_k:\\\mathrm{supp}(P)\cap \overline{T} \neq \emptyset}} |\widehat{U}_{P_Q}|^2 \geq 1- \frac{\varepsilon^2}{8},
\end{align}
where the inequality is a direct application of Eq.~\ref{eq:inf-junta}. We can now prove the desired result
\begin{align}
   D^2(U,V\otimes I^{\otimes (n-\ell)}) =  D^2(U_Q,V\otimes I^{\otimes (k-\ell)}) = 1 - |\braket{v(V)}|{v(U_Q)}|^2\leq \frac{\varepsilon^2}{4},
\end{align}
where we used the stability of $D(\cdot,\cdot)$ under tensor product.
\end{proof}
\begin{proof}[Proof of Claim~\ref{claim2}]
We just need to ensure the following:
\begin{equation}
    \|\widehat{\rho}_{T_2} - \rho_{T_2}\|_2 \leq \frac{\epsilon}{2}.
\end{equation}
We first make a preliminary observation. Let $c_P := \Tr[P \rho_{T_2}]$. Then,
\begin{equation}
    (x_p - c_p)^2 \leq  \left(c_p \frac{\varepsilon^2}{8} + \tau\right)^2 
\end{equation}
This allow to upper bound the distance between the partial state $\rho_{T_2}$ and its estimate $\widehat{\rho}_{T_2}$. 
\begin{align}
    \|\rho_{T_2} - \widehat{\rho}_{T_2}\|_2^2 = &\Tr[(\rho_T - \widehat{\rho}_T)^2] 
     = \frac{1}{
    16^\ell} \Tr\left[\left(\sum_{\substack{P \in \mathcal{P}_{2\ell}\setminus I^{\otimes 2\ell}}} (c_P- x_P)P\right)^2\right] 
    \\ = &\frac{1}{4^\ell} \sum_{\substack{P \in  \mathcal{P}_{2\ell}\setminus I^{\otimes 2\ell}}} (c_P - x_P)^2
    \leq  \frac{2}{4^\ell} \left(\sum_{\substack{P \in  \mathcal{P}_{2\ell}\setminus I^{\otimes 2\ell}}}  \frac{\varepsilon^4}{64} c_p^2 + \tau^2 \right)
    \leq \frac{\varepsilon^4}{32} + 4^\ell \tau^2,
\end{align}
where we used the inequality $(x+y)^2 \leq 2(x^2 + y^2)$ and the fact that the purity $\Tr\left[\rho_{T_2}^2\right] = 4^{-\ell}\sum_{P\in\mathcal{P}_{2\ell}} c_P^2$ is bounded by 1.
Then picking $\tau =  2^{-\ell} \epsilon/3$ ensures the desired upper bound.
By proceeding as in the proof of Lemma~\ref{lem:tomography}, we have
\begin{equation}
    D(V,W) \leq \frac{\epsilon}{2},
\end{equation}
as desired.
\end{proof}

\section{Constant-depth circuits}
As a first application of the tools introduced before, we show that very shallow circuits are learnable query-efficiently with QSQs according to a locally scrambled distribution. We will rely on the following recent result of~\cite{yu2023learning}, which essentially shows that ``learning marginal suffices'', i.e. learning the $k$-reduced density matrices of a state produced by a shallow circuit allows to perform a state tomography.
\begin{thm}[Adapted from~\cite{yu2023learning}, Theorem 4.3]
\label{thm:marginal}
Let $\psi = \ketbra{\psi}{\psi}$ a state produced by a circuit of depth at most $D$. For any state $\rho$, one of the following conditions
must be satisfied: either $\|\rho-\psi\|_{\tr} < \epsilon$; or $\|\rho_s-\psi_s\|_{\tr} > {\epsilon^2}/{n}$ for some $s\subseteq \{0,1,\dots,n-1\}$ with $|s|= 2^D$.
\end{thm}
An application of this result was also given in Ref.~\cite{arunachalam2023on}, where the authors showed that the class of $n$-qubit trivial states is learnable with $\poly(n)$ quantum statistical queries.
We now extend their result from states to unitaries.
\begin{thm}[Learning constant-depth circuits via QSQs]
\label{thm:constant}
Let $\mathcal{C}$ the class of $\mathcal{O}(1)$-depth circuits. Then for all $U\in\mathcal{C}$, there exists an algorithm that makes $\poly(n)$ queries to $\mathsf{QStat}_U$ with tolerance at least $\frac{\epsilon^2}{4  n}\cdot2^{-D/2}$ and returns a unitary $W\in\mathcal{U}_n$ such that
\begin{equation}
\label{eq:risk-shallow}
    D(U,W)\leq \epsilon.
\end{equation}
\end{thm}
\begin{proof}
Let $D$ be the depth of the circuit.
First, we consider the Choi state $\ket{v(U)} = I \otimes U \ket{\Omega}$ and recall that $\ket{\Omega}$ can be produced with a circuit of depth 2 over $2n$ qubits. Then we have $\ket{v(U)} = V\ket{0^{2n}} $ for a suitable unitary $V \in \mathcal{U}_{2n}$ implemented by a circuit of depth $D+2$. Let $k = 2^{D+2}$. Then it suffices to learn all the $k$-local reduced density matrices of the states $\ket{v(U)}$. There are $\binom{2n}{k} =\mathcal{O}\left(n^{2^D}\right)$ of them and each of them is learnable in trace distance with accuracy $\frac{\epsilon^2}{2n}$ by performing $4^{D+2}$ quantum statistical queries with tolerance $\frac{\epsilon^2}{4n} \cdot 2^{-D/2}$ by means of Lemma~\ref{lem:tomography}. 
We can thus determine thanks to Theorem~\ref{thm:marginal} a state $\ket{v(W)}$ such that $D(U,W) \coloneqq \|\ketbra{v(W)}{v(W)} -  \ketbra{v(U)}{v(U)}\|_{\tr} \leq \epsilon$. 
\end{proof}

\section{Exponential separations between QSQs and Choi state access}
We will now prove a lower bound for learning Choi states with QSQs, and derive from it an exponential separation between learning unitaries from QSQs and learning unitaries with Choi state access.
To this end, we combine Lemma~\ref{lem:equiv} with an argument based on the following concept class (of classical functions):
\begin{equation}
    \mathcal{C} = \left\{f_A : \{0,1\}^n \rightarrow \{0,1\}, f_A(x) =x^\top A x \mod 2 \;\; |\;\;  A \in \mathbb{F}_2^{n\times n}\right\}.
\end{equation}
As shown in Ref.~\cite{arunachalam2023on}, learning these functions via quantum statistical queries to associated \emph{quantum examples} is hard.
{\color{black}
\begin{lem}[Hardness of learning phase states, Theorem 17 in Ref.~\cite{arunachalam2023on}]
The concept class of phase oracle unitaries states $\ket{\psi_A}$, i.e.
\begin{equation}
    \left\{\ket{\psi_A} \coloneqq \frac{1}{\sqrt{2^n}} \sum_{x\in\{0,1\}^n} \ket{x,f(x)}\;\; \bigg|\;\;  f \in \mathcal{C}\right\}
\end{equation} 
requires $2^{\Omega(n)}$ many quantum statistical queries to $\mathsf{QStat}_{{f_A}}$ of tolerance $1/\poly(n)$  to be learnt below distance $D < 0.05$ with high probability.    
\end{lem}
In this section, we will establish an analogous result for learning phase-oracle unitaries encoding the functions in the above class. To this end, we first state some preliminary lemmas.

\begin{lem}[Quantum Statistical Dimension, Lemma 15 in Ref.\ \cite{arunachalam2023on}]
\label{lem:qsd}
    Let $\tau, \epsilon \in [0,1]$ such that $\epsilon \geq \tau$, $\mathcal{C}$ be a set of $n$-qubit quantum states and $\sigma \not \in \mathcal{C}$ be an $n$-qubit state. The associated quantum statistical dimension is 
    \begin{align}
        \mathsf{QSD}_\tau(\calC , \sigma) \coloneqq \sup_\mu \min_{M: \norm{M}\leq 1} \left(\Pr_{\rho \sim \mu}[\Tr[M(\rho - \sigma)] > \tau]\right)^{-1}.
    \end{align}
    where the supremum is taken over all the distribution over the set $\calC$.
    Moreover, assume that $\min_{\rho \in \calC} \norm{\rho - \sigma} > 2(\tau + \epsilon)$. Then learning a state $\rho \in \mathcal{C}$ in trace distance with precision $\epsilon$ and success probability $1-\delta$ requires at least
    \begin{align}
        (1-2\delta)\mathsf{QSD}_\tau(\calC , \sigma) 
    \end{align}
    queries to $\mathsf{QStat}_\rho$ with tolerance $\tau$.
\end{lem}

\begin{lem}[Variance bound, Theorem 16 in Ref.\ \cite{arunachalam2023on}]
\label{lem:variance}
Let $\tau > 0$ and let $\calC$ be a set of $n$-qubit quantum states. Let $\mu$ be a distribution over $\mathcal{C}$ such that $\mathbb{E}_{\rho \sim \mu} [\rho] \not\in \calC$. Then:
\begin{align}
    \mathsf{QSD}_\tau(\calC , \mathbb{E}_{\rho \sim \mu} [\rho] )\geq \tau^2 \cdot  \min_{M: \norm{M}\leq 1} \left(\mathbb{V}_{\rho \sim \mu}\Tr[\rho M]\right)^{-1},
\end{align}
where
\begin{align}
    \mathbb{V}_{\rho \sim \mu}\Tr[\rho M] = \mathbb{E}_{\rho \sim \mu}\left\{ \Tr[\rho M]^2\right\} - \left\{\mathbb{E}_{\rho \sim \mu}\Tr[\rho M]\right\}^2.
\end{align}
\end{lem}

\begin{lem}[Schwartz-Zippel lemma, Lemma 2.6 in Ref.~\cite{nisan1994degree}]
\label{lem:sch-zipp}
Let $p : \{x_1,x_2,\dots, x_n\} \rightarrow\{0,1\}$ be a non-zero multilinear polynomial of degree $d$. If we choose $x_1,x_2,\dots, x_n$, independently and uniformly at random in $\{0,1\}^n$,
then the following inequality holds:    
\begin{align}
    \Pr[p(x_1,x_2,\dots, x_n) \neq 0] \geq 2^{-d}
\end{align}
\end{lem}

\begin{coro}
\label{cor:zs}
For distinct $A, B \in \mathbb{F}^{n\times n}_2$, we have that 
\begin{align}
    \Pr [f_A(x) \neq f_B(x) ]\in \left[\frac{1}{4}, \frac{3}{4}\right],
\end{align}
where $x_1,x_2,\dots, x_n$ are chosen independently and uniformly at random in $\{0,1\}^n$,
\end{coro}
\begin{proof}
Recall that $f_A(x) \coloneqq x^\top A x \mod 2$, and therefore 
\begin{align}
    p_{A,B} (x) \coloneqq f_A(x) + f_B(x) \mod 2 = \sum_{i,j=1}^n (A_{ij} + B_{ij})x_i x_j \mod 2.  
\end{align}   
In order to prove the desired statement, it suffices to show that
\begin{align}
    & \Pr[p_{A,B} (x) = 0]  \geq \frac{1}{4} \label{eq:sz1}
   \\ \text{and } \quad & \Pr[p_{A,B} (x)  = 1]  \coloneqq \Pr[q_{A,B} (x) = 0]  \geq \frac{1}{4} \label{eq:sz2},
\end{align}
where we denoted $q_{A,B} \coloneqq p_{A,B} (x) + 1 \mod 2$.
Since both $p_{A,B}$ and $q_{A,B}$ are polynomials of degree $2$, Eqs.~\ref{eq:sz1} and~\ref{eq:sz2} follow from Lemma~\ref{lem:sch-zipp}.
\end{proof}

}

We can finally state the main result of this section.
\begin{thm}[Hardness of learning phase oracles]
\label{thm:hardness}
The concept class of phase oracle unitaries $V_{f_A}$, i.e.
\begin{equation}
    \{V_{f_A}\;\; |\;\;  A \in \mathbb{F}_2^{n\times n}\}
\end{equation} 
requires $2^{\Omega(n)}$ many quantum statistical queries to $\mathsf{QStat}_{V_{f_A}}$ of tolerance $1/\poly(n)$  to be learnt below distance $D < 0.05$ with high probability.
\end{thm}
\begin{proof}
Let $\mathcal{D}$ be the uniform distribution over $\mathbb{F}_2^{n \times n}$, i.e. in order to sample $B \sim \mathcal{D}$ we sample the entries $B_{ij}$ independently and uniformly from $\mathbb{F}_2$.
We will prove the theorem leveraging Lemmas~\ref{lem:qsd} and \ref{lem:variance}. 
Thus, we first need to lower bound the trace distance between $\ketbra{v(V_{f_A})}{v(V_{f_A})}$ and the first moment $\mathbb{E}_{B\sim \calD}  \ketbra{v(V_{f_B})}{v(V_{f_B})}$.
We obtain
\begin{align}
    &\norm{\ketbra{v(V_{f_A})}{v(V_{f_A})} - \mathbb{E}_{B\sim \calD}  \ketbra{v(V_{f_B})}{v(V_{f_B})}}_{\tr}
    \\ \geq &1 - \sqrt{\mathbb{E}_B \big|\left\langle v(V_{f_A}) \big| v(V_{f_B})\right\rangle\big|^2}
    \\ = & 1 -\sqrt{\mathbb{E}_B \left( 1 -2\Pr_{x \in \{0,1\}^n} [f_A(x) \neq f_B(x)]\right)^2}
    \\ \geq &1 - \sqrt{\frac{(2^{n^2} -1)\frac{1}{2} + 1 }{2^{n^2}}} \geq 1 - \sqrt{\frac{3}{4}} \geq 0.13,
\end{align}
where the first inequality follows from the lower bound on trace distance by fidelity~\cite{fuchs1999cryptographic} and the second by Corollary~\ref{cor:zs}.
Fix $\epsilon = 0.05$ and assume without loss of generality that $\tau \leq 0.015$. Then by Lemma\ \ref{lem:variance}, the number of required quantum statistical queries is at least
\begin{align}
    (1-2\delta)\mathsf{QSD}_\tau(\mathcal{C}, \mathbb{E}_{B\sim \calD}  \ketbra{v(V_{f_B})}{v(V_{f_B})}).
\end{align}
We can upper bound $\mathsf{QSD}_\tau(\mathcal{C}, \mathbb{E}_{B\sim \calD}  \ketbra{v(V_{f_B})}{v(V_{f_B})})$ via Lemma\ \ref{lem:variance}.
By Lemma~\ref{lem:equiv}, for any $2n$-qubit observable $M$, there's a $(n+1)$-qubit observable $\Phi(M)$ such that
\begin{align}
    \bra{v(f_A)} M \ket{v(f_A)} = \bra{\psi_A} \Phi(M) \ket{\psi_A} 
\end{align}
and therefore,
\begin{align}
    \max_M \mathbb{V}_A \bra{v(f_A)} M \ket{v(f_A)} =\max_{M} \mathbb{V}_A \bra{\psi_A} \Phi(M) \ket{\psi_A} 
    \leq \max_{M'} \mathbb{V}_A \bra{\psi_A} M' \ket{\psi_A}  \in 2^{-\Omega(n)},
\end{align}
where the final inclusion follows again from Eq. 36 in Ref.~\cite{arunachalam2023on}.
Thus, the theorem follows by the variance bound in Lemma~\ref{lem:variance}.
\end{proof}
On the other hand, the unitary $V_{f_A}$ is efficiently learnable from separable measurements to Choi states. This is an immediate consequence of a result given in Ref.~\cite{arunachalam2023optimal}, which establishes that phase oracle states with constant degree $d$ are efficiently learnable. Therefore, the function ${f_A}$ is efficiently learnable from separable measurements to phase states, defined as $\ket{\phi_{f_A}} = {2^{-n/2}} \sum_{x\in\{0,1\}^n} (-1)^{f_A(x)} \ket{x}$. We observe the $ \ket{v(V_{f_A})}  = 2^{-n/2} \sum_{x\in\{0,1\}^n} (-1)^{f_A(x)} \ket{x,x}$, then adapting the argument to Choi states is straightforward.

We also provide a double exponential lower bound for testing properties of channels, hinging on a lower bound for testing purity of states given in Ref.~\cite{arunachalam2023on}.
{\color{black}
\begin{lem}[Hardness of testing purity, Theorem 25 in Ref.~\cite{arunachalam2023on}]
\label{lem:purity}
Let $\mathcal{A}$ be an algorithm that upon an input of quantum state $\rho$
estimates the purity of $\rho$ with error at most $ 1/4$ using $\mathsf{QStat}_\rho$ queries with tolerance at least $\tau$. Then $\mathcal{A}$ must make at least $2^{\Omega(\tau^2 2^n)}$
such queries.
\end{lem}
}
Exploiting the relation between unitarity and purity of the Choi state, we derive the following Corollary.
\begin{coro}[Hardness of testing unitarity]
\label{coro:unitarity}
Let $\mathcal{A}$ be an algorithm that upon an input of quantum channel $\mathcal{N}$
estimates the unitarity of $\mathcal{N}$ with at most $ 1/4 - 1/4^n$ using $\mathsf{QStat}_\mathcal{N}$ queries with tolerance at least $\tau$. Then $\mathcal{A}$ must make at least $2^{\Omega(\tau^2 2^n)}$ such queries.
\end{coro}
\begin{proof}
By Lemma\ \ref{lem:unitarity}, the unitarity of a channel $\mathcal{N}$ and the purity of its Choi state are within an inverse exponentially small additive factor:
\begin{align}
    &\frac{4^n}{4^n -1} \Tr[\calJ(\mathcal{N})^2] - \frac{1}{4^n - 1}\leq  u (\mathcal{N}) \leq \frac{4^n}{4^n -1} \Tr[\calJ(\mathcal{N})^2]
   \\ \implies &\Tr[\calJ(\mathcal{N})^2] - \frac{1}{4^n}\leq \left(\frac{4^n - 1}{4^n}\right) u (\mathcal{N}) \leq \Tr[\calJ(\mathcal{N})^2]
\end{align}

Let $\rho$ be an arbitrary state. By the Choi-Jamiolkowski isomorphism, there exists a channel $\mathcal{N}_\rho$ such that $\mathcal{J}(\mathcal{N}_\rho) = \rho$.
Assume that an algorithm $\mathcal{A}$ can estimate the unitarity of  channel $\mathcal{N}_\rho$  with error $\epsilon$ using $\mathsf{QStat}_{\mathcal{N}_\rho}$ queries with tolerance at least $\tau$. Then $\mathcal{A}$ produces an estimate of the purity of $\calJ(\mathcal{N}_\rho) = \rho$ with additive error at most $\epsilon + \frac{1}{4^n}$.
Therefore, if $\epsilon = \frac{1}{4} - \frac{1}{4^n}$, $\mathcal{A}$ produces an estimate of the purity of $\rho$ with additive error at most $\frac{1}{4}$, and  by Lemma\ \ref{lem:purity} $\mathcal{A}$ must make at least $2^{\Omega(\tau'^2 2^n)} = 2^{\Omega((\tau^2 + ) 2^n)}$ queries.

\end{proof}

It's easy to see that the unitarity can be estimated with $\mathcal{O}(1)$ joint measurements to the Choi state or $\mathcal{O}(2^n)$ separable measurements to the Choi state. This can be shown invoking previous upper bounds for purity estimation~\cite{montanaro2013survey, chen2021exponential} and exploiting again the connection between unitarity and the purity of the Choi state.

\section{Applications}
{
\color{black}

Given its flexibility and noise-tolerance, we expect that the framework of quantum statistical query can be applied to a variety of unitary learning tasks. In this section, we will showcase some selected applications in various fields of quantum information processing.

}

\subsection{Classical surrogates for quantum learning models}
\label{sec:surrogates}
In this section we discuss a potential application of our results to quantum machine learning. We will consider particularly variational quantum algorithms for approximating a classical function $f : \mathcal{X} \rightarrow \mathbb{R}$. For a broad class of such algorithms~\cite{schuld2019quantum, schuld2021supervised, schuld2021quantummodels}, the prediction phase can be cast as follows: the input $\boldsymbol{x}\in\mathcal{X}$ is encoded into a quantum state with a suitable feature map $\boldsymbol{x} \mapsto \rho(\boldsymbol{x})$, which
evolves according to a parametric channel $\mathcal{U}_{\boldsymbol{\theta}}$ and subsequently is measured with a local observable $O$. Hence, the parametric circuit induces a hypothesis function $h(\cdot)$, which associates $\boldsymbol{x}$  to the following label
\begin{equation}
    h(\boldsymbol{x}) = \Tr[O \mathcal{U}_{\boldsymbol{\theta}}(\rho(\boldsymbol{x}))].
\end{equation}
Thus, given a distribution $\mathcal{D}$ over $\mathcal{X}$,  the goal is to find a parameter $\boldsymbol{\theta}^*$ satisfying the following:
\begin{equation}
    \mathbb{E}_{\boldsymbol{x}\sim \mathcal{D}} |h(\boldsymbol{x}) - f(\boldsymbol{x}) | = \mathbb{E}_{\boldsymbol{x}\sim \mathcal{D}}  |\Tr[O \mathcal{U}_{\boldsymbol{\theta}^*}(\rho(\boldsymbol{x}))]- f(\boldsymbol{x}) | \leq \varepsilon,
\end{equation}
where $\varepsilon$ is a small positive constant. Given a set of examples $(\boldsymbol{x}_1,f(\boldsymbol{x}_1)), (\boldsymbol{x}_2,f(\boldsymbol{x}_2)),\dots, (\boldsymbol{x}_m,f(\boldsymbol{x}_m))$ one can then train this model in a hybrid fashion and select a parameter $\boldsymbol{\theta}$. Then the label of an unseen instance $\boldsymbol{x}_{m+1}$ can be predicted with accuracy $\varepsilon$ preparing $\mathcal{O}(\varepsilon^{-2})$ copies of the state $\mathcal{U}_{\boldsymbol{\theta}}(\rho(\boldsymbol{x}))$ and measuring the observable $O$.

A recent line of research showed that, in some cases, one can fruitfully perform the prediction phase with a purely classical algorithm, that goes under the name of \emph{classical surrogate}~\cite{schreiber2023classical}. So far, the proposed approaches rely on the classical shadow tomography~\cite{jerbi2023shadows} and the Fourier analysis of real functions~\cite{landman2022classically, schreiber2023classical}, which can be applied to the general expression of quantum models as trigonometric polynomials. Here we argue that the QSQ learning framework can find application in the quest for surrogate models, introducing more flexibility in the surrogation process.
Particularly,~\cite{jerbi2023shadows} resorts to a flipped model of quantum circuit where the parameter $\boldsymbol{\theta}$ is encoded in a quantum state, subsequently measured by a variational measurements depending on the $\boldsymbol{x}$. While this model can provide quantum advantage for specific tasks, it would be interesting to obtain similar results beyond the flipped circuit model, and specifically for the setting where the instance $\boldsymbol{x}$ is encoded before the parameter $\boldsymbol{\theta}$. This goal can be achieved through the algorithms discussed in the present paper, since they do not require the unitary to be a flipped a circuit. However, the distance over unitaries we adopted brings accuracy guarantees for the prediction only when the input state is sampled from a locally scrambled ensemble.
Thus, we need to extend the definition given in Ref.~\cite{schreiber2023classical} to incorporate the input distribution $\mathcal{D}$.
\begin{definition}[Average-case surrogate models]
Let $\varepsilon \geq0$ and $0\leq \delta \leq 1$.
A hypothesis class of quantum learning models $\mathcal{F}$ has 
an average-case $(\varepsilon,\delta)$-classical surrogate if there
exists a process $\mathcal{S}$ that upon input of a learning model $f \in \mathcal{F}$
produces a classical model $g \in \mathcal{G}$ such that
\begin{equation}
    \Pr[\mathbb{E}_{\boldsymbol{x}\sim \mathcal{D}} \|f(\boldsymbol{x}) - g(\boldsymbol{x})\| \leq \varepsilon] \geq 1-\delta,
\end{equation}
for a suitable norm on the output space $\mathcal{Y}$.
The process $\mathcal{S}$ must be efficient in the size of the quantum learning
model, the error bound $\varepsilon$ and the failure probability $\delta$.
\end{definition}
In particular, it is easy to see that if the conditional distribution of the states $\rho(\boldsymbol{x})$ is low-average (cf.~Equation~\ref{eq:low-a}), then we can produce an average-case classical surrogate of $f(\boldsymbol{x}) = \Tr[O \mathcal{U}_{\boldsymbol{\theta}}\rho(\boldsymbol{x})]$ via QSQs by means of the algorithms discussed in the present paper.
We conclude this discussion with a concrete example for shallow-depth geometrically local circuits.
{\color{black}
\begin{example}[Shallow-depth geometrically local circuit]
To showcase the applicability of the proposed surrogate, we consider the following setting:
\begin{itemize}
    \item$O$ is a single-qubit Pauli operator.
    \item $U_{\boldsymbol{\theta}}$ is a parametrized circuit of depth $L$ with geometric dimension $D$. Specifically, we assume that the evolved observable $U_{\boldsymbol{\theta}}^\dag O U_{\boldsymbol{\theta}}$ acts non trivially on at most $\calO(L^D)$ qubits.
    \item  $\calD$ is the uniform distribution over $\{0,1\}^n$, and we consider the quantum feature map $\boldsymbol{x} \mapsto \rho(\boldsymbol{x})$ obtained by encoding $\boldsymbol{x}$ into the computational basis state $\ketbra{\boldsymbol{x}}{\boldsymbol{x}}$. The derived distribution onto quantum states forms a state 1-design, which is a low-average ensemble with purity 1. 
\end{itemize} 
Let $f(\boldsymbol{x}) \coloneqq \Tr\left[\left(U_{\boldsymbol{\theta}}^\dag O U_{\boldsymbol{\theta}}\right) \rho(\boldsymbol{x})\right]$ be the target function we seek to surrogate. As the Heisenberg-evolved Pauli observable $U_{\boldsymbol{\theta}}^\dag O U_{\boldsymbol{\theta}}$ contains only Pauli terms of weight $\calO(L^D)$, by Theorem\ \ref{thm:GL-low-d}, there is an algorithm running in time $n\cdot \poly\left(\epsilon^{-L^D}, 2^{L^D} \right)$ that access the unitary $\widetilde{A} \coloneqq U^\dag \exp\left(i\frac{\pi}{6}O\right) U$ via $\mathsf{QStat}_{\widetilde{A}}$ queries with tolerance at least $\poly\left( \epsilon^{L^D}, 2^{-L^D}\right)$ and outputs a function $g$ such that
\begin{align}
    \mathbb{E}_{\boldsymbol{x}\sim \mathcal{D}} \abs{f(\boldsymbol{x}) - g(\boldsymbol{x})} \leq \epsilon,
\end{align}
where $g(\boldsymbol{x})$ is of the form $g(\boldsymbol{x}) \coloneqq \Tr\left[B \rho(\boldsymbol{x})\right] $, for a suitable Hermitian operator $B$ found via Theorem\ \ref{thm:GL-low-d}.
\end{example}
Thus, the surrogation can be done efficiently whenever $L^{D} \in \calO(\log(n))$.

}

{\color{black}
\subsection{Estimating correlations in quantum many-body systems}
Another useful application of our results is in quantum many-body physics. Specifically, we show how the Goldreich-Levin algorithm can be leveraged to estimate two-point functions\ \cite{starykh1997dynamics, sirker2005real, steinigeweg2009density}, which are fundamental in the study of quantum many-body systems -- with applications ranging from diagnosing quantum chaos\ \cite{gharibyan2020characterization} to analyzing thermalization effects\ \cite{luitz2016anomalous, venuti2019ergodicity, alhambra2020time, von2022operator, zhang2024thermalization}.

Given an Hamiltonian $H$ and a local observable $A$, we consider the Heisenberg-evolved observable $A(t)$:
\begin{align}
    A(t) \coloneqq e^{iHt}AP e^{-iHt} = \sum_{P\in\calP_n} a_P(t)P.
\end{align}
Given a second local operator $B$, we compute the following \emph{two-point} correlation function (also referred as dynamical response or Green’s function):
\begin{align}
 C^{AB}(t) \coloneqq {\frac{1}{2^n}\Tr[A(t)B]}. 
\end{align}
In many-body physics, a quantum dynamics is referred as \emph{dissipative} if
the function $C^{AB}(t)$ satisfies the following
\begin{align}
    C^{AB}(t)  \xrightarrow{t\rightarrow \infty} \frac{1}{4^n}\Tr[A]\Tr[B] .
\end{align}
In the following, we assume that $A$ is a non-identity Pauli operator. Since Pauli matrices are traceless, the above expression simplifies to
\begin{align}
    C^{AB}(t)  \xrightarrow{t\rightarrow \infty} 0.
\end{align}

Using the Goldreich-Levin algorithm (Algorithm~\ref{alg:GL}), we can find all the Pauli operators such that the associated correlation functions exceeds a given threshold $\tau$. Specifically, we can determine the following set
\begin{align}
    \left\{ B \in \calP_n | \abs{C^{AB}(t)} \geq \tau\right\}.
\end{align}
Thus, if such sets does not contain local Pauli operators, we can interpret this as a signature of dissipative dynamics.
We can further determine the values of $C^{AB}(t)$ for the $B$ within the above set via Lemma\ \ref{lem:coeff2}, and use those values for estimating the local observable with maximal correlation function.

\subsection{Unitary cross-platform verification}

A major challenge in near-term quantum computing involves directly comparing two quantum states $\rho_1$ and $\rho_2$ produced by two separate devices at different locations and times while executing a specific quantum computation or simulation. This process, known as cross-platform verification, assess the performance of these experimental quantum systems by measuring the overlap $\Tr[\rho_1 \rho_2]$.

Previous works~\cite{elben2020cross, anshu2022distributed, gong2024sample, arunachalam2024distributed, hinsche2024efficient} have explored various distributed approaches to this task, demonstrating that under certain constraints on state complexity, the overlap $\Tr[\rho_1 \rho_2]$ can be estimated with limited quantum communication -- or even purely classical communication -- between the two parties.
Specifically, one of the state-of-the-art {approaches} employs a subroutine named Pauli sampling~\cite{flammia2011direct}, which consists {of drawing samples} from the distribution induced by the (squared) coefficients of the representation of the state in the Pauli basis. As shown in Refs.~\cite{hinsche2024efficient}, {these methods} establish an efficient protocol for cross-platform verification under the assumption that the states $\rho, \sigma$ have both stabilizer 2-Rényi entropy and 2-Rényi entropy of entanglement at most logarithmic in system size.

Here, we consider a distinct but related problem, which we dub \emph{unitary} cross-platform verification. We consider two separate parties that attempt to implement the same quantum computation; however, they implement two different unitaries $U_1$ and $U_2$ due to their experimental constraints.
A natural question is whether 
average-case input states, when evolved by the unitaries $U_1$ and $U_2$, are mapped to output states with a substantial overlap, that is, whether the following quantity is sufficiently large:
\begin{align}
    \mathbb{E}_{\rho \sim \calD} \Tr[U\rho U^\dag V\rho V^\dag ]
    = 1 - \mathbb{E}_{\rho \sim \calD}  \norm{U\rho U^\dag - V\rho V^\dag}_{\tr}^2 = 1- \calL^{(\mathrm{tr})}_\calD (U,V).
\end{align}
If we further assume that $\calD$ is a locally scrambled distribution (e.g., $\rho$ is a Haar-random state, or the tensor product of random single-qubit stabilizer states), then we have:
\begin{align}
    \mathbb{E}_{\rho \sim \calD} \Tr[U_1\rho U^\dag_1 U_2\rho U^\dag_2 ]
    = 1 - \frac{2^n}{2^n+1} D^2(U,V).
\end{align}
While this quantity could, in principle, be estimated by repeatedly applying $U$ and $V$ on randomly sampled stabilizer states and performing a Pauli sampling subroutine akin to Ref.~\cite{hinsche2024efficient}, we note that our results encompass unitaries producing high magic and entanglement {for which the cross-platform verification protocol of Ref.~\cite{hinsche2024efficient} becomes inefficient.} 
As a toy example, even a depth-1 circuit consisting of T gates applied on each qubit would create states with stabilizer 2-Rényi entropy linear in system size, far surpassing the logarithmic threshold under which the results of Ref.~\cite{hinsche2024efficient} hold.
{Since arbitrary constant-depth unitaries are learnable within our framework (Theorem~\ref{thm:constant}), we expect that our results can serve as a valuable tool for benchmarking near-term devices, enabling cross-platform verification on a broader class of circuits.}

}

\label{sec:app-QML}

\section*{Acknowledgments}
The author thanks the anonymous reviewers for their useful feedback. He also thanks Chirag Wadhwa and Mina Doosti for helpful discussions and comments on the draft of the paper, and for sharing the draft of their related work on quantum statistical queries, and Elham Kashefi, Daniel Stilck França, Alex B. Grilo, Tom Gur, Zoë Holmes, Shao-Hen Chiew, Yao Ma, Dominik Leichtle and Sean Thrasher for helpful discussions at different stages of this project.
The author acknowledges financial support from the QICS (Quantum Information Center Sorbonne) and from the Sandoz Family Foundation-Monique de Meuron program for Academic Promotion. 


\bibliography{quantum2}

\begin{thebibliography}{76}
\providecommand{\natexlab}[1]{#1}
\providecommand{\url}[1]{\texttt{#1}}
\expandafter\ifx\csname urlstyle\endcsname\relax
  \providecommand{\doi}[1]{doi: #1}\else
  \providecommand{\doi}{doi: \begingroup \urlstyle{rm}\Url}\fi

\bibitem[Chuang and Nielsen(1997)]{chuang1997prescription}
Isaac~L Chuang and Michael~A Nielsen.
\newblock Prescription for experimental determination of the dynamics of a quantum black box.
\newblock \emph{Journal of Modern Optics}, 44\penalty0 (11-12):\penalty0 2455--2467, 1997.
\newblock \doi{10.1080/09500349708231894}.
\newblock URL \url{https://www.tandfonline.com/doi/abs/10.1080/09500349708231894}.

\bibitem[Bisio et~al.(2010)Bisio, Chiribella, D’Ariano, Facchini, and Perinotti]{bisio2010optimal}
Alessandro Bisio, Giulio Chiribella, Giacomo~Mauro D’Ariano, Stefano Facchini, and Paolo Perinotti.
\newblock Optimal quantum learning of a unitary transformation.
\newblock \emph{Physical Review A}, 81\penalty0 (3):\penalty0 032324, 2010.
\newblock \doi{10.1103/PhysRevA.81.032324}.
\newblock URL \url{https://journals.aps.org/pra/abstract/10.1103/PhysRevA.81.032324}.

\bibitem[Caro(2024)]{caro2022learning}
Matthias~C Caro.
\newblock Learning quantum processes and hamiltonians via the pauli transfer matrix.
\newblock \emph{ACM Transactions on Quantum Computing}, 5\penalty0 (2):\penalty0 1--53, 2024.
\newblock \doi{10.1145/3670418}.
\newblock URL \url{https://dl.acm.org/doi/full/10.1145/3670418}.

\bibitem[Gutoski and Johnston(2014)]{gutoski2014process}
Gus Gutoski and Nathaniel Johnston.
\newblock Process tomography for unitary quantum channels.
\newblock \emph{Journal of Mathematical Physics}, 55\penalty0 (3), 2014.
\newblock \doi{10.1063/1.4867625}.
\newblock URL \url{https://pubs.aip.org/aip/jmp/article-abstract/55/3/032201/233162/Process-tomography-for-unitary-quantum-channels}.

\bibitem[Montanaro and Osborne(2010)]{montanaro2010quantum}
Ashley Montanaro and Tobias~J Osborne.
\newblock Quantum boolean functions.
\newblock \emph{Chicago Journal OF Theoretical Computer Science}, 1:\penalty0 1--45, 2010.
\newblock \doi{10.4086/cjtcs.2010.001}.
\newblock URL \url{http://cjtcs.cs.uchicago.edu/articles/2010/1/contents.html}.

\bibitem[Chen et~al.(2023)Chen, Nadimpalli, and Yuen]{chen2023testing}
Thomas Chen, Shivam Nadimpalli, and Henry Yuen.
\newblock Testing and learning quantum juntas nearly optimally.
\newblock In \emph{Proceedings of the 2023 Annual ACM-SIAM Symposium on Discrete Algorithms (SODA)}, pages 1163--1185. SIAM, 2023.
\newblock \doi{10.1137/1.9781611977554.ch43}.
\newblock URL \url{https://epubs.siam.org/doi/abs/10.1137/1.9781611977554.ch43}.

\bibitem[Bao and Yao(2025)]{bao2025testing}
Zongbo Bao and Penghui Yao.
\newblock On testing and learning quantum junta channels.
\newblock \emph{IEEE Transactions on Pattern Analysis \& Machine Intelligence}, 47\penalty0 (04):\penalty0 2991--3002, 2025.
\newblock \doi{10.1109/TPAMI.2025.3528648}.
\newblock URL \url{https://www.computer.org/csdl/journal/tp/2025/04/10839063/23th0T1C1Py}.

\bibitem[Fanizza et~al.(2024)Fanizza, Quek, and Rosati]{fanizza2024learning}
Marco Fanizza, Yihui Quek, and Matteo Rosati.
\newblock Learning quantum processes without input control.
\newblock \emph{PRX Quantum}, 5\penalty0 (2):\penalty0 020367, 2024.
\newblock \doi{10.1103/PRXQuantum.5.020367}.
\newblock URL \url{https://journals.aps.org/prxquantum/abstract/10.1103/PRXQuantum.5.020367}.

\bibitem[Zhao et~al.(2024)Zhao, Lewis, Kannan, Quek, Huang, and Caro]{zhao2024learning}
Haimeng Zhao, Laura Lewis, Ishaan Kannan, Yihui Quek, Hsin-Yuan Huang, and Matthias~C Caro.
\newblock Learning quantum states and unitaries of bounded gate complexity.
\newblock \emph{PRX Quantum}, 5\penalty0 (4):\penalty0 040306, 2024.
\newblock \doi{10.1103/PRXQuantum.5.040306}.
\newblock URL \url{https://link.aps.org/doi/10.1103/PRXQuantum.5.040306}.

\bibitem[Nadimpalli et~al.(2024)Nadimpalli, Parham, Vasconcelos, and Yuen]{nadimpalli2024pauli}
Shivam Nadimpalli, Natalie Parham, Francisca Vasconcelos, and Henry Yuen.
\newblock On the pauli spectrum of qac0.
\newblock In \emph{Proceedings of the 56th Annual ACM Symposium on Theory of Computing}, pages 1498--1506, 2024.
\newblock \doi{10.1145/3618260.3649662}.
\newblock URL \url{https://dl.acm.org/doi/abs/10.1145/3618260.3649662}.

\bibitem[Huang et~al.(2024)Huang, Liu, Broughton, Kim, Anshu, Landau, and McClean]{huang2024learning}
Hsin-Yuan Huang, Yunchao Liu, Michael Broughton, Isaac Kim, Anurag Anshu, Zeph Landau, and Jarrod~R McClean.
\newblock Learning shallow quantum circuits.
\newblock In \emph{Proceedings of the 56th Annual ACM Symposium on Theory of Computing}, pages 1343--1351, 2024.
\newblock \doi{10.1145/3618260.3649722}.
\newblock URL \url{https://dl.acm.org/doi/10.1145/3618260.3649722}.

\bibitem[Vasconcelos and Huang(2024)]{vasconcelos2024learning}
Francisca Vasconcelos and Hsin-Yuan Huang.
\newblock Learning shallow quantum circuits with many-qubit gates.
\newblock \emph{arXiv preprint arXiv:2410.16693}, 2024.
\newblock \doi{10.48550/arXiv.2410.16693}.
\newblock URL \url{https://arxiv.org/abs/2410.16693}.

\bibitem[Lai and Cheng(2022)]{lai2022learning}
Ching-Yi Lai and Hao-Chung Cheng.
\newblock Learning quantum circuits of some t gates.
\newblock \emph{IEEE Transactions on Information Theory}, 68\penalty0 (6):\penalty0 3951--3964, 2022.
\newblock \doi{10.1109/TIT.2022.3151760}.
\newblock URL \url{https://ieeexplore.ieee.org/abstract/document/9714418}.

\bibitem[Xue et~al.(2022)Xue, Liu, Wang, Zhu, Guo, and Wu]{xue2022variational}
Shichuan Xue, Yong Liu, Yang Wang, Pingyu Zhu, Chu Guo, and Junjie Wu.
\newblock Variational quantum process tomography of unitaries.
\newblock \emph{Physical Review A}, 105\penalty0 (3):\penalty0 032427, 2022.
\newblock \doi{10.1103/PhysRevA.105.032427}.
\newblock URL \url{https://journals.aps.org/pra/abstract/10.1103/PhysRevA.105.032427}.

\bibitem[Holmes et~al.(2021)Holmes, Arrasmith, Yan, Coles, Albrecht, and Sornborger]{holmes2021barren}
Zo\"{e} Holmes, Andrew Arrasmith, Bin Yan, Patrick~J. Coles, Andreas Albrecht, and Andrew~T Sornborger.
\newblock Barren plateaus preclude learning scramblers.
\newblock \emph{Physical Review Letters}, 126\penalty0 (19):\penalty0 190501, 2021.
\newblock \doi{10.1103/PhysRevLett.126.190501}.
\newblock URL \url{https://doi.org/10.1103/PhysRevLett.126.190501}.

\bibitem[Torlai et~al.(2023)Torlai, Wood, Acharya, Carleo, Carrasquilla, and Aolita]{torlai2023quantum}
Giacomo Torlai, Christopher~J Wood, Atithi Acharya, Giuseppe Carleo, Juan Carrasquilla, and Leandro Aolita.
\newblock Quantum process tomography with unsupervised learning and tensor networks.
\newblock \emph{Nature Communications}, 14\penalty0 (1):\penalty0 2858, 2023.
\newblock \doi{10.1038/s41467-023-38332-9}.
\newblock URL \url{https://www.nature.com/articles/s41467-023-38332-9}.

\bibitem[Huang et~al.(2023)Huang, Chen, and Preskill]{huang2022learning}
Hsin-Yuan Huang, Sitan Chen, and John Preskill.
\newblock Learning to predict arbitrary quantum processes.
\newblock \emph{PRX Quantum}, 4\penalty0 (4):\penalty0 040337, 2023.
\newblock \doi{10.1103/PRXQuantum.4.040337}.
\newblock URL \url{https://journals.aps.org/prxquantum/abstract/10.1103/PRXQuantum.4.040337}.

\bibitem[Montanaro and de~Wolf(2016)]{montanaro2013survey}
Ashley Montanaro and Ronald de~Wolf.
\newblock A survey of quantum property testing.
\newblock \emph{Theory of Computing}, pages 1--81, 2016.
\newblock \doi{10.4086/toc.gs.2016.007}.
\newblock URL \url{https://theoryofcomputing.org/articles/gs007/}.

\bibitem[Preskill(2018)]{preskill2018quantum}
John Preskill.
\newblock Quantum computing in the {NISQ} era and beyond.
\newblock \emph{Quantum}, 2:\penalty0 79, 2018.
\newblock \doi{10.22331/q-2018-08-06-79}.
\newblock URL \url{https://quantum-journal.org/papers/q-2018-08-06-79/}.

\bibitem[Arunachalam et~al.(2020)Arunachalam, Grilo, and Yuen]{arunachalam2020quantum}
Srinivasan Arunachalam, Alex~B Grilo, and Henry Yuen.
\newblock Quantum statistical query learning.
\newblock \emph{arXiv preprint arXiv:2002.08240}, 2020.
\newblock \doi{https://doi.org/10.48550/arXiv.2002.08240}.
\newblock URL \url{https://doi.org/10.48550/arXiv.2002.08240}.

\bibitem[Kearns(1998)]{kearns1998efficient}
Michael Kearns.
\newblock Efficient noise-tolerant learning from statistical queries.
\newblock \emph{Journal of the ACM (JACM)}, 45\penalty0 (6):\penalty0 983--1006, 1998.
\newblock \doi{10.1145/293347.293351}.
\newblock URL \url{https://dl.acm.org/doi/abs/10.1145/293347.293351}.

\bibitem[Arunachalam et~al.(2023{\natexlab{a}})Arunachalam, Havlicek, and Schatzki]{arunachalam2023on}
Srinivasan Arunachalam, Vojtech Havlicek, and Louis Schatzki.
\newblock On the role of entanglement and statistics in learning.
\newblock In \emph{Advances in Neural Information Processing Systems}, volume~36, pages 55064--55076. Curran Associates, Inc., 2023{\natexlab{a}}.
\newblock \doi{10.5555/3666122.3668526}.
\newblock URL \url{https://dl.acm.org/doi/10.5555/3666122.3668526}.

\bibitem[Caro et~al.(2024)Caro, Hinsche, Ioannou, Nietner, and Sweke]{caro2024classical}
Matthias~C. Caro, Marcel Hinsche, Marios Ioannou, Alexander Nietner, and Ryan Sweke.
\newblock {Classical Verification of Quantum Learning}.
\newblock \emph{15th Innovations in Theoretical Computer Science Conference (ITCS 2024)}, 287:\penalty0 24:1--24:23, 2024.
\newblock ISSN 1868-8969.
\newblock \doi{10.4230/LIPIcs.ITCS.2024.24}.
\newblock URL \url{https://drops.dagstuhl.de/entities/document/10.4230/LIPIcs.ITCS.2024.24}.

\bibitem[Du et~al.(2021)Du, Hsieh, Liu, You, and Tao]{du2020learnability}
Yuxuan Du, Min-Hsiu Hsieh, Tongliang Liu, Shan You, and Dacheng Tao.
\newblock On the learnability of quantum neural networks.
\newblock \emph{PRX Quantum}, 2\penalty0 (4):\penalty0 040337, 2021.
\newblock \doi{10.1103/PRXQuantum.2.040337}.
\newblock URL \url{https://journals.aps.org/prxquantum/abstract/10.1103/PRXQuantum.2.040337}.

\bibitem[Quek et~al.(2024)Quek, Stilck~Fran{\c{c}}a, Khatri, Meyer, and Eisert]{quek2022exponentially}
Yihui Quek, Daniel Stilck~Fran{\c{c}}a, Sumeet Khatri, Johannes~Jakob Meyer, and Jens Eisert.
\newblock Exponentially tighter bounds on limitations of quantum error mitigation.
\newblock \emph{Nature Physics}, 20\penalty0 (10):\penalty0 1648--1658, 2024.
\newblock \doi{10.1038/s41567-024-02536-7}.
\newblock URL \url{https://www.nature.com/articles/s41567-024-02536-7}.

\bibitem[Wadhwa and Doosti(2025)]{wadhwa2023learning}
Chirag Wadhwa and Mina Doosti.
\newblock Learning quantum processes with quantum statistical queries.
\newblock \emph{Quantum}, 9:\penalty0 1739, 2025.
\newblock \doi{10.22331/q-2025-05-12-1739}.
\newblock URL \url{https://quantum-journal.org/papers/q-2025-05-12-1739/}.

\bibitem[Caro et~al.(2023)Caro, Huang, Ezzell, Gibbs, Sornborger, Cincio, Coles, and Holmes]{caro2022outofdistribution}
Matthias~C Caro, Hsin-Yuan Huang, Nicholas Ezzell, Joe Gibbs, Andrew~T Sornborger, Lukasz Cincio, Patrick~J Coles, and Zo{\"e} Holmes.
\newblock Out-of-distribution generalization for learning quantum dynamics.
\newblock \emph{Nature Communications}, 14\penalty0 (1):\penalty0 3751, 2023.
\newblock \doi{10.1038/s41467-023-39381-w}.
\newblock URL \url{https://www.nature.com/articles/s41467-023-39381-w}.

\bibitem[At{\i}c{\i} and Servedio(2007)]{atici2007quantum}
Alp At{\i}c{\i} and Rocco~A Servedio.
\newblock Quantum algorithms for learning and testing juntas.
\newblock \emph{Quantum Information Processing}, 6\penalty0 (5):\penalty0 323--348, 2007.
\newblock \doi{10.1007/s11128-007-0061-6}.
\newblock URL \url{https://link.springer.com/article/10.1007/s11128-007-0061-6}.

\bibitem[Nietner(2023)]{nietner2023unifying}
Alexander Nietner.
\newblock Unifying (quantum) statistical and parametrized (quantum) algorithms.
\newblock \emph{arXiv preprint arXiv:2310.17716}, 2023.
\newblock \doi{10.48550/arXiv.2310.17716}.
\newblock URL \url{https://arxiv.org/abs/2310.17716}.

\bibitem[Caro(2020)]{caro2020quantumlearning}
Matthias~C. Caro.
\newblock Quantum learning boolean linear functions w.r.t.~product distributions.
\newblock \emph{Quantum Information Processing}, 19:\penalty0 172, 2020.
\newblock \doi{10.1007/s11128-020-02661-1}.
\newblock URL \url{https://link.springer.com/article/10.1007/s11128-020-02661-1}.

\bibitem[Kanade et~al.(2019)Kanade, Rocchetto, and Severini]{kanade2019learning}
V~Kanade, A~Rocchetto, and S~Severini.
\newblock Learning dnfs under product distributions via $\mu$-biased quantum fourier sampling.
\newblock \emph{Quantum Information and Computation}, 19\penalty0 (15\&16), 2019.
\newblock \doi{10.26421/QIC19.15-16-1}.
\newblock URL \url{https://www.rintonpress.com/journals/doi/QIC19.15-16-1.html}.

\bibitem[Hinsche et~al.(2023)Hinsche, Ioannou, Nietner, Haferkamp, Quek, Hangleiter, Seifert, Eisert, and Sweke]{hinsche2023one}
Marcel Hinsche, Marios Ioannou, Alexander Nietner, Jonas Haferkamp, Yihui Quek, Dominik Hangleiter, J-P Seifert, Jens Eisert, and Ryan Sweke.
\newblock One t gate makes distribution learning hard.
\newblock \emph{Physical review letters}, 130\penalty0 (24):\penalty0 240602, 2023.
\newblock \doi{10.1103/PhysRevLett.130.240602}.
\newblock URL \url{https://journals.aps.org/prl/abstract/10.1103/PhysRevLett.130.240602}.

\bibitem[Nietner et~al.(2023)Nietner, Ioannou, Sweke, Kueng, Eisert, Hinsche, and Haferkamp]{nietner2023average}
Alexander Nietner, Marios Ioannou, Ryan Sweke, Richard Kueng, Jens Eisert, Marcel Hinsche, and Jonas Haferkamp.
\newblock On the average-case complexity of learning output distributions of quantum circuits.
\newblock \emph{arXiv preprint arXiv:2305.05765}, 2023.
\newblock \doi{https://doi.org/10.48550/arXiv.2305.05765}.
\newblock URL \url{https://arxiv.org/abs/2305.05765}.

\bibitem[Coles et~al.(2019)Coles, Cerezo, and Cincio]{coles2019strong}
Patrick~J. Coles, M.~Cerezo, and Lukasz Cincio.
\newblock Strong bound between trace distance and hilbert-schmidt distance for low-rank states.
\newblock \emph{Physical Review A}, 100\penalty0 (2):\penalty0 022103, 2019.
\newblock \doi{10.1103/PhysRevA.100.022103}.
\newblock URL \url{https://journals.aps.org/pra/abstract/10.1103/PhysRevA.100.022103}.

\bibitem[Mele(2024)]{mele2023introduction}
Antonio~Anna Mele.
\newblock Introduction to haar measure tools in quantum information: A beginner's tutorial.
\newblock \emph{Quantum}, 8:\penalty0 1340, 2024.
\newblock \doi{10.22331/q-2024-05-08-1340}.
\newblock URL \url{https://quantum-journal.org/papers/q-2024-05-08-1340/}.

\bibitem[Schuster et~al.(2024)Schuster, Yin, Gao, and Yao]{schuster2024polynomial}
Thomas Schuster, Chao Yin, Xun Gao, and Norman~Y Yao.
\newblock A polynomial-time classical algorithm for noisy quantum circuits.
\newblock \emph{arXiv preprint arXiv:2407.12768}, 2024.
\newblock \doi{https://doi.org/10.48550/arXiv.2407.12768}.
\newblock URL \url{https://arxiv.org/abs/2407.12768}.

\bibitem[Choi(1975)]{choi1975completely}
Man-Duen Choi.
\newblock Completely positive linear maps on complex matrices.
\newblock \emph{Linear algebra and its applications}, 10\penalty0 (3):\penalty0 285--290, 1975.
\newblock \doi{10.1016/0024-3795(75)90075-0}.
\newblock URL \url{https://doi.org/10.1016/0024-3795(75)90075-0}.

\bibitem[Jamio{\l}kowski(1972)]{jamiolkowski1972linear}
Andrzej Jamio{\l}kowski.
\newblock Linear transformations which preserve trace and positive semidefiniteness of operators.
\newblock \emph{Reports on mathematical physics}, 3\penalty0 (4):\penalty0 275--278, 1972.
\newblock \doi{10.1016/0034-4877(72)90011-0}.
\newblock URL \url{https://www.sciencedirect.com/science/article/abs/pii/0034487772900110}.

\bibitem[Low(2009)]{low2009learning}
Richard~A Low.
\newblock Learning and testing algorithms for the clifford group.
\newblock \emph{Physical Review A}, 80\penalty0 (5):\penalty0 052314, 2009.
\newblock \doi{10.1103/PhysRevA.80.052314}.
\newblock URL \url{https://journals.aps.org/pra/abstract/10.1103/PhysRevA.80.052314}.

\bibitem[Wang(2011)]{wang2011property}
Guoming Wang.
\newblock Property testing of unitary operators.
\newblock \emph{Physical Review A—Atomic, Molecular, and Optical Physics}, 84\penalty0 (5):\penalty0 052328, 2011.
\newblock \doi{10.1103/PhysRevA.84.052328}.
\newblock URL \url{https://journals.aps.org/pra/abstract/10.1103/PhysRevA.84.052328}.

\bibitem[Knill and Laflamme(1998)]{knill1998power}
Emanuel Knill and Raymond Laflamme.
\newblock Power of one bit of quantum information.
\newblock \emph{Physical Review Letters}, 81\penalty0 (25):\penalty0 5672, 1998.
\newblock \doi{10.1103/PhysRevLett.81.5672}.
\newblock URL \url{https://journals.aps.org/prl/abstract/10.1103/PhysRevLett.81.5672}.

\bibitem[Rouz{\'e} et~al.(2024)Rouz{\'e}, Wirth, and Zhang]{rouze2024quantum}
Cambyse Rouz{\'e}, Melchior Wirth, and Haonan Zhang.
\newblock Quantum talagrand, kkl and friedgut’s theorems and the learnability of quantum boolean functions.
\newblock \emph{Communications in Mathematical Physics}, 405\penalty0 (4):\penalty0 95, 2024.
\newblock \doi{10.1007/s00220-024-04981-0}.
\newblock URL \url{https://link.springer.com/article/10.1007/s00220-024-04981-0}.

\bibitem[O'Donnell(2021)]{o2021analysis}
Ryan O'Donnell.
\newblock Analysis of boolean functions.
\newblock \emph{arXiv preprint arXiv:2105.10386}, 2021.
\newblock \doi{10.48550/arXiv.2105.10386}.
\newblock URL \url{https://arxiv.org/abs/2105.10386}.

\bibitem[Bshouty and Jackson(1995)]{bshouty1995learning}
Nader~H Bshouty and Jeffrey~C Jackson.
\newblock Learning dnf over the uniform distribution using a quantum example oracle.
\newblock In \emph{Proceedings of the eighth annual conference on Computational learning theory}, pages 118--127, 1995.
\newblock \doi{10.1145/225298.225312}.
\newblock URL \url{https://dl.acm.org/doi/10.1145/225298.225312}.

\bibitem[Wallman et~al.(2015)Wallman, Granade, Harper, and Flammia]{wallman2015estimating}
Joel Wallman, Chris Granade, Robin Harper, and Steven~T Flammia.
\newblock Estimating the coherence of noise.
\newblock \emph{New Journal of Physics}, 17\penalty0 (11):\penalty0 113020, 2015.
\newblock \doi{10.1088/1367-2630/17/11/113020}.
\newblock URL \url{https://iopscience.iop.org/article/10.1088/1367-2630/17/11/113020}.

\bibitem[Carignan-Dugas et~al.(2019)Carignan-Dugas, Wallman, and Emerson]{carignan2019bounding}
Arnaud Carignan-Dugas, Joel~J Wallman, and Joseph Emerson.
\newblock Bounding the average gate fidelity of composite channels using the unitarity.
\newblock \emph{New Journal of Physics}, 21\penalty0 (5):\penalty0 053016, 2019.
\newblock \doi{10.1088/1367-2630/ab1800}.
\newblock URL \url{https://iopscience.iop.org/article/10.1088/1367-2630/ab1800}.

\bibitem[Melkani et~al.(2020)Melkani, Gneiting, and Nori]{melkani2020eigenstate}
Abhijeet Melkani, Clemens Gneiting, and Franco Nori.
\newblock Eigenstate extraction with neural-network tomography.
\newblock \emph{Phys. Rev. A}, 102:\penalty0 022412, Aug 2020.
\newblock \doi{10.1103/PhysRevA.102.022412}.
\newblock URL \url{https://link.aps.org/doi/10.1103/PhysRevA.102.022412}.

\bibitem[Goldreich and Levin(1989)]{goldreich1989hard}
Oded Goldreich and Leonid~A Levin.
\newblock A hard-core predicate for all one-way functions.
\newblock In \emph{Proceedings of the twenty-first annual ACM symposium on Theory of computing}, pages 25--32, 1989.
\newblock \doi{10.1145/73007.73010}.
\newblock URL \url{https://dl.acm.org/doi/10.1145/73007.73010}.

\bibitem[Volberg and Zhang(2024)]{volberg2024noncommutative}
Alexander Volberg and Haonan Zhang.
\newblock Noncommutative bohnenblust--hille inequalities.
\newblock \emph{Mathematische Annalen}, 389\penalty0 (2):\penalty0 1657--1676, 2024.
\newblock \doi{10.1007/s00208-023-02680-0}.
\newblock URL \url{https://link.springer.com/article/10.1007/s00208-023-02680-0}.

\bibitem[Yu and Wei(2023)]{yu2023learning}
Nengkun Yu and Tzu-Chieh Wei.
\newblock Learning marginals suffices!
\newblock \emph{arXiv preprint arXiv:2303.08938}, 2023.
\newblock \doi{10.48550/arXiv.2303.08938}.
\newblock URL \url{https://arxiv.org/abs/2303.08938}.

\bibitem[Nisan and Szegedy(1994)]{nisan1994degree}
Noam Nisan and Mario Szegedy.
\newblock On the degree of boolean functions as real polynomials.
\newblock \emph{Computational complexity}, 4:\penalty0 301--313, 1994.
\newblock \doi{10.1007/BF01263419}.
\newblock URL \url{https://link.springer.com/article/10.1007/BF01263419}.

\bibitem[Fuchs and Van De~Graaf(1999)]{fuchs1999cryptographic}
Christopher~A Fuchs and Jeroen Van De~Graaf.
\newblock Cryptographic distinguishability measures for quantum-mechanical states.
\newblock \emph{IEEE Transactions on Information Theory}, 45\penalty0 (4):\penalty0 1216--1227, 1999.
\newblock \doi{10.1109/18.761271}.
\newblock URL \url{https://ieeexplore.ieee.org/document/761271}.

\bibitem[Arunachalam et~al.(2023{\natexlab{b}})Arunachalam, Bravyi, Dutt, and Yoder]{arunachalam2023optimal}
Srinivasan Arunachalam, Sergey Bravyi, Arkopal Dutt, and Theodore~J Yoder.
\newblock Optimal algorithms for learning quantum phase states.
\newblock \emph{18th Conference on the Theory of Quantum Computation, Communication and Cryptography (TQC 2023)}, pages 3--1, 2023{\natexlab{b}}.
\newblock \doi{10.4230/LIPIcs.TQC.2023.3}.
\newblock URL \url{https://drops.dagstuhl.de/entities/document/10.4230/LIPIcs.TQC.2023.3}.

\bibitem[Chen et~al.(2022)Chen, Cotler, Huang, and Li]{chen2021exponential}
Sitan Chen, Jordan Cotler, Hsin-Yuan Huang, and Jerry Li.
\newblock Exponential separations between learning with and without quantum memory.
\newblock In \emph{2021 IEEE 62nd Annual Symposium on Foundations of Computer Science (FOCS)}, pages 574--585. IEEE, 2022.
\newblock \doi{10.1109/FOCS52979.2021.00063}.
\newblock URL \url{https://ieeexplore.ieee.org/document/9719827}.

\bibitem[Schuld and Killoran(2019)]{schuld2019quantum}
Maria Schuld and Nathan Killoran.
\newblock Quantum machine learning in feature hilbert spaces.
\newblock \emph{Physical Review Letters}, 122\penalty0 (4):\penalty0 040504, 2019.
\newblock \doi{10.1103/PhysRevLett.122.040504}.
\newblock URL \url{https://journals.aps.org/prl/abstract/10.1103/PhysRevLett.122.040504}.

\bibitem[Schuld(2021)]{schuld2021supervised}
Maria Schuld.
\newblock Supervised quantum machine learning models are kernel methods.
\newblock \emph{arXiv preprint arXiv:2101.11020}, 2021.
\newblock \doi{10.48550/arXiv.2101.11020}.
\newblock URL \url{https://arxiv.org/abs/2101.11020}.

\bibitem[Schuld and Petruccione(2021)]{schuld2021quantummodels}
Maria Schuld and Francesco Petruccione.
\newblock Quantum models as kernel methods.
\newblock \emph{Machine Learning with Quantum Computers}, pages 217--245, 2021.
\newblock \doi{10.1007/978-3-030-83098-4_6}.
\newblock URL \url{https://link.springer.com/content/pdf/10.1007/978-3-030-83098-4_6.pdf}.

\bibitem[Schreiber et~al.(2023)Schreiber, Eisert, and Meyer]{schreiber2023classical}
Franz~J Schreiber, Jens Eisert, and Johannes~Jakob Meyer.
\newblock Classical surrogates for quantum learning models.
\newblock \emph{Physical Review Letters}, 131\penalty0 (10):\penalty0 100803, 2023.
\newblock \doi{10.1103/PhysRevLett.131.100803}.
\newblock URL \url{https://journals.aps.org/prl/abstract/10.1103/PhysRevLett.131.100803}.

\bibitem[Jerbi et~al.(2024)Jerbi, Gyurik, Marshall, Molteni, and Dunjko]{jerbi2023shadows}
Sofiene Jerbi, Casper Gyurik, Simon~C Marshall, Riccardo Molteni, and Vedran Dunjko.
\newblock Shadows of quantum machine learning.
\newblock \emph{Nature Communications}, 15\penalty0 (1):\penalty0 5676, 2024.
\newblock \doi{10.1038/s41467-024-49877-8}.
\newblock URL \url{https://www.nature.com/articles/s41467-024-49877-8}.

\bibitem[Landman et~al.(2022)Landman, Thabet, Dalyac, Mhiri, and Kashefi]{landman2022classically}
Jonas Landman, Slimane Thabet, Constantin Dalyac, Hela Mhiri, and Elham Kashefi.
\newblock Classically approximating variational quantum machine learning with random fourier features.
\newblock \emph{arXiv preprint arXiv:2210.13200}, 2022.
\newblock \doi{https://doi.org/10.48550/arXiv.2210.13200}.
\newblock URL \url{https://arxiv.org/abs/2210.13200}.

\bibitem[Starykh et~al.(1997)Starykh, Sandvik, and Singh]{starykh1997dynamics}
OA~Starykh, AW~Sandvik, and RRP Singh.
\newblock Dynamics of the spin-heisenberg chain at intermediate temperatures.
\newblock \emph{Physical Review B}, 55\penalty0 (22):\penalty0 14953, 1997.
\newblock \doi{10.1103/PhysRevB.55.14953}.
\newblock URL \url{https://journals.aps.org/prb/abstract/10.1103/PhysRevB.55.14953}.

\bibitem[Sirker and Kl{\"u}mper(2005)]{sirker2005real}
Jesko Sirker and Andreas Kl{\"u}mper.
\newblock Real-time dynamics at finite temperature by the density-matrix renormalization group: A path-integral approach.
\newblock \emph{Physical Review B—Condensed Matter and Materials Physics}, 71\penalty0 (24):\penalty0 241101, 2005.
\newblock \doi{10.1103/PhysRevB.71.241101}.
\newblock URL \url{https://journals.aps.org/prb/abstract/10.1103/PhysRevB.71.241101}.

\bibitem[Steinigeweg and Gemmer(2009)]{steinigeweg2009density}
Robin Steinigeweg and Jochen Gemmer.
\newblock Density dynamics in translationally invariant spin-1 2 chains at high temperatures: A current-autocorrelation approach to finite time and length scales.
\newblock \emph{Physical Review B—Condensed Matter and Materials Physics}, 80\penalty0 (18):\penalty0 184402, 2009.
\newblock \doi{10.1103/PhysRevB.80.184402}.
\newblock URL \url{https://journals.aps.org/prb/abstract/10.1103/PhysRevB.80.184402}.

\bibitem[Gharibyan et~al.(2020)Gharibyan, Hanada, Swingle, and Tezuka]{gharibyan2020characterization}
Hrant Gharibyan, Masanori Hanada, Brian Swingle, and Masaki Tezuka.
\newblock Characterization of quantum chaos by two-point correlation functions.
\newblock \emph{Physical Review E}, 102\penalty0 (2):\penalty0 022213, 2020.
\newblock \doi{10.1103/PhysRevE.102.022213}.
\newblock URL \url{https://journals.aps.org/pre/abstract/10.1103/PhysRevE.102.022213}.

\bibitem[Luitz and Lev(2016)]{luitz2016anomalous}
David~J Luitz and Yevgeny~Bar Lev.
\newblock Anomalous thermalization in ergodic systems.
\newblock \emph{Physical review letters}, 117\penalty0 (17):\penalty0 170404, 2016.
\newblock \doi{10.1103/PhysRevLett.117.170404}.
\newblock URL \url{https://journals.aps.org/prl/abstract/10.1103/PhysRevLett.117.170404}.

\bibitem[Venuti and Liu(2019)]{venuti2019ergodicity}
Lorenzo~Campos Venuti and Lawrence Liu.
\newblock Ergodicity, eigenstate thermalization, and the foundations of statistical mechanics in quantum and classical systems.
\newblock \emph{arXiv preprint arXiv:1904.02336}, 2019.
\newblock \doi{10.48550/arXiv.1904.02336}.
\newblock URL \url{https://arxiv.org/abs/1904.02336}.

\bibitem[Alhambra et~al.(2020)Alhambra, Riddell, and Garc{\'\i}a-Pintos]{alhambra2020time}
{\'A}lvaro~M Alhambra, Jonathon Riddell, and Luis~Pedro Garc{\'\i}a-Pintos.
\newblock Time evolution of correlation functions in quantum many-body systems.
\newblock \emph{Physical Review Letters}, 124\penalty0 (11):\penalty0 110605, 2020.
\newblock \doi{10.1103/PhysRevLett.124.110605}.
\newblock URL \url{https://journals.aps.org/prl/abstract/10.1103/PhysRevLett.124.110605}.

\bibitem[Von~Keyserlingk et~al.(2022)Von~Keyserlingk, Pollmann, and Rakovszky]{von2022operator}
Curt Von~Keyserlingk, Frank Pollmann, and Tibor Rakovszky.
\newblock Operator backflow and the classical simulation of quantum transport.
\newblock \emph{Physical Review B}, 105\penalty0 (24):\penalty0 245101, 2022.
\newblock \doi{10.1103/PhysRevB.105.245101}.
\newblock URL \url{https://journals.aps.org/prb/abstract/10.1103/PhysRevB.105.245101}.

\bibitem[Zhang et~al.(2024)Zhang, Nie, and von Keyserlingk]{zhang2024thermalization}
Carolyn Zhang, Laimei Nie, and Curt von Keyserlingk.
\newblock Thermalization rates and quantum ruelle-pollicott resonances: insights from operator hydrodynamics.
\newblock \emph{arXiv preprint arXiv:2409.17251}, 2024.
\newblock \doi{10.48550/arXiv.2409.17251}.
\newblock URL \url{https://arxiv.org/abs/2409.17251}.

\bibitem[Elben et~al.(2020)Elben, Vermersch, Van~Bijnen, Kokail, Brydges, Maier, Joshi, Blatt, Roos, and Zoller]{elben2020cross}
Andreas Elben, Beno{\^\i}t Vermersch, Rick Van~Bijnen, Christian Kokail, Tiff Brydges, Christine Maier, Manoj~K Joshi, Rainer Blatt, Christian~F Roos, and Peter Zoller.
\newblock Cross-platform verification of intermediate scale quantum devices.
\newblock \emph{Physical review letters}, 124\penalty0 (1):\penalty0 010504, 2020.
\newblock \doi{10.1103/PhysRevLett.124.010504}.
\newblock URL \url{https://journals.aps.org/prl/abstract/10.1103/PhysRevLett.124.010504}.

\bibitem[Anshu et~al.(2022)Anshu, Landau, and Liu]{anshu2022distributed}
Anurag Anshu, Zeph Landau, and Yunchao Liu.
\newblock Distributed quantum inner product estimation.
\newblock In \emph{Proceedings of the 54th Annual ACM SIGACT Symposium on Theory of Computing}, pages 44--51, 2022.
\newblock \doi{10.1145/3519935.3519974}.
\newblock URL \url{https://dl.acm.org/doi/10.1145/3519935.3519974}.

\bibitem[Gong et~al.(2024)Gong, Haferkamp, Ye, and Zhang]{gong2024sample}
Weiyuan Gong, Jonas Haferkamp, Qi~Ye, and Zhihan Zhang.
\newblock On the sample complexity of purity and inner product estimation.
\newblock \emph{arXiv preprint arXiv:2410.12712}, 2024.
\newblock \doi{10.48550/arXiv.2410.12712}.
\newblock URL \url{https://arxiv.org/abs/2410.12712}.

\bibitem[Arunachalam and Schatzki(2024)]{arunachalam2024distributed}
Srinivasan Arunachalam and Louis Schatzki.
\newblock Distributed inner product estimation with limited quantum communication.
\newblock \emph{arXiv preprint arXiv:2410.12684}, 2024.
\newblock \doi{10.48550/arXiv.2410.12684}.
\newblock URL \url{https://arxiv.org/abs/2410.12684}.

\bibitem[Hinsche et~al.(2024)Hinsche, Ioannou, Jerbi, Leone, Eisert, and Carrasco]{hinsche2024efficient}
Marcel Hinsche, Marios Ioannou, Sofiene Jerbi, Lorenzo Leone, Jens Eisert, and Jose Carrasco.
\newblock Efficient distributed inner product estimation via pauli sampling.
\newblock \emph{arXiv preprint arXiv:2405.06544}, 2024.
\newblock \doi{10.48550/arXiv.2405.06544}.
\newblock URL \url{https://arxiv.org/abs/2405.06544}.

\bibitem[Flammia and Liu(2011)]{flammia2011direct}
Steven~T Flammia and Yi-Kai Liu.
\newblock Direct fidelity estimation from few pauli measurements.
\newblock \emph{Physical review letters}, 106\penalty0 (23):\penalty0 230501, 2011.
\newblock \doi{10.1103/PhysRevLett.106.230501}.
\newblock URL \url{https://journals.aps.org/prl/abstract/10.1103/PhysRevLett.106.230501}.

\bibitem[Higham(1988)]{higham1988computing}
Nicholas~J Higham.
\newblock Computing a nearest symmetric positive semidefinite matrix.
\newblock \emph{Linear algebra and its applications}, 103:\penalty0 103--118, 1988.
\newblock \doi{10.1016/0024-3795(88)90223-6}.
\newblock URL \url{https://www.sciencedirect.com/science/article/pii/0024379588902236}.

\end{thebibliography}
\bibliographystyle{unsrtnat}

\newpage

\appendix
{\color{black}

\section{Useful lemmas}

\begin{lem}[Adapted from Theorem 2.1 in Ref.\ \cite{higham1988computing}]
Let $\Lambda$ be a Hermitian matrix with spectral decomposition  $\Lambda = \sum_{i\in\{0,1\}^n} \lambda_{i} \ketbra{\lambda_i}{\lambda_i}$.
Let $\Lambda_+$  and $\Lambda_-$ be the positive part and negative part of $\Lambda$ respectively:
\begin{align}
    &\Lambda_+ = \sum_{\substack{i\in\{0,1\}^n \\ \lambda_i >0}} \lambda_{i} \ketbra{\lambda_i}{\lambda_i},
    &\Lambda_- = \sum_{\substack{i\in\{0,1\}^n \\ \lambda_i <0}} \abs{\lambda_{i}} \ketbra{\lambda_i}{\lambda_i}.
\end{align}
Then $\Lambda_+$ is the closest positive semi-definite Hermitian matrix to $\Lambda$ with respect to the Frobenius norm, i.e.
\begin{align}
    \min_{\substack{M : \\ M = M^\dag \\ M \geq 0}} \norm{M - \Lambda}_2^2 = \norm{\Lambda_+ - \Lambda}_2^2= \norm{\Lambda_-}_2^2.
\end{align}
\end{lem}
\begin{proof}
Let $H = \sum_{i,j \in\{0,1\}^n} h_{ij} \ketbra{\lambda_i}{\lambda_j}$ an arbitrary positive semidefinite Hermitian operator.
We have
\begin{align}
     \norm{H - \Lambda}_2^2 = \sum_{\substack{i,j\in\{0,1\}^n \\i\neq j}} \abs{h_{ij}}^2 +  \sum_{i\in\{0,1\}^n} (h_{ii} - \lambda_i)^2
     \geq \sum_{\substack{i\in\{0,1\}^n \\ \lambda_i < 0} }  (h_{ii} - \lambda_i)^2 \geq \sum_{\substack{i\in\{0,1\}^n \\ \lambda_i < 0} }  \lambda_i^2,
\end{align}
where we used that fact that $h_{ii}\geq 0$ as $H$ is positive semi-semidefinite.
On the other hand, it is easy to see that $\Lambda_+$ saturates this lower bound:
\begin{align}
    \norm{\Lambda - \Lambda_+}_2^2 = \norm{\Lambda_-}_2^2 = \sum_{\substack{i\in\{0,1\}^n \\ \lambda_i < 0} }  \lambda_i^2.
\end{align}
This concludes the proof.
\end{proof}

\begin{lem}[Proposition 2 in Ref.\ \cite{melkani2020eigenstate}]
\label{lem:eigenstate}
Let $\rho$ be a quantum state. The unique closest pure state to $\rho$ with respect to the trace distance is its dominant eigenstate.
\end{lem}

\end{document}